\documentclass[]{article}

\usepackage{arxiv}
\usepackage[utf8]{inputenc} 
\usepackage[T1]{fontenc}    
\usepackage{hyperref}       
\usepackage{url}            
\usepackage{booktabs}       
\usepackage{amsfonts}       
\usepackage{nicefrac}       
\usepackage{microtype}      
\usepackage{lipsum}             
\usepackage{graphicx}
\usepackage{subfigure}
\usepackage{graphicx}
\usepackage{wrapfig}
\usepackage{fullpage}
\usepackage{multirow}
\usepackage{amsmath}
\usepackage{bm}
\usepackage{hyperref}
\usepackage{fancyhdr}
\usepackage{enumitem}
\usepackage{color}
\usepackage{framed}
\usepackage{nicefrac}
\usepackage{units}
\usepackage{amstext}
\usepackage{amsmath}
\usepackage{amssymb}
\usepackage{cite}
\usepackage{color}
\usepackage[usenames,dvipsnames,svgnames,table]{xcolor}
\usepackage{psfrag}
\usepackage{subfigure}
\usepackage{array}
\usepackage{threeparttable}
\usepackage{dcolumn}
\newcolumntype{d}{D{.}{.}{-1}}
\usepackage{enumitem}
\usepackage{titlesec}
\usepackage{placeins}
\usepackage{soul}
\usepackage{comment}
\usepackage[sort,numbers]{natbib}
\usepackage{multirow}
\usepackage{bm}

\title{A Moving Embedded Boundary Approach For The Compressible Navier-Stokes Equations In A Block-Structured Adaptive Refinement Framework}

\author{ {\hspace{4mm}Mahesh~Natarajan}\thanks{Corresponding author} \\
	\hspace{4mm}Computational Aerosciences\\
	\hspace{4mm}NASA Ames Research Center, M/S 258-2,\\
	\hspace{4mm}Moffett Field, CA 94035 \\
	\hspace{4mm}\texttt{mahesh.natarajan@nasa.gov} \\
	\And
	{\hspace{8mm}Ray Grout} \\
        \hspace{8mm}Computational Science Center\\
        \hspace{8mm}National Renewable Energy Laboratory\\
        \hspace{8mm}Golden, CO 80401 \\
        \hspace{8mm}\texttt{Ray.Grout@nrel.gov} \\
	\And
	{Weiqun Zhang} \\
        Center for Computational Sciences and Engineering\\
        Lawrence Berkeley National Laboratory\\
        Berkeley, CA 94720\\
        \texttt{WeiqunZhang@lbl.gov} \\
	\And
        {Marc Day} \\
        Computational Science Center\\
        National Renewable Energy Laboratory\\
        Golden, CO 80401 \\
        \texttt{Marc.Day@nrel.gov} \\
}

\hypersetup{
pdftitle={A template for the arxiv style},
pdfsubject={q-bio.NC, q-bio.QM},
pdfauthor={David S.~Hippocampus, Elias D.~Striatum},
pdfkeywords={Embedded boundary, moving bodies, cut-cell, block-structured adaptive refinement, AMReX},
}

\graphicspath{{./Images/}}

\begin{document}
\maketitle

{\color{black}
\begin{abstract}
A computational technique has been developed to perform compressible flow simulations involving moving boundaries using an embedded boundary approach within the block-structured 
adaptive mesh refinement framework of AMReX. A conservative, unsplit, cut-cell approach is utilized and a ghost-cell approach is developed for computing the flux on the moving, 
embedded boundary faces. Various test cases are performed to validate the method, and compared with analytical, experimental, and other numerical results in literature. Inviscid and 
viscous test cases are performed that span a wide regime of flow speeds -- acoustic (harmonically pulsating sphere), smooth flows (expansion fan created by a receding piston) and flows 
with shocks (shock-cylinder interaction, shock-wedge interaction, pitching NACA 0012 airfoil and shock-cone interaction). A closed system 
with moving boundaries -- an  oscillating piston in a cylinder, showed that the percentage error in mass within the system decreases with refinement, demonstrating the conservative 
nature of the moving boundary algorithm. Viscous test cases involve that of a horizontally moving cylinder at $Re=40$, an inline oscillating cylinder at 
$Re=100$, and a transversely oscillating cylinder at $Re=185$. The judicious use of adaptive mesh refinement with appropriate 
refinement criteria to capture the regions of interest leads to well-resolved flow features, and good quantitative comparison is observed with the results available in literature. 
\end{abstract}
}


\section{Introduction}

\par Flows with moving boundaries are of high significance for a variety of applications in a wide range of flow regimes -- low Reynolds number flows such as swimming fish
\citep{kern2006simulations,borazjani2009numerical}, moderate speed flows such as wind turbines \citep{arrigan2011control}, and high Mach number flows such as store separation 
from a tactical aircraft. 
\par The term immersed boundary (IB) method encompasses all such methods that simulate viscous flows with immersed (or embedded) boundaries on grids that do not conform to the shape of these boundaries. A variety of numerical approaches have been used for computing the flow around moving bodies, and they can be broadly classified as --  
body-fitted mesh methods, and embedded boundary (EB) methods. Body-fitted mesh approaches such as the overset/Chimera approach \citep{steger1983chimera} and the Arbitrary Lagrangian-Eulerian (ALE) approach \citep{donea1982arbitrary,hughes1981lagrangian} require expensive regeneration of the mesh as the body moves, which becomes cumbersome with complex body motion \citep{liu1998numerical,sahin2009arbitrary}. The embedded boundary method places the body within a Cartesian mesh which does not conform with the geometry, and hence does not require complex grid (re)generation. But such an approach requires sophisticated numerical schemes for computing the terms in the governing equations on and close to the moving body, and the imposition of boundary conditions at the embedded boundary.

\par The earliest work in immersed boundary methods is the diffuse interface approach by \citet{peskin1972flow}, in which a two dimensional simulation of flow in a heart valve 
was performed. The boundary was replaced by a force field defined on the mesh points of the rectangular domain which was calculated from the configuration of the boundary. 
In order to link the representations of the boundary and fluid, since boundary points and mesh points need not coincide, a semi-discrete analog of the delta function 
was introduced. Later, \citet{goldstein1993modeling} used the idea to impose a force field along a surface and can vary in space and time, with a magnitude and direction opposing the 
local flow was applied to bring the flow to rest. Two dimensional flow around cylinders and three dimensional turbulent channel flow in riblet-covered surface were simulated. Later, a direct forcing 
approach was developed for rigid body problems, in which the forces at immersed boundaries were calculated based on the temporally discretized momentum equation \citep{mohd1997combined,fadlun2000combined,
uhlmann2005immersed,su2007immersed}. Another class of immersed boundary methods are the sharp-interface methods that have no smearing, and thus have an accurate representation of the geometry. The Ghost fluid Method (GFM) belongs to the class of sharp interface methods. \citet{fedkiw1999non} and \citet{fedkiw2002coupling} developed the ghost fluid method for multiphase flows, in which the interface is tracked with a level set function, which gives the exact sub-cell interface location, and at the interface, an approximate Riemann problem is solved. Further improvements were made in the method by  \citet{liu2003ghost} and \citet{terashima2009front}. \citet{tseng2003ghost} developed a ghost cell IB method for flows with  complex geometries using a reconstruction procedure to determine the values in the ghost cells for enforcing the boundary conditions. \citet{mittal2008versatile} used the ghost-cell approach to develop a sharp interface method for incompressible flows with moving and deforming bodies. More recently, \citet{brahmachary2018sharp}  developed a sharp interface method for high-speed, compressible, inviscid flows, which imposed the boundary conditions on the body geometry, and 
used a novel reconstruction procedure to compute the solution in the vicinity of the solid-fluid interface. \citet{al2017versatile} developed a multidimensional partial differential extrapolation approach to reconstruct the solution in the ghost fluid regions and imposing boundary conditions on the fluid-solid interface, coupled with a multi-dimensional algebraic interpolation for freshly cleared cells.

\par  In contrast to the diffuse interface methods, cut-cell methods are piecewise linear "accurate" and are strictly conservative. 
One of the early studies of the cut-cell approach \citep{clarke1986euler} computed the flow over single- and multi-element airfoils on two-dimensional 
Cartesian grids using a finite volume approach. Later, \citet{pember1995adaptive} developed a Godunov method which used a volume-of-fluid approach 
for the fluid-body interface. \citet{yang1997cartesian1} developed a cut-cell method for static boundaries and later extended to 
moving boundaries \citep{yang1997cartesian2,yang2000calculation}, in which the upwind fluxes on the interfaces of static cells were 
updated using an HLLC approximate Riemann solver and an exact Riemann solution for a moving piston is used to update moving solid boundaries. One of the earliest studies 
that employed adaptive mesh refinement in a cut-cell, Cartesian framework was by \citet{dezeeuw1993adaptively}. They performed compressible flow simulations on single- 
and multi-element airfoils using Roe's approximate Riemann solver and a solution-adaptive refinement to resolve the 
high gradient regions. The cut-cell approach was used by \citet{hu2006conservative} to solve multi-fluid and complex moving geometry problems for compressible flows 
in two dimensions. Other early studies that used the cut-cell approach involve simulations of viscous incompressible flows with complex boundaries \citep{ye1999accurate}, forced and 
natural convection problems in an incompressible framework \citep{udaykumar1996elafint} and computation of solid-liquid phase fronts \citep{udaykumar1999computation}. 
There are a number of studies that used the approach for simulations of incompressible, two- and three-dimensional flows over complex geometries \citep{almgren1997cartesian,kirkpatrick2003representation,popinet2003gerris,meyer2010conservative}. \citet{hartmann2009general} developed a cut-cell approach using ghost cells which can be freely 
positioned in space, hence making the approach flexible in terms of the shape and size of the embedded boundaries. A linear least-squares method is used to reconstruct the cell center 
gradients in irregular regions of the mesh to compute the flux at the surface. \citet{cheny2010ls} developed a Cartesian grid/immersed boundary method 
for incompressible viscous flows in two-dimensions for moving complex geometries.

\par The use of explicit time discretization schemes in a cut-cell approach leads to the classical small-cell issue. The Courant-Friedrichs-Lewy  (CFL) restriction is based on the fluid volume in a cell, and hence will 
cause the admissible time step to be extremely small for cells with low volume fractions. \citet{noh1963cel} did some of the earliest work on this 
using a cell-merging and redistribution technique. The cell-merging technique identifies a cluster of cells around a cut-cell and merges them to form a larger control volume, 
and computes the flux update for this newly formed, merged control volume. The cell-merging technique was used by \citet{bayyuk1993simulation} for moving and deforming bodies, and by \citet{quirk1994cartesian} in a block-structured  adaptive framework. This technique has been used for compressible flows with moving complex geometries \citep{yang2000calculation}, and compressible, multi-solid/fluid 
systems \citep{barton2011conservative}. Though widely used, \citet{schneiders2013accurate} showed that the technique can lead to unphysical oscillations for moving boundary problems, and developed a 
method that used a smooth discrete formulation when cells are freshly cleared or covered by the moving boundary. \citet{kirkpatrick2003representation} developed a novel cell-linking algorithm, which avoids the complexities involved with the cell-merging approach. Another technique to address the small-cell issue is the $h$-box method 
\citep{berger1989adaptive,leveque1988cartesian,berger2003h}, which approximates the numerical fluxes at the interfaces of a small cell based on initial values specified over regions of length $h$ - size of a regular grid cell, which will allow for the time step to be based on the regular grid size, rather than the small cell. In the current work, we use the flux redistribution technique  \citep{pember1995adaptive,colella2006cartesian}.  This involves updating the cut-cells using a hybrid divergence -- a linear combination of the flux divergence of the cut-cell and the non-conservative divergence computed using the neighboring cell divergences, and then redistributing the ``excess" quantity of the conserved variables to a neighborhood region of the cut-cell to ensure global conservation. \citet{ji2010numerical} used a cut-cell approach for detonation simulations with a cell-merging technique to avoid the small-cell issue. \citet{muralidharan2016high} developed a novel cell clustering approach to 
treat the small-cell issue and maintain stability. The central idea was to employ a $k$-order, polynomial piecewise approximation of the flow solution to a cluster of cells, and also extended it 
for moving boundary problems \citep{muralidharan2018simulation}. More recently, \citet{sharan2020stable} have developed a method for deriving higher-order, provably stable schemes that avoids the small cell issue in a finite difference, cut-cell framework. 

\par Immersed boundary methods offer a significant advantage in the simulation of flows with moving boundaries, and there have been a number of studies in this context \citep{gilmanov2005hybrid,yang2006embedded,khalili2018immersed,mittal2008versatile}. But the cut-cell approach presents additional challenges when applied to moving boundary problems. In particular, 
cut-cell based approach to high-speed, moving body problems require special care, and hence, the number of studies is limited. One of the earliest studies was by \citet{yang1997cartesian2,yang2000calculation}, in which the finite volume, unsplit MUSCL--Hancock method of the Godunov type was modified for moving boundaries, in conjunction with a  cell-merging technique to maintain numerical stability in the presence of arbitrarily small cut cells to ensure strict conservation at the moving boundaries. \citet{schneiders2013accurate} showed that the widely used cell-merging technique creates unphysical oscillations for moving boundary problems, and developed an accurate moving boundary formulation based on the varying discretization operators which avoids the oscillations. \citet{muralidharan2018simulation} developed a second-order cut-cell approach for flows with moving boundaries enforcing strict conservation using the small-cell clustering algorithm \citep{muralidharan2016high}. The cell clustering algorithm also preserves the smoothness of solution near moving surfaces. \citet{bennett2018moving} employed a directional operator splitting method by extending the cut-cell approach for static walls from \citet{klein2009well}. The scheme calculates the fluxes needed for a conservative update of the near-wall cut-cells as linear combinations of fluxes, which were obtained without regard to the small sub-cell problem, from a one-dimensional extended stencil. \citet{tan2011high} developed a high order numerical boundary condition for compressible inviscid flows involving complex moving geometries. Their methodology was based on finite difference methods which was an extension of the inverse Lax–Wendroff procedure \citep{tan2010inverse} for conservation laws in static geometries.

\par Fluid flow simulations with predictive capability require high resolution and superior parallel performance of the flow solvers, and this inevitably leads to the need for adaptive mesh 
refinement (AMR) strategies. In particular, for moving body problems, where the flow features that are of interest constantly change with time, solution-adaptive refinement can result in a 
significant reduction in compute time and memory. However, issues of deciding when and how to adapt, and keeping track of the evolving mesh, have to be addressed carefully for scalable performance. The cell-based tree data structure, which is very widely used for AMR, is very flexible, and provides a systematic way to keep track of the mesh. However, since each node of a cell-based tree is a single cell, computations suffer from significant overhead due to indirect addressing and lower FLOP rates are achieved \citep{stout1997adaptive}. Similar performance issues arise with general unstructured grids as well. Block-structured adaptive mesh refinement (SAMR) is more advantageous in many respects compared to the cell-based tree and unstructured grids \citep{stout1997adaptive}. Loop and cache optimizations can be performed over the arrays of cells when using adaptive blocks. The cost of neighbor pointers are amortized over entire arrays, and their ghost cell to computational cell ratio is superior to other data structures. Since the blocks permit refinement of larger multi-cell regions at a time, mesh adaptation is required less frequently than other data structures, which reduce computational cost. A number of frameworks exist for SAMR --  BoxLib, Cactus, Chombo, Enzo, FLASH, and Uintah, are some of the publicly available frameworks. A survey of these frameworks can be found in \citet{dubey2014survey}.
\par In the current work, we use AMReX \citep{amrex,AMReX_JOSS} -- a publicly available software framework for building massively parallel SAMR applications with C\texttt{++} and Fortran interfaces. The features include parallelization via flat MPI, OpenMP, hybrid MPI/OpenMP, or MPI/MPI, GPUs, logical tiling of grids, support for multilevel mesh operations such as coarsening/interpolation between different levels and ghost cell filling, multigrid solvers for Poisson and Helmholtz equations, sub-cycling time-stepping algorithm, and support for particles and particle-mesh operations. The block-structured adaptive refinement strategy is based on the work of \citet{berger1982adaptive}, and has been subsequently employed in various studies \citep{berger1984adaptive,berger1989local,bell1994three,berger1991algorithm,brown2000adaptive}.

\par In this paper we devise a strategy for moving boundary problems for the compressible, Navier-Stokes equations within the finite-volume, block-structured adaptive mesh refinement 
framework of AMReX. The paper is organized as follows. Section~\ref{sec:flow_solver} describes the algorithm for the cut-cell approach with static boundaries and its extension to
moving boundaries. The numerical results for one-, two-, and three-dimensional compressible flow problems and comparison with experiments and other cases in literature are demonstrated in Section~\ref{sec:testcases}. Conclusion are given in Section~\ref{sec:conclusions}.

\section{The flow solver}\label{sec:flow_solver}
The flow solver employs a finite volume, second order method to solve the compressible Navier-Stokes equations, given by 
\begin{eqnarray*}
\frac{\partial \bm{Q}}{\partial t} + \frac{1}{V}\int\limits_A\bm{F}\cdot\bm{n}\,dA = S,
\end{eqnarray*}
where $\bm{Q}\equiv(\rho,\rho \bm{u},\rho E$) is the vector of intensive conserved quantities -- density, momentum  and total energy, averaged over the cell with volume $V$, $\bm{F}$ 
is the flux vector (including both inviscid and viscous fluxes), $A$ denotes the surface of the finite volume region, $\bm{n}$ is the outward normal to the surface and $S$ denotes the source terms. 
A Godunov approach is used to discretize the advection terms and Riemann solver evaluates the single-valued conservative fluxes on each cell face based on Van Leer-limited solution 
gradients of the transformed characteristic variables. The temporal discretization uses the second-order Runge-Kutta method. 
The flow solver \citep{amrexcns} is implemented within the block-structured adaptive mesh refinement (AMR) framework of AMReX \citep{AMReX_JOSS}. An embedded boundary (EB) approach is used to 
modify the finite volume discretization near complex geometries \citep{colella2006cartesian,modiano2000higher}. The EB approach uses the volume fraction of the cut-cells, the area  
fraction of the cut-cell faces, the face normals, and the fluid volumetric centroid for the flux computation. In a naive formulation of the embedded boundary approach, the update in a cut-cell is given by (the source terms are omitted)
\begin{eqnarray*}
\bm{Q}^{n+1} = \bm{Q}^{n} - \frac{\Delta t}{\alpha V}\int\limits_A\bm{F}\cdot\bm{n}\,dA, 
\end{eqnarray*}
where $\alpha$ is the volume fraction of the cut-cell. If we use an explicit time advancement scheme, then this leads to the classical small cut-cell issue. As $\alpha\rightarrow0$, the CFL restriction results in the admissible time step 
$\Delta t\rightarrow0$. The technique of flux redistribution is utilized to treat the issue \citep{colella2006cartesian,pember1995adaptive}, and is described here for completeness. 
This involves a two-step procedure -- a hybrid divergence update of the cut-cells, and a redistribution of ``excess'' in the conserved quantity to the neighboring cells. The  hybrid divergence is a volume fraction weighted average of the 
conservative ($c$) and non-conservative divergences ($nc$), and the update with the hybrid divergence is given by
\begin{eqnarray*}
\bm{Q}^{n+1} = \bm{Q}^{n} - \Delta t(\alpha (\nabla\cdot\bm{F})_c +  (1-\alpha)(\nabla\cdot\bm{F})_{nc}).
\end{eqnarray*}
The conservative divergence is given by the standard finite-volume expression
\begin{eqnarray*}
(\nabla\cdot\bm{F})_c=\frac{1}{\alpha V}\int\limits_A\bm{F}\cdot\bm{n}\,dA,
\end{eqnarray*}
and hence the update can be written as
\begin{eqnarray*}
\bm{Q}^{n+1} = \bm{Q}^{n} - \Delta t\Bigg(\frac{1}{V}\int\limits_A\bm{F}\cdot\bm{n}\,dA + (1-\alpha)(\nabla\cdot\bm{F})_{nc}\Bigg),
\end{eqnarray*}
thereby avoiding the volume fraction $\alpha$ appear explicitly in the denominator. Note that this approach circumvents the CFL restriction that leads to vanishing small time steps 
for small $\alpha$, however it is not strictly conservative. The non-conservative divergence contribution is computed as a weighted average of the  conservative divergences of the neighboring cells as 
\begin{eqnarray*}
(\nabla\cdot\bm{F})_{nc} = \cfrac{\sum\limits_{i\epsilon\text{N}}\alpha_iV_i(\nabla\cdot\bm{F})_c}{\sum\limits_{i\epsilon\text{N}}\alpha_iV_i},
\end{eqnarray*} 
where $N$ is the set of all reachable cells containing fluid in a 3$\times$3$\times$3 cell neighborhood. A flux redistribution technique following that described in \citet{colella2006cartesian} 
is used to modify this update and ensure conservation. Had we only used the conservative divergence for the cut-cell $i$, a conservative update would be
\begin{eqnarray*}
\bm{Q}_i^{n+1} = \bm{Q}_i^n - \Delta t (\nabla\cdot\bm{F})_c.
\end{eqnarray*}
The hybrid update instead is
\begin{eqnarray*}
\bm{Q}_i^{n+1} = \bm{Q}_i^n - \Delta t (\alpha_i(\nabla\cdot\bm{F})_c + (1-\alpha_i)(\nabla\cdot\bm{F})_{nc}).
\end{eqnarray*}
The latter expression leads to  excess ``mass'' (mass refers to any of the conserved variables) in the EB cell given by
\begin{eqnarray*}
\delta \bm{M}_\mathrm{excess} = -\Delta t(1-\alpha_i)((\nabla\cdot\bm{F})_{nc}-(\nabla\cdot\bm{F})_c).
\end{eqnarray*}
This excess mass is subtracted from the neighbors of the cut-cell. Let $\delta \bm{M}_i=-\alpha_i\delta \bm{M}_\mathrm{excess}$, and the total mass to be redistributed is
\begin{eqnarray}\label{eqn:redistribution}
\delta \bm{M}_i V_i = \sum\limits_{j\epsilon N} \alpha_j\delta \bm{M}_{ij} V_j,
\end{eqnarray}
where $N$ is the set of neighboring cells and  $\delta \bm{M}_{ij}$ is the portion of redistributed mass that is to be added to the $j^\mathrm{th}$ neighbor. Now, let us assign weights $w_j$ to the cells, which will determine the amount of redistributed quantity it gets. A cell with weight $w_j$ gets a volume averaged redistributed quantity $\delta\bm{M}_{ij}=w_j\delta \bm{M}_i$, and 
Eq.~\ref{eqn:redistribution} implies $\sum\limits_{j\epsilon N} \alpha_jw_j=1$. There are a number of choices for how to partition the redistribution.  If, for example, the redistribution is volume-weighted (assuming equal volumes for all cells), and 
\begin{eqnarray*}
w_j = \frac{\alpha_j}{\sum\limits_{k\epsilon N} \alpha_k^2},
\end{eqnarray*}
note that $\sum\limits_{j\epsilon N} \alpha_jw_j=1$. Other strategies include upwind and/or mass weighting the distribution.  For simplicity here, we select the volume-weighting scheme,
\begin{eqnarray*}
\delta \bm{M}_{ij} = \frac{\alpha_j}{\sum\limits_{k\epsilon N} \alpha_k^2}\delta \bm{M}_i.
\end{eqnarray*}
Hence, the final update for every cell $p$ in the domain (fluid cells and cut-cells) is given by
\begin{eqnarray*}
{\bm{Q}_p^{n+1}}_\mathrm{final} = \bm{Q}_p^{n+1} + \sum\limits_{i\epsilon  \mathrm{N}_\mathrm{cut-cells}}\delta\bm{M}_{ip},
\end{eqnarray*}
where $\bm{Q}_p^{n+1}$ is the update obtained using the divergence (conservative divergence for regular fluid cells and hybrid divergence for cut-cells), and N$_\mathrm{cut-cells}$ is the set of all neighboring cut-cells  of cell  $p$, which contributes a redistributed mass of $\delta\bm{M}_{ip}$ to cell $p$.

\subsection{Moving boundary method}\label{sec:movingEB}
In this section, we develop the moving boundary formulation for the compressible Navier-Stokes equations. From the Reynolds transport theorem for 
the general case of moving/deformable control volumes, we have
\begin{eqnarray}
\frac{d}{dt}\Bigg(\int\limits_{\Omega(t)}f\,dV\Bigg) = \int\limits_{\Omega(t)}\frac{\partial f}{\partial t}\,dV + \int\limits_{\partial\Omega(t)}f\bm{u}\cdot\bm{n}\,dA = \text{RHS},
\end{eqnarray}
where $f$ is a conserved variable, $\Omega(t)$ and $\partial\Omega(t)$ are the temporally varying volume and surface of the control volume respectively, $\bm{u}$ is the velocity of the fluid on the control surface, $\bm{n}$ is the outward unit normal to the control surface, and RHS is the contribution of the inviscid (pressure part) and viscous fluxes including source terms. For a control volume with volume fraction $\alpha$ (the fraction of cell volume occupied by the fluid), we have
\begin{eqnarray}
\overline{\frac{\partial f}{\partial t}}\alpha\Delta V + \int\limits_{\partial\Omega(t)}f\bm{u}\cdot\bm{n}\,dA = \text{RHS}.
\end{eqnarray}
where $\Delta V=\Delta x\Delta y\Delta z$ is the cell volume, and $\overline{(\cdot)}$ denotes the cell averaged value. First-order discretization in time (for simplicity) gives
\begin{eqnarray}\label{eqn:Update}
\frac{\overline{f}^{n+1}-\overline{f}^n}{\Delta t}\alpha\Delta V &=& - \int\limits_{\partial\Omega(t)}f\bm{u}\cdot\bm{n}\,dA + \text{RHS}\nonumber, \\
\overline{f}^{n+1} &=& \overline{f}^{n} + \frac{\Delta t}{\alpha^{n}\Delta V}\Bigg(-\int\limits_{\partial\Omega(t)}f\bm{u}\cdot\bm{n}\,dA + \text{RHS}\Bigg),
\end{eqnarray}
where we have chosen $\alpha=\alpha^{n}$, though other approximations are possible. In our simulations a second-order Runge-Kutta scheme 
is employed, and the above time discretization simplifies the presentation.

\begin{figure}
\subfigure[]
{
\includegraphics[trim=1.0cm 6.0cm 18cm 1.0cm,clip=true,scale=0.5]{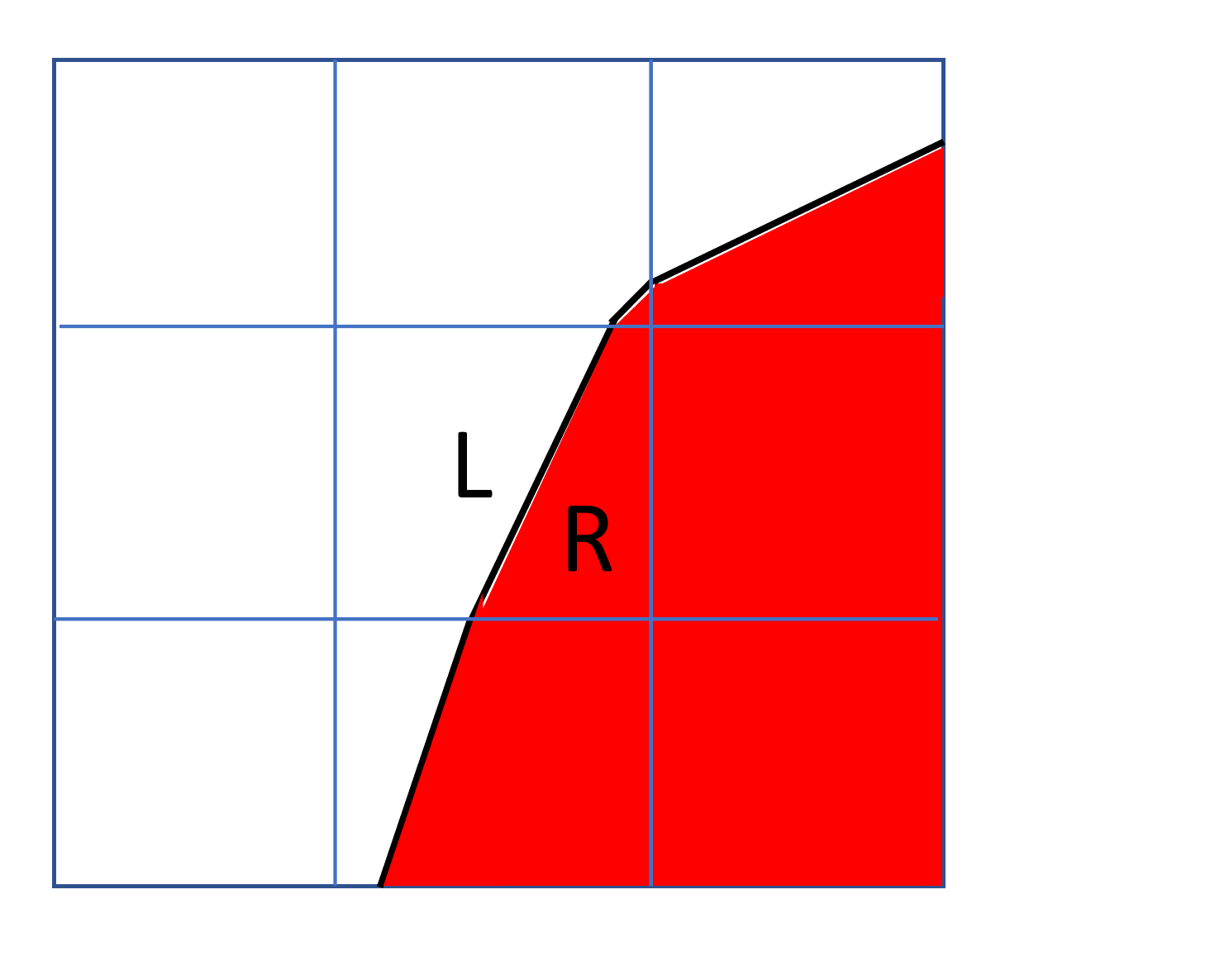}
\label{fig:EB1}
}
\subfigure[]
{
\includegraphics[trim=1.0cm 6.0cm 15cm 1.0cm,clip=true,scale=0.5]{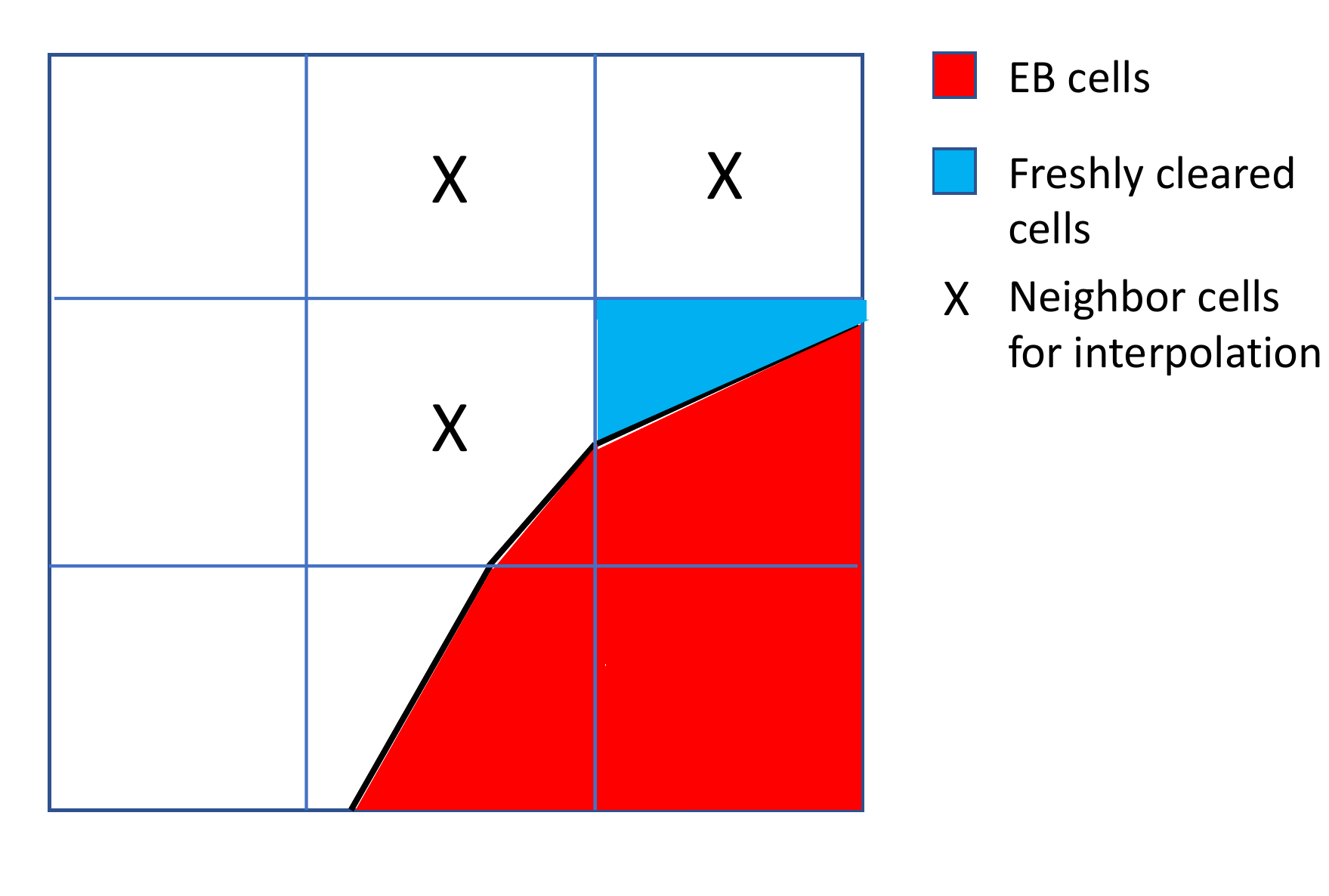}
\label{fig:EB2}
}
\caption{(a) The EB surface at $t^n$ and (b) $t^{n+1}$ showing the EB cells (red), freshly cleared cells (blue) and the neighbors of the freshly cleared cell ($\times$).}
\label{fig:EB}
\end{figure}

The compressible flow equations (viscous terms omitted for simplicity) in the finite volume formulation are
\begin{eqnarray}
\frac{\partial\rho}{\partial t} + \frac{1}{\alpha\Delta V}\int\limits_{\partial\Omega(t)}\rho\bm{u}\cdot\bm{n} \,dA&=& 0\nonumber\\
\frac{\partial\rho\bm{u}}{\partial t} + \frac{1}{\alpha\Delta V}\int\limits_{\partial\Omega(t)}(p\bm{n}+\rho\bm{u}\bm{u}\cdot\bm{n})\,dA &=& 0\nonumber\\
\frac{\partial\rho E}{\partial t} + \frac{1}{\alpha\Delta V}\int\limits_{\partial\Omega(t)}(p+\rho E)\bm{u}\cdot\bm{n}\, dA &=& 0,
\end{eqnarray}
where the conservative variables $(\rho,\rho\bm{u},\rho E$) are cell-averaged. The finite volume solver computes the flux at every face of the control volume. Special care is needed for treating the EB faces especially when the surface is moving. In the present implementation, a Riemann solver takes in the left and right states for a face, and computes the upwind flux on that face. 
For a stationary EB, a Riemann-like problem is constructed consistent with the no-slip boundary condition. "Left" and "right" states are generated using the pressure and density from the 
cut fluid cell, and the normal velocity at the EB surface is set to satisfy the inviscid no-penetration condition ($u_n$ (body) = $-u_n$ (cut-cell)). For an EB that moves at a velocity 
$\bm{u}_w$, the the velocity on the EB surface $\bm{u}_s$ should satisfy the inviscid no-penetration condition given by
\begin{eqnarray}\label{eqn:NoPenetration}
\bm{u}_s\cdot\bm{n}=\bm{u}_w\cdot\bm{n}.
\end{eqnarray}
The velocity for the ghost point inside the EB is given by a reflective or mirroring condition as \citep{bennett2018moving,toro2013riemann,forrer1999flow}
\begin{eqnarray*}
\bm{u}_g = \bm{u}-2(\bm{u}\cdot\bm{n})\bm{n} + 2(\bm{u}_w\cdot\bm{n})\bm{n},
\end{eqnarray*}
and hence the velocity at the EB surface which moves at a velocity $\bm{u}_w$ is given by
\begin{eqnarray}\label{eqn:us}
\bm{u}^s = \frac{\bm{u} + \bm{u_g}}{2} = \bm{u}-(\bm{u}\cdot\bm{n})\bm{n} + (\bm{u}_w\cdot\bm{n})\bm{n},
\end{eqnarray}
which satisfies the no-penetration condition at the EB boundary given by Eqn.~\ref{eqn:NoPenetration}. With $\bm{u}^s$ defined by Eqn.\ref{eqn:us}, 
the flux on the moving EB face is
\begin{eqnarray}\label{eqn:MovingEBFlux}
F_\rho &=&\rho \bm{u}^s\cdot\bm{n}\nonumber\\
F_{\rho\bm{u}} &= &p_{gd}\bm{n} + (\rho u)_{gd}\bm{u}^s\cdot\bm{n}\nonumber\\
F_{\rho E} &=& (p_{gd}+\rho E)\bm{u}^s\cdot\bm{n} = (p_{gd}+p_{gd}+\frac{1}{2}\rho|\bm{u}^s|^2)\bm{u}^s\cdot\bm{n}.
\end{eqnarray}
where $p_{gd}$ and $(\rho u)_{gd}$ are the pressure and momentum flux evaluated by the Riemann solver using the left and right hand states at the EB surface (Fig.~\ref{fig:EB1}).
\par For flows with a moving EB, we have to deal with the issue of freshly-cleared cells 
(FC) i.e. a cell covered by the EB at $t^n$ becomes a cut-cell at $t^{n+1}$. Such cells do not have a history of fluid data, and hence special care needs to be 
taken to compute the data on these cells. Fig.~\ref{fig:EB1} and 
Fig.~\ref{fig:EB2} show the position of the boundary at $t^n$ and $t^{n+1}$ respectively. 
The conserved state of a FC cell at $t^{n+1}$ (shown in blue in Fig.~\ref{fig:EB2}) is initialized using a volume-weighted average of the neighboring valid cells at $t^n$ as
\begin{eqnarray*}\label{eqn:fcc_interp}
f_{\text{FC}} = \cfrac{\sum\limits_{i\epsilon\text{N}}\alpha_if_iV_i}{\sum\limits_{i\epsilon\text{N}}\alpha_iV_i},
\end{eqnarray*}
where $N$ is the set of all valid cells (cut-cells and fluid cells) at $t^n$. To justify this approach, consider a system with fluid moving at a uniform velocity of $U\bm{s}$, where $\bm{s}$ is the unit normal in the flow direction, and a body moving at the same velocity as the fluid. In this case, the flow field should remain unchanged with time. Fluid cell updates are trivial for this case, and hence we consider the fluxes on a cut-cell. 
Using $\bm{u}=U\bm{s}$, and since the flow quantities are constant everywhere in the domain (since density and pressure on the EB face are the same as the adjoining cut-cell fluid), we have the flux contribution of the density, momentum and energy fluxes as 
\begin{eqnarray*}
\text{F}_\rho &=& \int\limits_{\partial\Omega(t)} \rho U\bm{s}\cdot\bm{n}\, dA = \rho  U\int\limits_{\Omega(t)}\nabla\cdot\bm{s} \,dV\\
\text{F}_{\rho \bm{u}} &=& \int\limits_{\Omega(t)}p\bm{n} + \rho U^2\bm{s}\bm{s}\cdot\bm{n} \, dA = p\int\limits_{\partial\Omega(t)}\bm{n}\,dA + \rho U^2\int\limits_{\Omega(t)}\nabla\cdot\bm{s}\bm{s} \, dV\\
\text{F}_{\rho E} &=& \int\limits_{\Omega(t)}(p+\rho E)U\bm{s}\cdot\bm{n} \, dA = (p+\rho E)U\int\limits_{\Omega(t)}\nabla\cdot \bm{s}\,dV.
\end{eqnarray*}
All the above integrals evaluate to exactly zero discretely, since $\bm{s}$ is a constant vector and $\int\limits_{\partial\Omega(t)}\bm{n}\,dA=\bm{0}$ for a closed control volume $\Omega(t)$. This ensures that the contribution of the fluxes to the field update is exactly zero, and hence the field remains unchanged.

\begin{figure}
\centering
\includegraphics[trim=15.0cm 3.0cm 1.0cm 9.0cm,clip=true,scale=0.5]{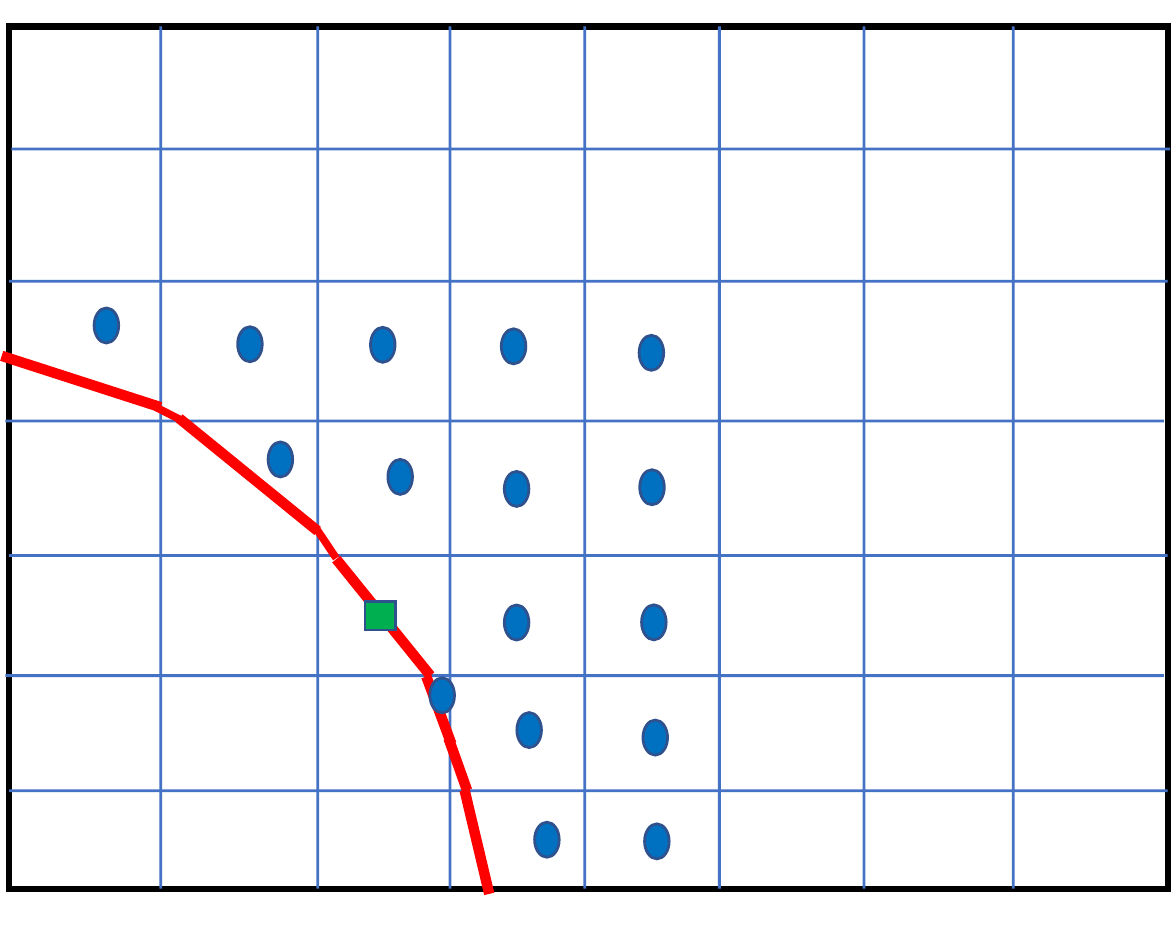}
\caption{ The neighborhood stencil for a least squares method that computes the gradients at the EB surface (shown by the green square) by minimizing an error norm (see Appendix) over 
a neighborhood (shown by blue circles) \citep{schwartz2006cartesian}. This stencil has a cluster size of 2, which means that the neighborhood region is given by the reachable 
fluid-containing cells in the index region $(i-2:i+2,j-2:j+2,k-2:k+2)$.}
\label{fig:EBGradient_LS}
\end{figure}

{\color{black}

\subsection{Treatment of the viscous terms at the embedded boundary}
A number of different approaches have been used in the literature for the gradient evaluation at the EB \citep{johansen1998cartesian,schwartz2006cartesian,muralidharan2017embedded}.
 For the time varying momentum equation
\begin{eqnarray*}
\frac{\partial\rho\bm{u}}{\partial t} + \frac{1}{\alpha\Delta V}\int\limits_{\partial\Omega(t)}(p\bm{n}+\rho\bm{u}\bm{u}\cdot\bm{n})\,dA = \int\limits_{\partial\Omega(t)} \bm{\tau}\cdot\bm{n}\,dA,
\end{eqnarray*}
we use  a 3$^\mathrm{rd}$ order least squares formulation to approximate the stress tensor, $\tau$, at the EB surface. The details of the formulation are given in the Appendix. To determine the gradients on the EB face (green square in Fig.~\ref{fig:EBGradient_LS}), the corresponding least squares formulation results in a linear system of equations at each EB face, given by 
\begin{eqnarray*}
A_\mathrm{LS} X = b_\mathrm{LS}(\phi_\mathrm{nei},\phi_0).
\end{eqnarray*}
For a 3$^\mathrm{rd}$ order formulation, $A_\mathrm{LS}$ is a 9$\times$9 matrix that is dependent only on the mesh in the neighborhood region, $X$ is the solution vector 
that contains the derivatives of variable $\phi$ -- first and second (including mixed derivatives), and $b_\mathrm{LS}$ is a 9$\times$1 vector 
which is a function of the values of $\phi_\mathrm{nei}$ in the neighborhood region (blue circles in Fig.~\ref{fig:EBGradient_LS}), and the value $\phi_0$ at the 
EB face (the green square in Fig.~\ref{fig:EBGradient_LS}). While evaluating gradients of velocity, $\phi_0$ will take the value of the 
velocity of the point on the moving body. Note that the values of $\phi_\mathrm{nei}$ are assumed to be positioned at the volumetric centroids of the fluid region of the cell (blue circles in 
Fig.~\ref{fig:EBGradient_LS}). The 9 $\times$ 9 system of equations is solved using LAPACK \citep{anderson1990lapack}. In the present work, we use a cluster size of 3, which means that for evaluating 
the gradients at the EB face on cell $(i,j,k)$, the neighborhood region is given by the reachable fluid-containing cells in the index region $(i-3:i+3,j-3:j+3,k-3:k+3)$.
}

\section{Test cases}\label{sec:testcases}
Several inviscid and viscous test cases of increasing complexity are performed to validate the numerical method for moving embedded boundaries.
\subsection{Harmonically pulsating sphere}
\begin{figure}[htpb!]
\centering
\subfigure[]
{
\includegraphics[trim=0.0cm 2.0cm 0.0cm 2.0cm, clip=true,scale=0.2]{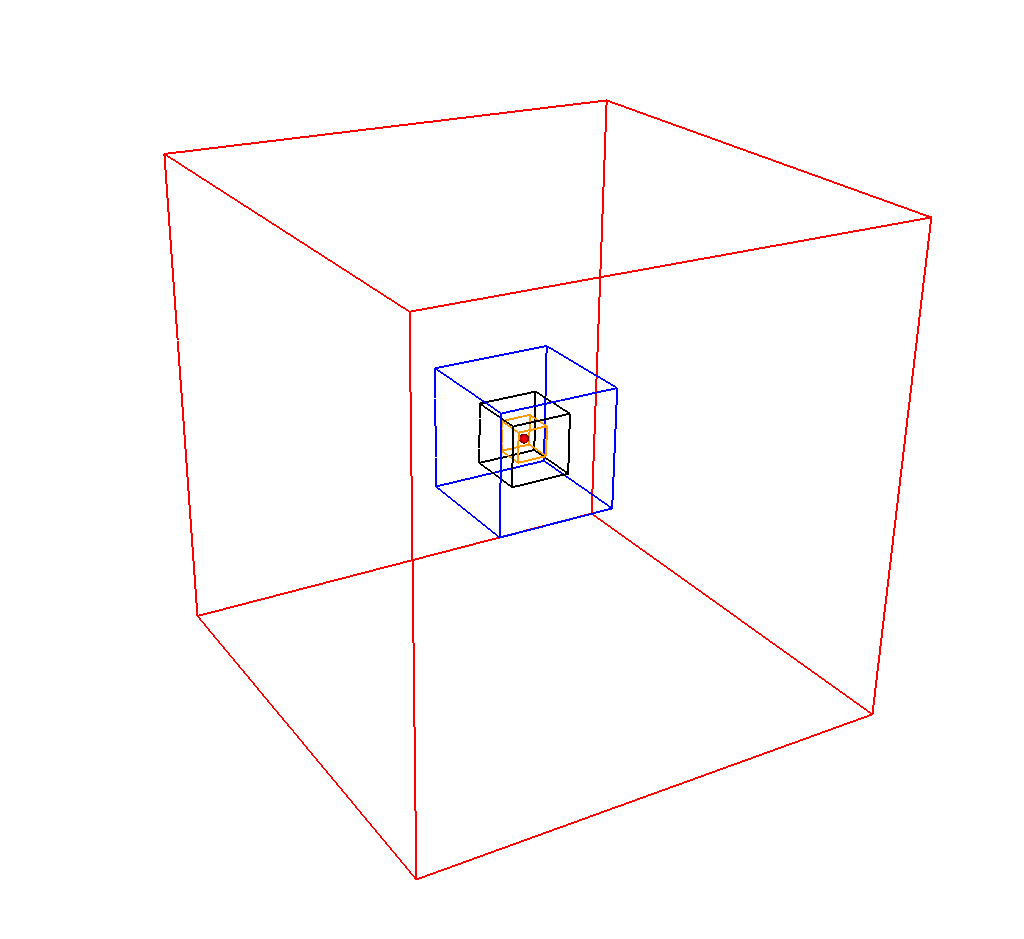}
\label{fig:PulsatingSphere_Levels}
}
\subfigure[]
{
\includegraphics[trim=0.0cm 2.0cm 0.0cm 2.0cm, clip=true,scale=0.2]{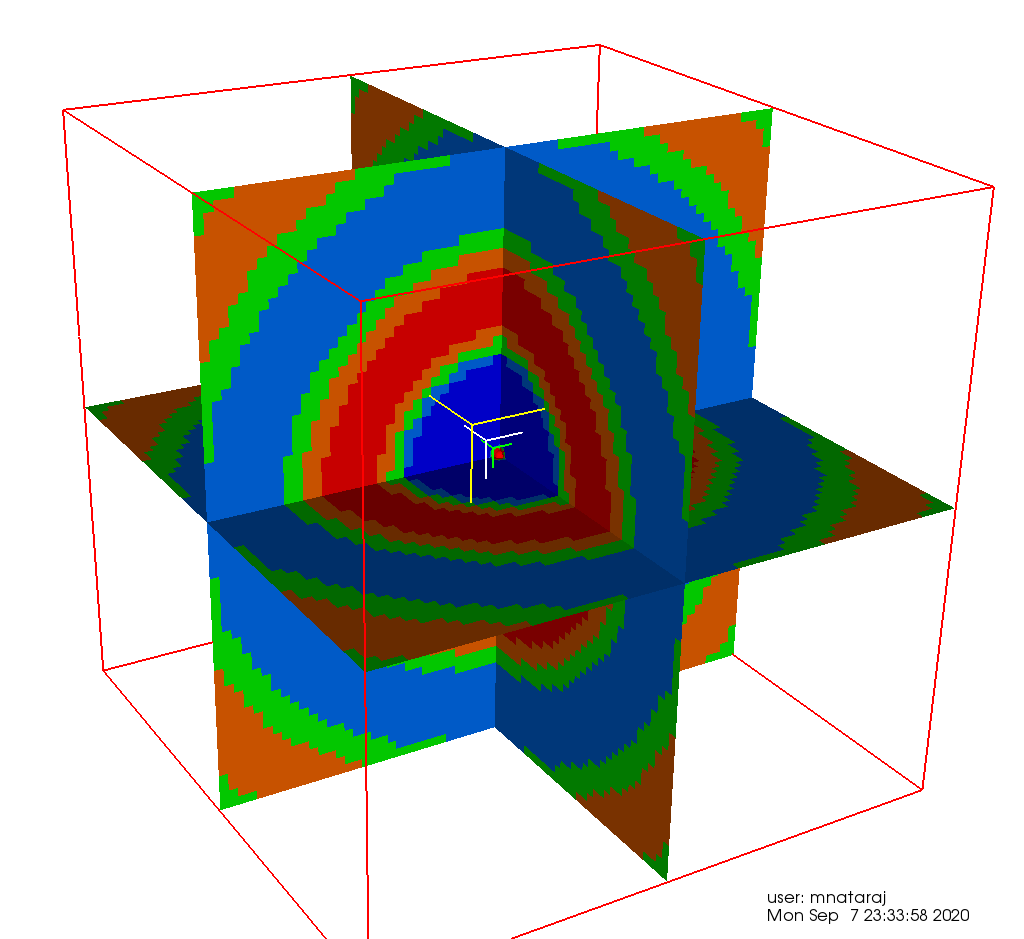}
\label{fig:PulsatingSphere_pprime_contours}
}
\subfigure[]
{
\includegraphics[trim=0.0cm 0.0cm 0.0cm 0.0cm, clip=true,scale=0.35]{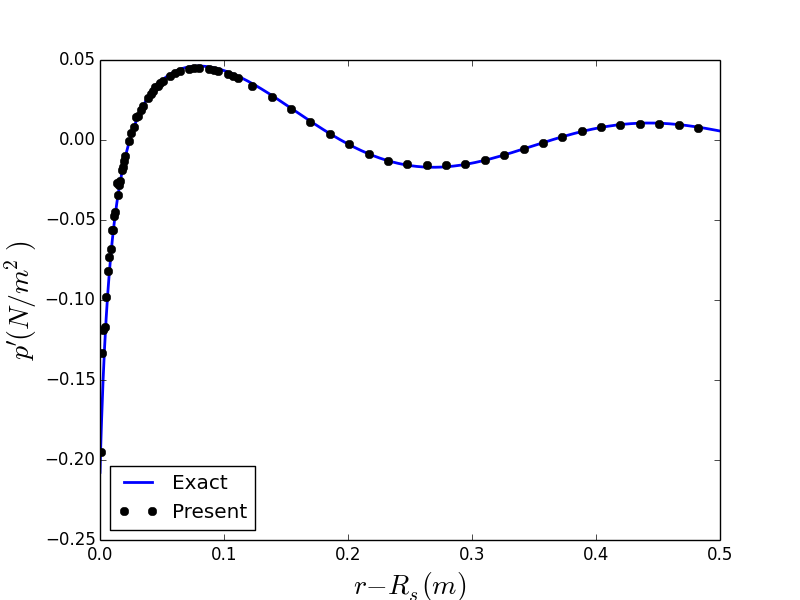}
\label{fig:PulsatingSphere0037}
}
\subfigure[]
{
\includegraphics[trim=0.0cm 0.0cm 0.0cm 0.0cm, clip=true,scale=0.35]{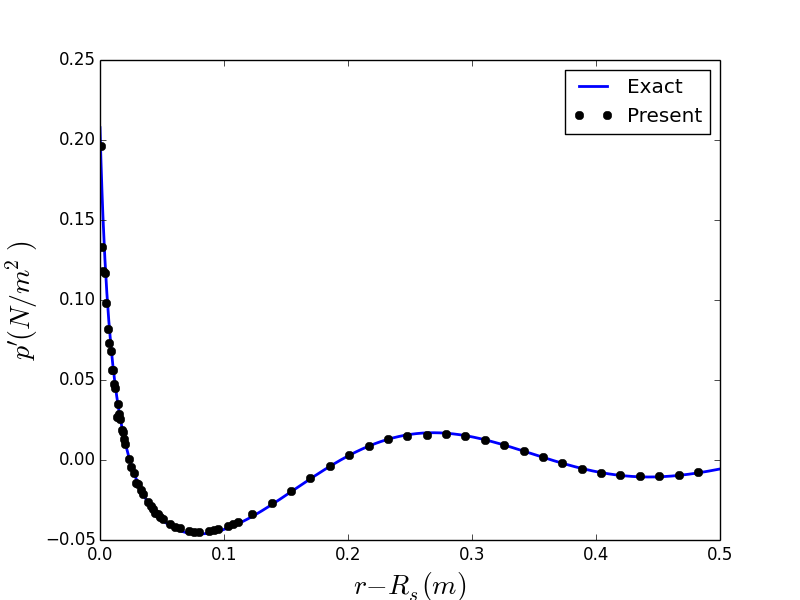}
\label{fig:PulsatingSphere0047}
}
\caption{(a) The 0.5 isocontour of the volume fraction and the 4-level refinement mesh boxes, (b) instantaneous contours of pressure perturbation 
$p'$, and comparison of the numerical and exact pressure perturbation along a radial line originating from the surface at 
(c) $t=1.85$ ms and (d) $t=2.35$ ms.}
\end{figure}

This test case is that of a sphere of mean radius $R_s=0.01$ m, with a harmonically pulsating surface with radius $r(t) = R_s + A\cos(2\pi ft)$, with $A = 10^{-6}$ m, and $f = 1000$ Hz .The pulsation of the sphere creates traveling pressure waves in the surrounding fluid. In the limit of the mean radius of the sphere being small compared to the wavelength corresponding to the acoustic wave with frequency $f$, i.e. $R_s \ll \lambda = \frac{c_0}{f}$, where $c_0$ is the ambient speed of sound, then the pressure perturbation in the domain is given by \citep{tsangaris2000analytical} as
\begin{eqnarray}
p'(r,t) \approx -\frac{4\pi^2\rho_0 R_s f^2}{r}\cos\Bigg(2\pi f\Bigg(t-\frac{r}{c_0}\Bigg)\Bigg).
\end{eqnarray}
The quiescent ambient condition of the surrounding air ($\gamma=1.4$, $R=287.0$ J/kg~K) is given by $\rho_0=1.226$ kg/m$^3$, $p_0=101325.0$ N/m$^2$. 
The domain size is $1$ m $\times$ $1$ m $\times$ $1$ m, with a base mesh size $64\times64\times64$, and three levels of refinement, that gives a 
resolution of $n_d=0.02/(1.0/(64\times2^3))=10$ points across the diameter of the sphere. The velocity of the sphere is surface is given by
\begin{eqnarray*}
\bm{u_s} = -2\pi Af\sin(2\pi ft)\bm{n}, 
\end{eqnarray*}
where $\bm{n}$ is the outward normal to the spherical surface. Fig.~\ref{fig:PulsatingSphere_Levels} shows the sphere and the three levels of refinement 
(note that in this and all subsequent test cases presented here, the refinement criteria is set to refine all cut cells in order to avoid intersecting the EB with 
coarse-fine boundaries. Such intersections are manageable, but would unnecessarily complicate the presentation here.). 
Fig.~\ref{fig:PulsatingSphere_pprime_contours} shows the instantaneous contours of pressure perturbation on three 
orthogonal planes through the center of the sphere. Figs.~\ref{fig:PulsatingSphere0037} and \ref{fig:PulsatingSphere0047} show the comparison of 
the numerical and exact solution of pressure perturbation along a radial line that originates at the surface. 
In this test case, the amplitude of oscillation is small ($A\ll\Delta x, \Delta y, \Delta z$), and hence the sphere surface does not cut across cells. This test case provides a 
verification of the correctness of the moving EB flux computation in three dimensions.

\subsection{An accelerating piston}
This test case demonstrates the classical introduction to shock waves and expansion fans -- an accelerating, 
advancing piston compressing the fluid creates a series of compression waves, the coalescence of which eventually creates 
a shock wave, and an accelerating, receding piston creates an expansion fan. 
\subsubsection{Advancing piston}
\begin{figure}[htpb!]
\begin{tabular}{cc}
&
\multirow{2}*{\includegraphics[trim=0.0cm 0.0cm 1.0cm 0.0cm, clip=true,scale=0.42]{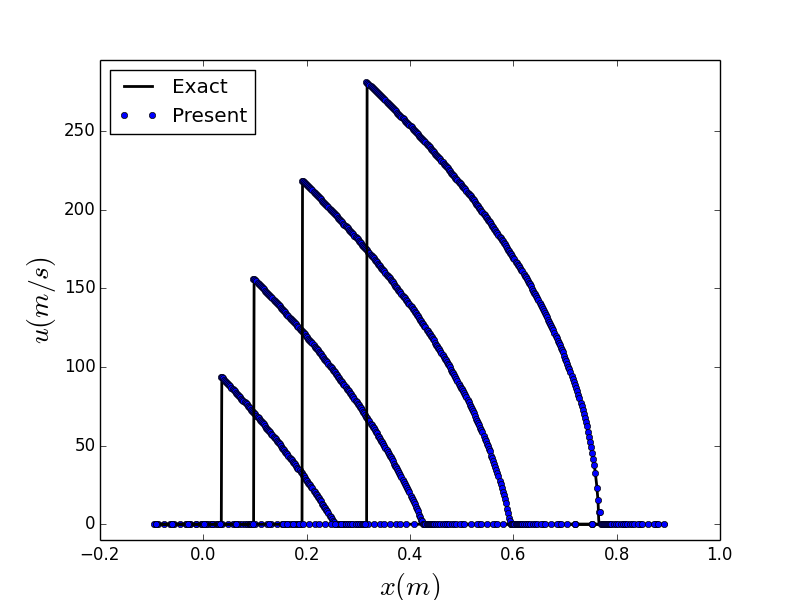}}\\
\includegraphics[trim=2.0cm 14.0cm 2.0cm 14.0cm, clip=true,scale=0.25]{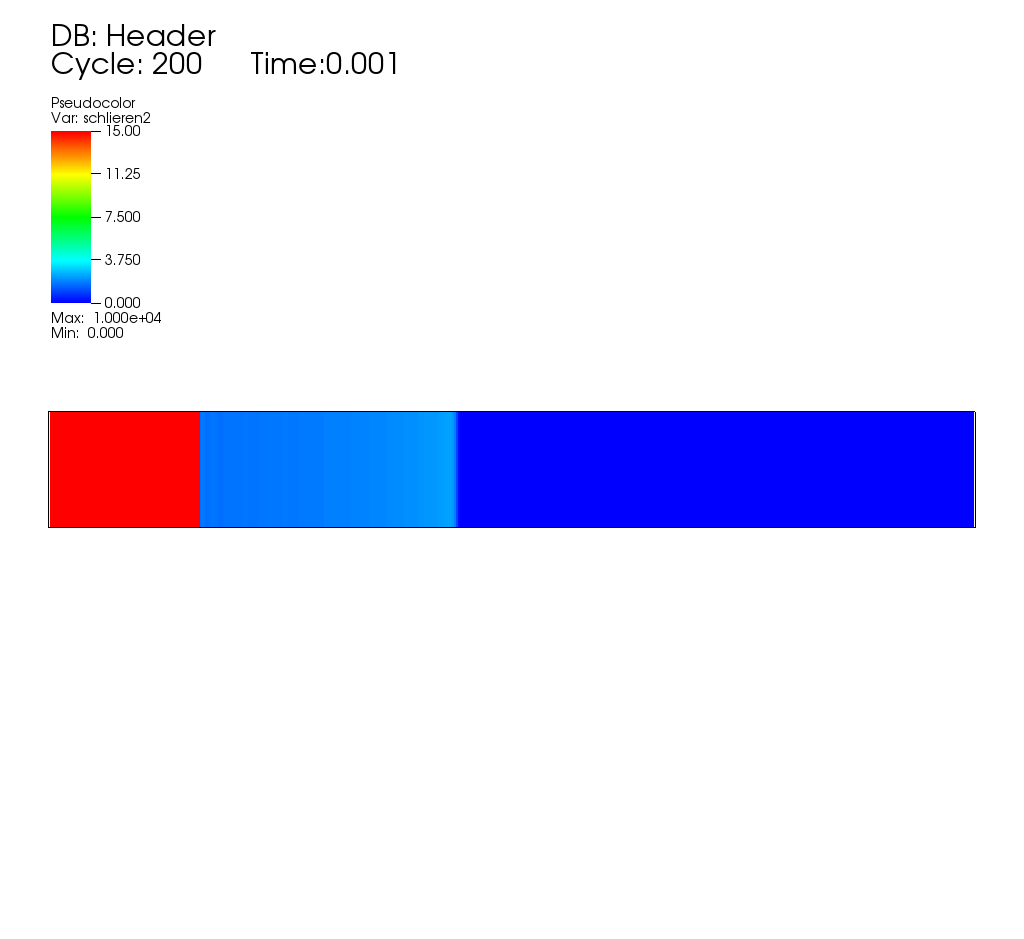}\\
\includegraphics[trim=2.0cm 14.0cm 2.0cm 14.0cm, clip=true,scale=0.25]{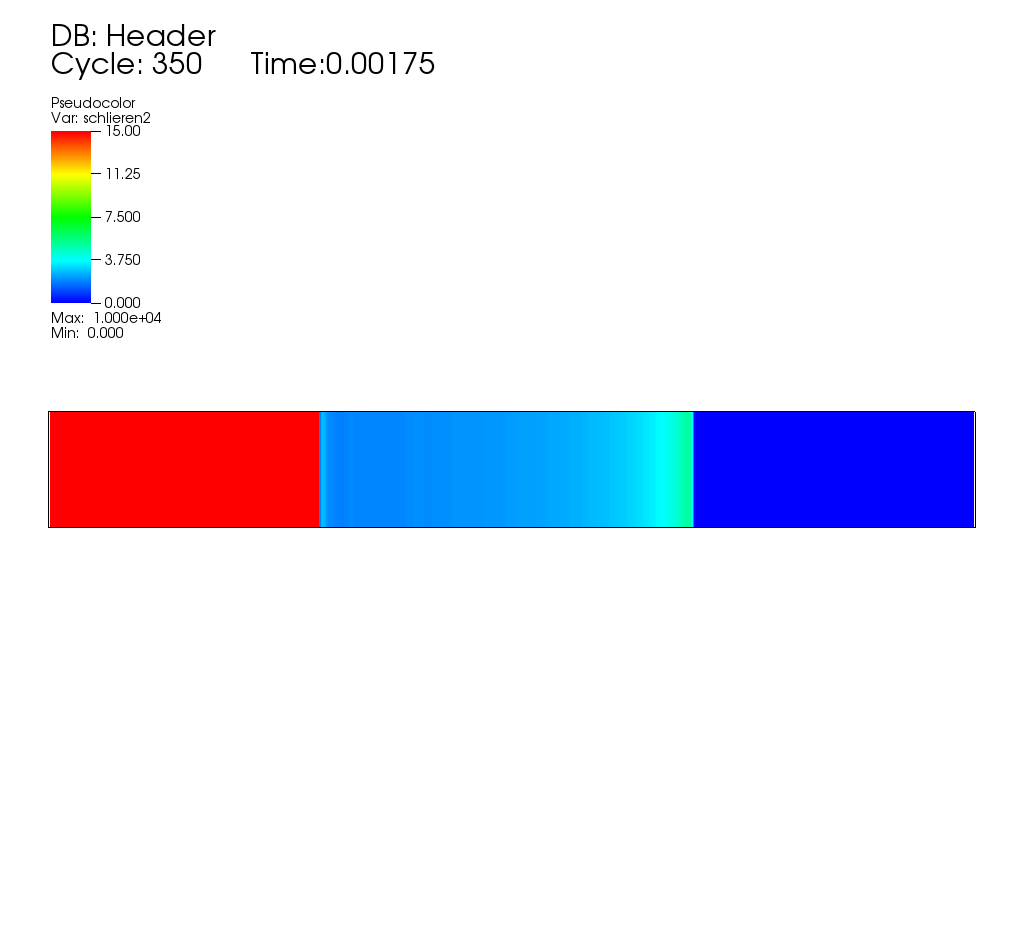}\\
\includegraphics[trim=2.0cm 14.0cm 2.0cm 14.0cm, clip=true,scale=0.25]{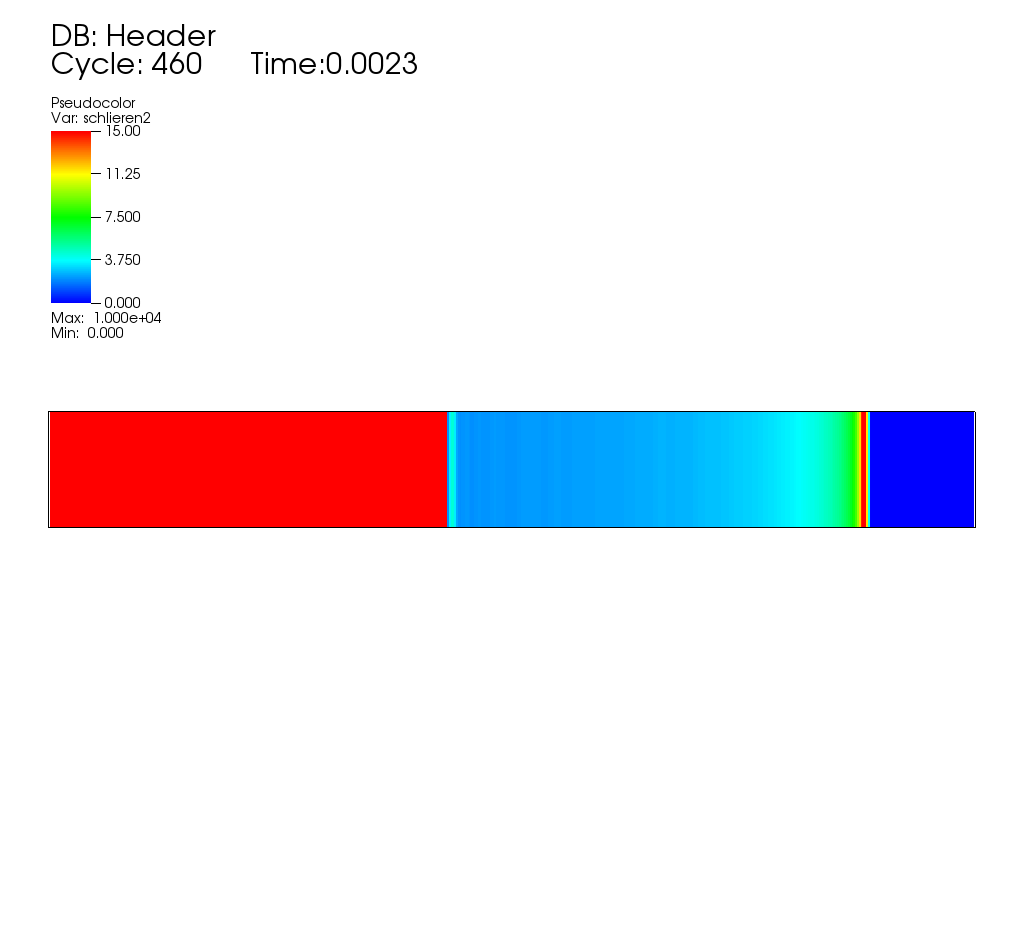}\\
\includegraphics[trim=2.0cm 14.0cm 2.0cm 14.0cm, clip=true,scale=0.25]{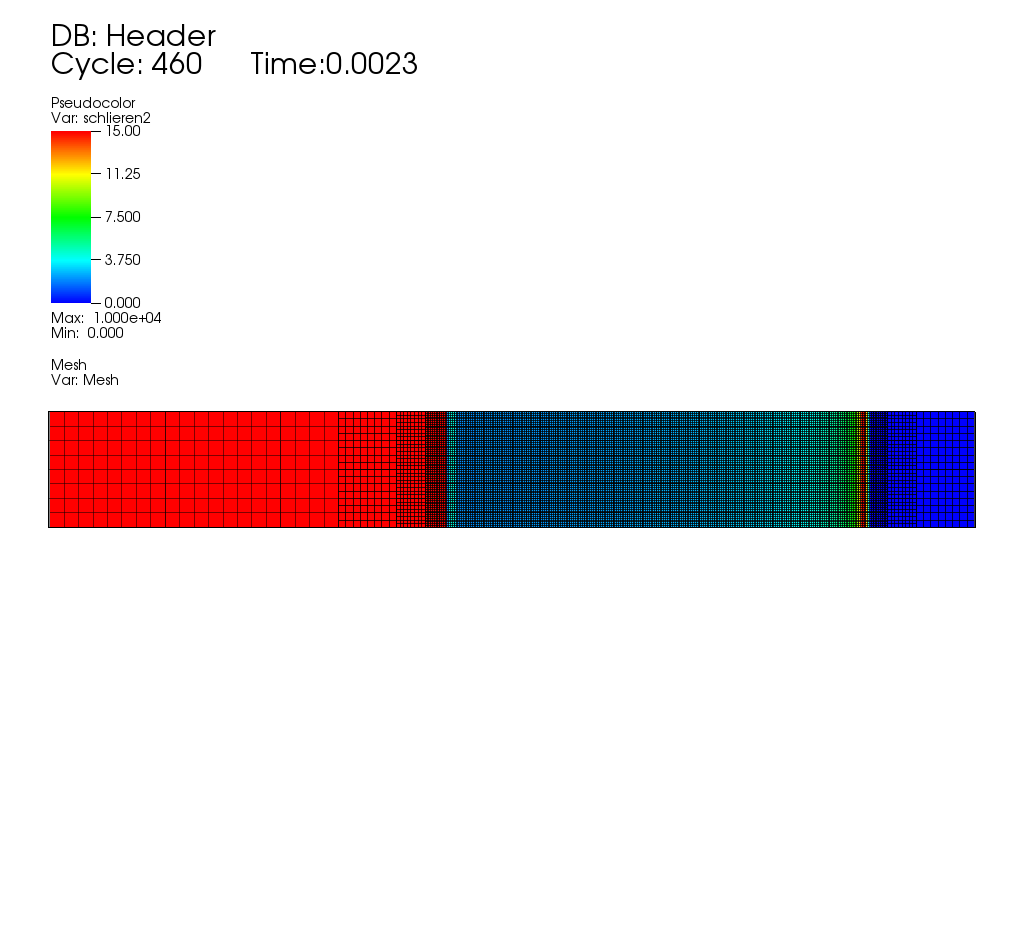}\\
&\\
(a)&(b)
\end{tabular}
\caption{(a) Numerical Schlieren ($\vert\nabla\rho\vert$) at $t=$ 1 ms, 1.75 ms, and at the shock formation time $t_c=$ 2.3 ms, and the 4-level mesh, and (b) the comparison of the velocity profiles with the exact solution at $t=$ 0.75 ms, 1.25 ms, 1.75 ms and $t_c=$ 2.3 ms.}
\label{fig:PistonShock}
\end{figure}

The initial condition consists of quiescent air ($\gamma=1.4$, $R=287.0$ J/kgK) with $\rho_0=1.226$ kg/m$^3$, and 
$p_0=101325.0$ N/m$^2$, with the piston located at $x=0.0$ m. The piston accelerates a constant rate $a=1.25e5$ m/s$^2$ at $t=0$, 
and hence, the velocity of the piston is given by 
\begin{eqnarray*}
u_p = at,
\end{eqnarray*}
which gives the piston motion as 
\begin{eqnarray*}
x(t) = \frac{1}{2}at^2.
\end{eqnarray*}
The velocity of the fluid is given as \citep{tsangaris2000analytical}
\begin{eqnarray}\label{eqn:PistonEqn}
u(x,t) =
  \begin{cases}
    -\frac{1}{\gamma}\Big(c_0-\frac{\gamma+1}{2}at\Big)+\sqrt{\frac{1}{\gamma^2}\Big(c_0-\frac{\gamma+1}{2}at\Big)^2-\frac{2}{\gamma}a(x-c_0t)} & \quad \text{if} \quad x>\frac{1}{2}at^2 \quad \text{and} \quad x-c_0t < 0\\
    0.0  & \quad \text{otherwise.}
  \end{cases}
\end{eqnarray}
The location and time of coalescence of compression waves to a shock wave are given by 
\begin{eqnarray*}
x_c = \frac{2c_0^2}{(\gamma+1)a}\quad\mathrm{and}\quad t_c = \frac{2c_0}{(\gamma+1)a}.
\end{eqnarray*}
The domain size is 1 m $\times$ 0.125 m and the base mesh size is 64$\times$8 with three levels of refinement. 
Fig.~\ref{fig:PistonShock} (a) shows the instantaneous Schlieren ($\vert\nabla\rho\vert$) images and the 4-level adaptive mesh.
The coalescence of the compression waves to form the shock wave  can be seen at $t=t_c=2.26$ ms. The location of the shock
$x\approx0.77$ m shows good quantitative agreement with the theory. The comparison of the numerical velocity profiles with the exact solution at different time instants is shown 
in Fig.~\ref{fig:PistonShock} (b). 

\subsubsection{Receding piston}
\begin{figure}
\begin{tabular}{cc}
&
\multirow{2}*{\includegraphics[trim=0.0cm 0.0cm 1.0cm 1.0cm, clip=true,scale=0.4]{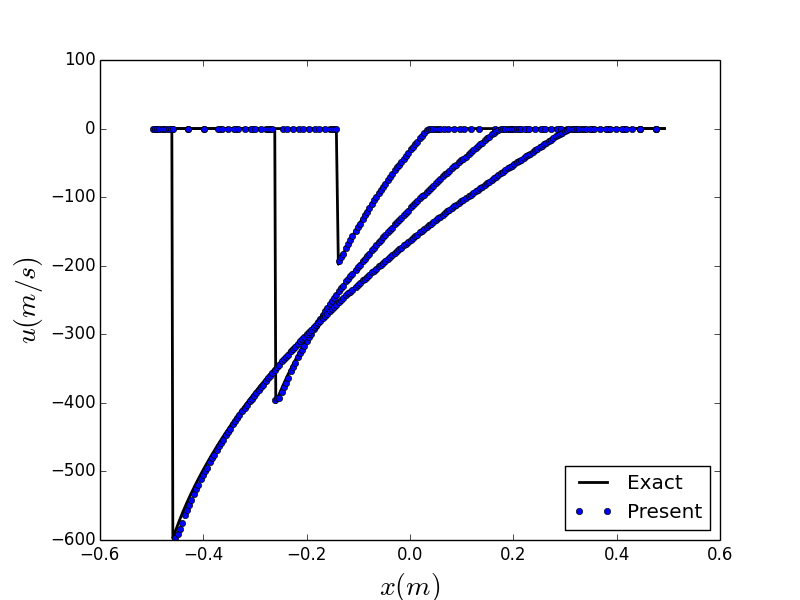}}\\
\includegraphics[trim=1.0cm 14.0cm 1.0cm 11.0cm, clip=true,scale=0.25]{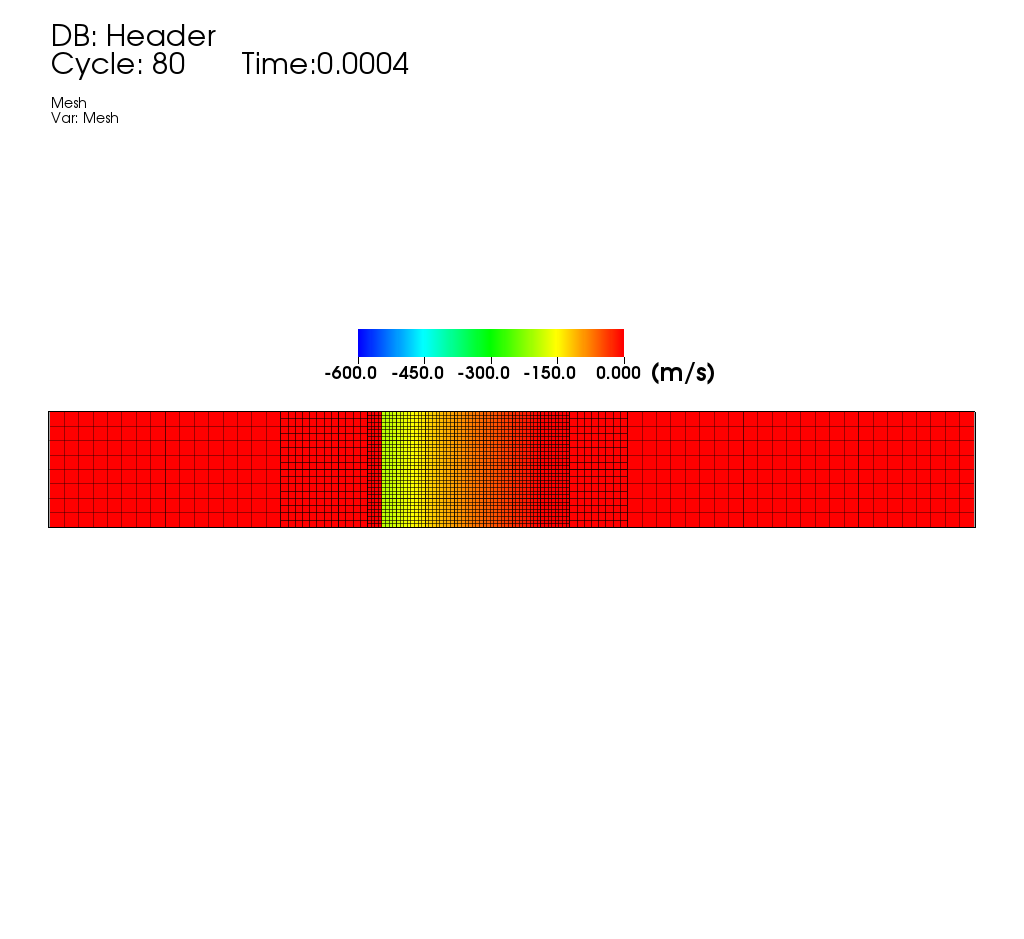}&\\
\includegraphics[trim=1.0cm 14.0cm 1.0cm 14.0cm, clip=true,scale=0.25]{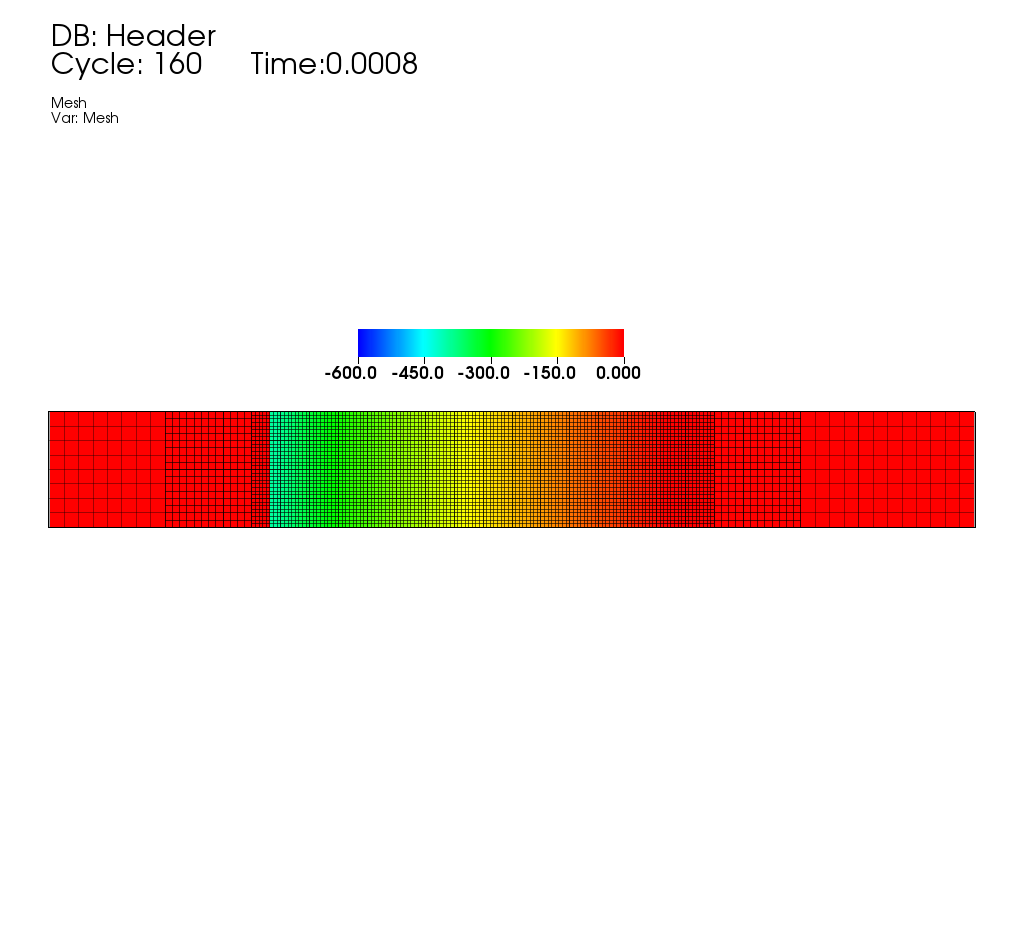}&\\
\includegraphics[trim=1.0cm 14.0cm 1.0cm 14.0cm, clip=true,scale=0.25]{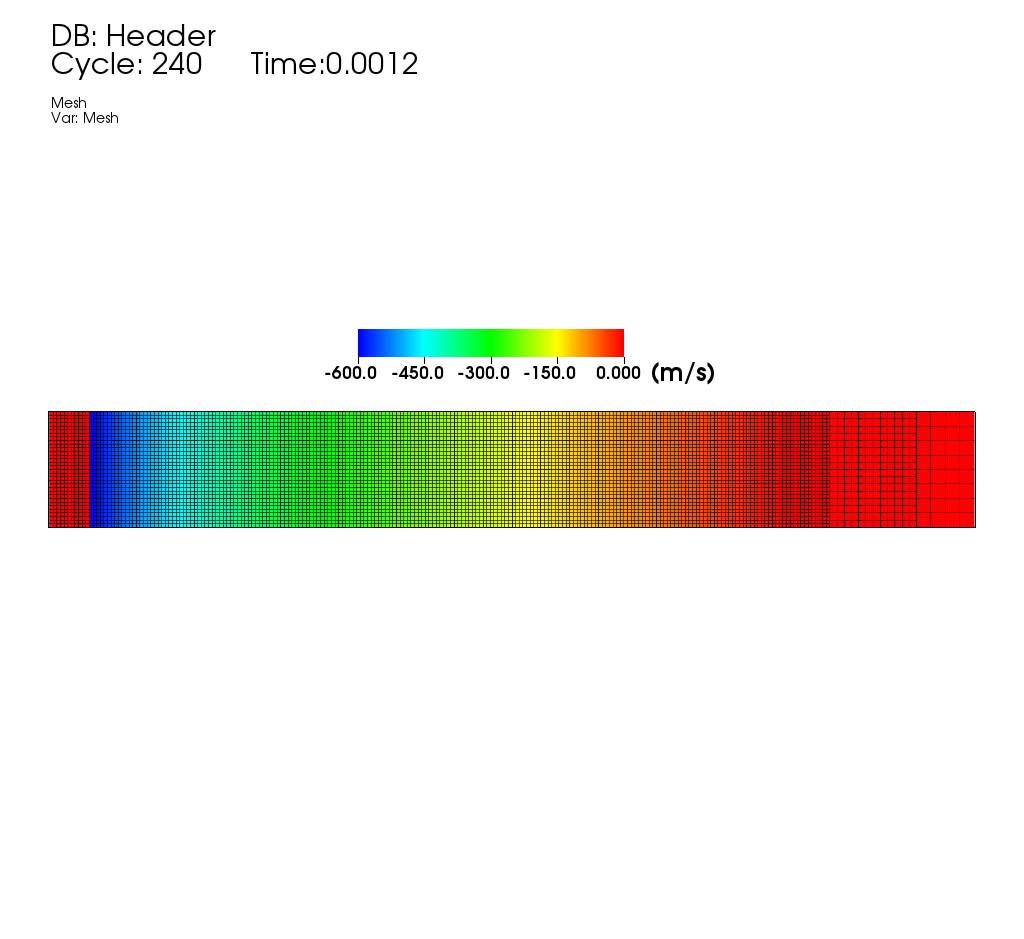}&\\
&\\&\\
(a)&(b)\\
\multicolumn{2}{c}{\multirow{2}{*}{\includegraphics[trim=0.0cm 0.0cm 1.0cm 1.0cm, clip=true,scale=0.4]{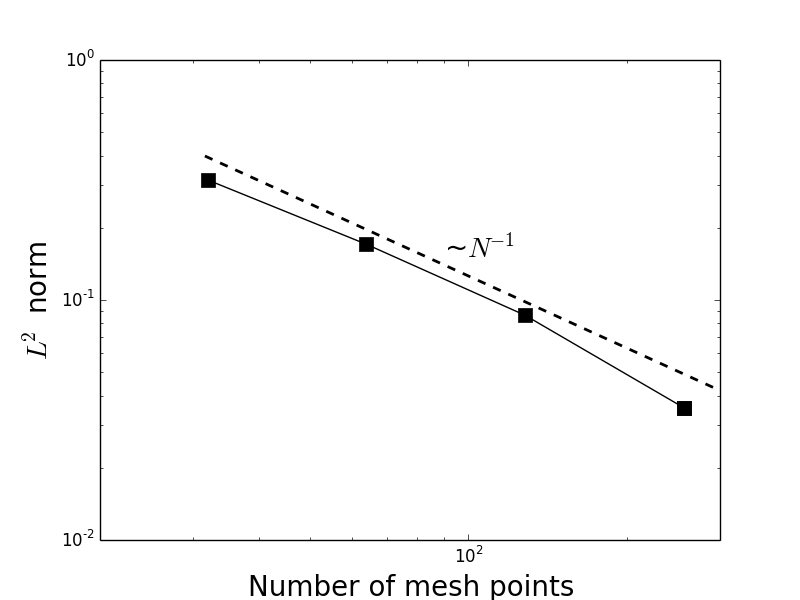}}}\\
&\\&\\&\\&\\&\\&\\&\\&\\&\\&\\&\\&\\&\\&\\
\multicolumn{2}{c}{\multirow{2}{*}{(c)}}\\
&\\
\end{tabular}
\caption{The contours of velocity and the 3-level mesh at (a) $t=$ 0.4, 0.8, 1.2 ms, (b) comparison of numerical and exact velocity profiles at $t=$ 0.4, 0.8, 1.2 ms and (c) $L^2$ norm of entropy showing that the scheme is first order accurate for moving boundary problems}
\label{fig:Expansion}
\end{figure}

The initial condition consists of quiescent air ($\gamma=1.4$, $R=287.0$ J/kgK), with $\rho_0=1.226$ kg/m$^3$, and
$p_0=101325.0$ N/m$^2$, with the piston located at $x=0.0$ m. The piston accelerates and recedes at a constant rate $a=-5e5$ m/s$^2$ at $t=0$.
The velocity of the fluid is given by Eqn.~\ref{eqn:PistonEqn}. The domain size is 1 m $\times$ 0.125 m and the base mesh size is 64$\times$8 with two levels of refinement.
Fig.~\ref{fig:Expansion}(a) shows the instantaneous velocity contours and the 3-level mesh at various time instants. 
The comparison of the numerical velocity profiles with the exact solution at different time instants is shown in Fig.~\ref{fig:Expansion} (b). This test case creates 
freshly cleared cells as the piston recedes, and demonstrates the efficiency of our moving boundary formulation.
\par An expansion fan is a smooth, isentropic flow, and hence the entropy should remain exactly zero at all times. We compute the order of accuracy of the numerical scheme 
for moving boundary problems using the $L^2$ norm of entropy computed as $s=\text{ln}\Bigg[\Bigg(\cfrac{p}{p_0}\Bigg)\Bigg(\cfrac{\rho}{\rho_0}\Bigg)^\gamma\Bigg]$. Fig.~\ref{fig:Expansion}(c) 
shows that the order of accuracy is 1. Although, the numerical scheme has an accuracy of 2 for smooth problems, for moving boundary problems, the order is found to be 1, and is attributed to the 
interpolation procedure for the freshly cleared cells given by Eqn.~\ref{eqn:fcc_interp}, as has been observed by \citet{muralidharan2018simulation} as well.

\subsection{Shock-cylinder interaction}\label{sec:shock-cylinder}
\begin{figure}[htpb!]
\subfigure[]
{
\includegraphics[trim=0.0cm 3.0cm 0.0cm 1.5cm, clip=true,scale=0.22]{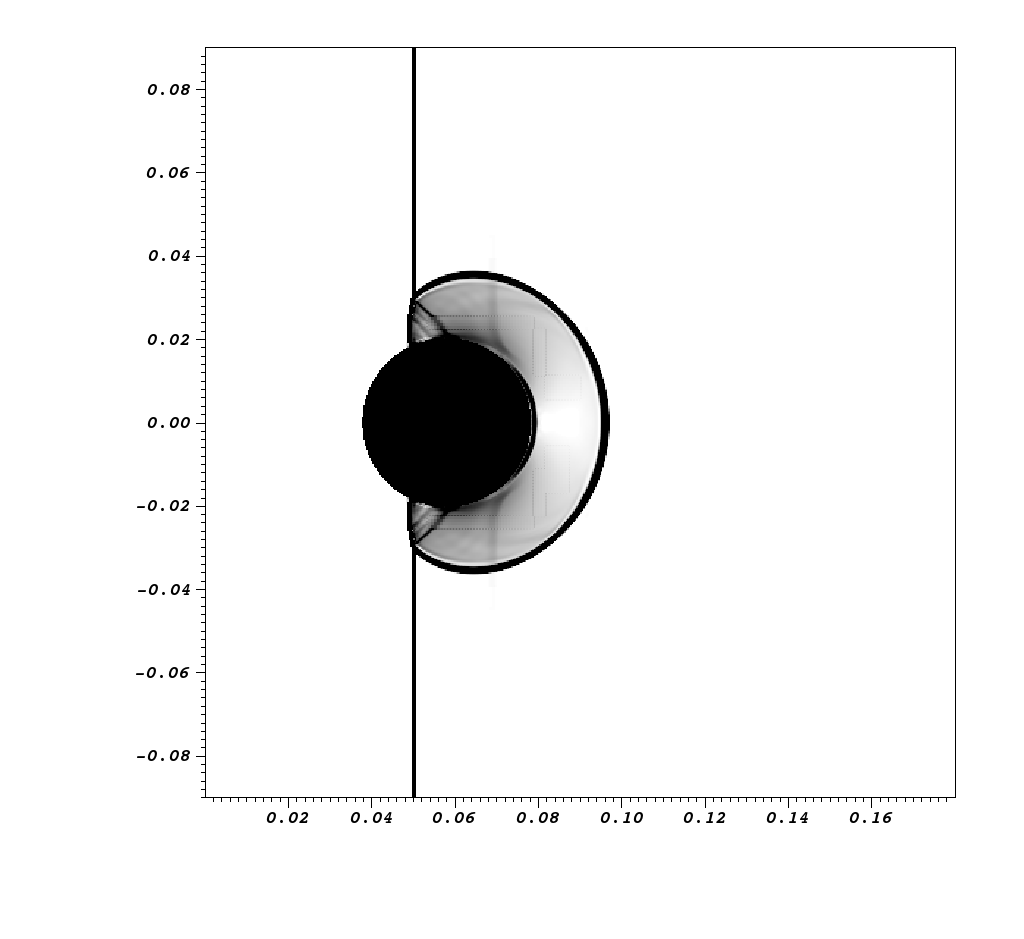}
}
\subfigure[]
{
\includegraphics[trim=0.0cm 3.0cm 0.0cm 1.5cm, clip=true,scale=0.22]{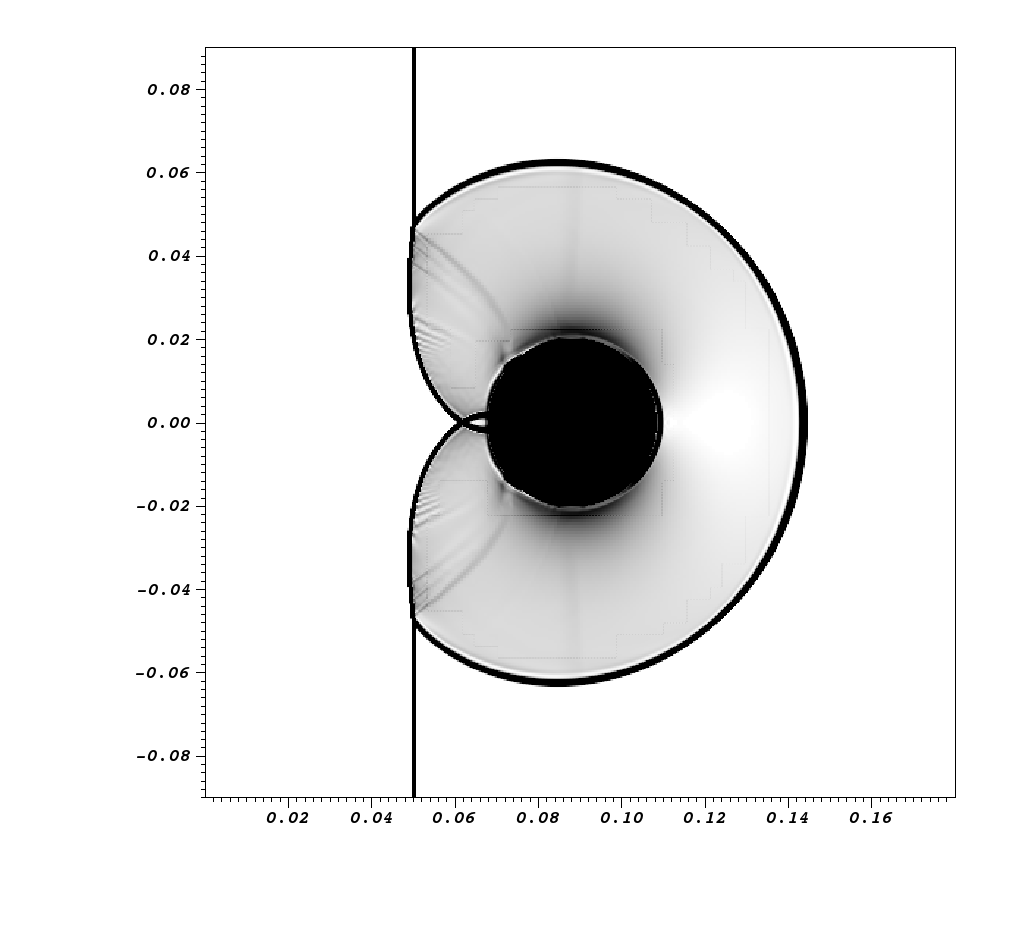}
}\\
\subfigure[]
{
\includegraphics[trim=0.0cm 3.0cm 0.0cm 1.5cm, clip=true,scale=0.22]{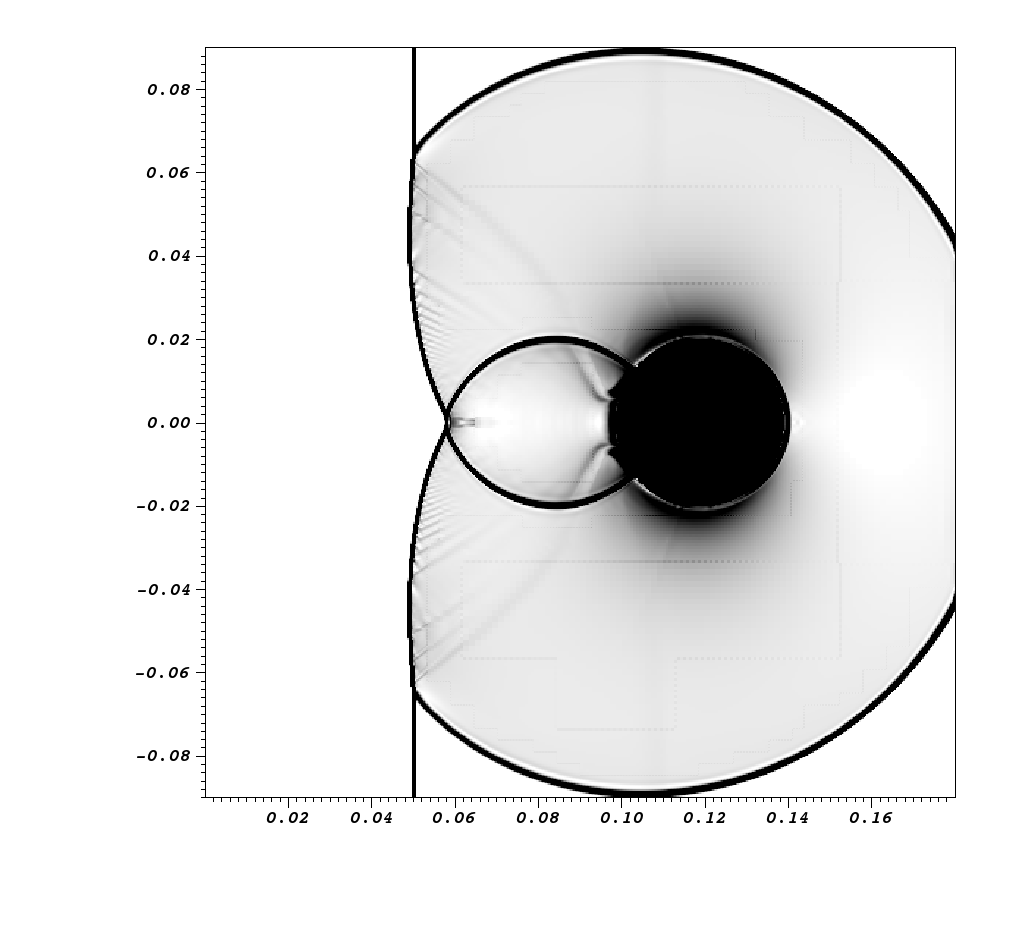}
}
\subfigure[]
{
\includegraphics[trim=0.0cm 3.0cm 0.0cm 1.5cm, clip=true,scale=0.22]{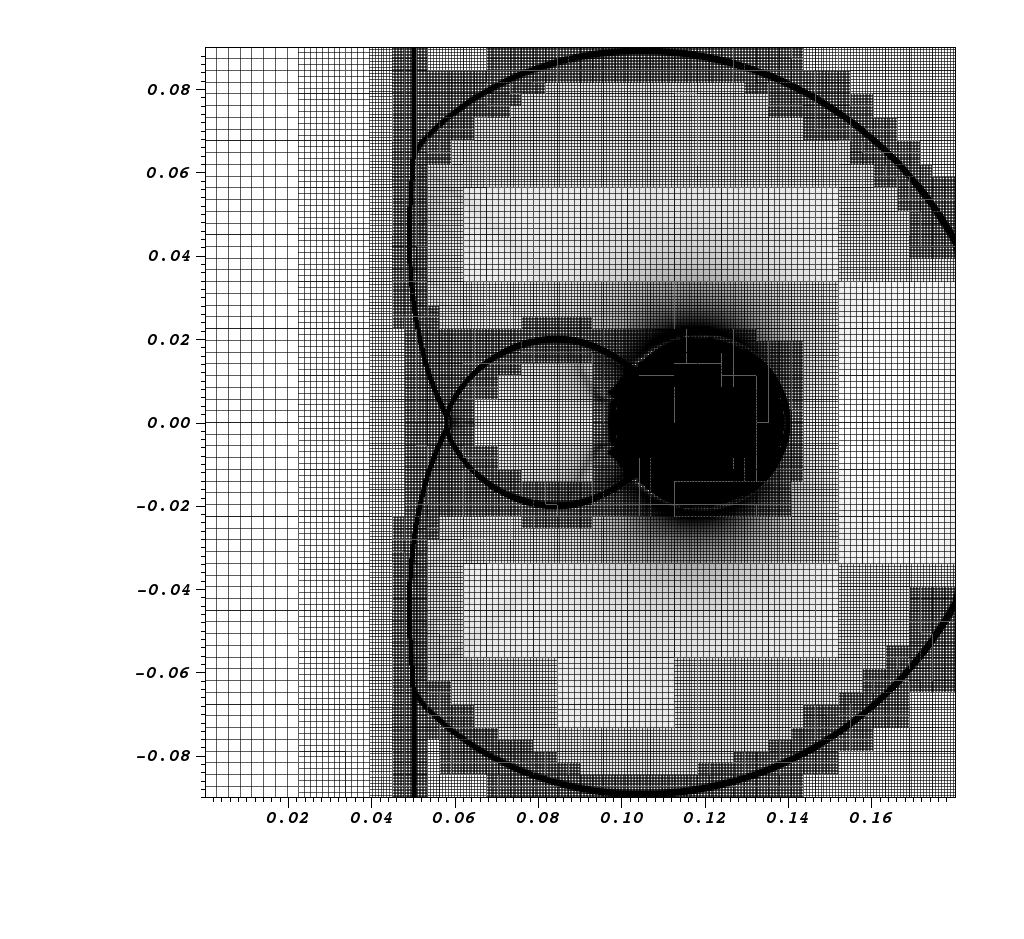}
\label{fig:ShockCylinder_Schlieren_4}
}
\begin{center}
\subfigure[]
{
\includegraphics[trim=6.0cm 3.5cm 6.0cm 3.0cm, clip=true,scale=0.5]{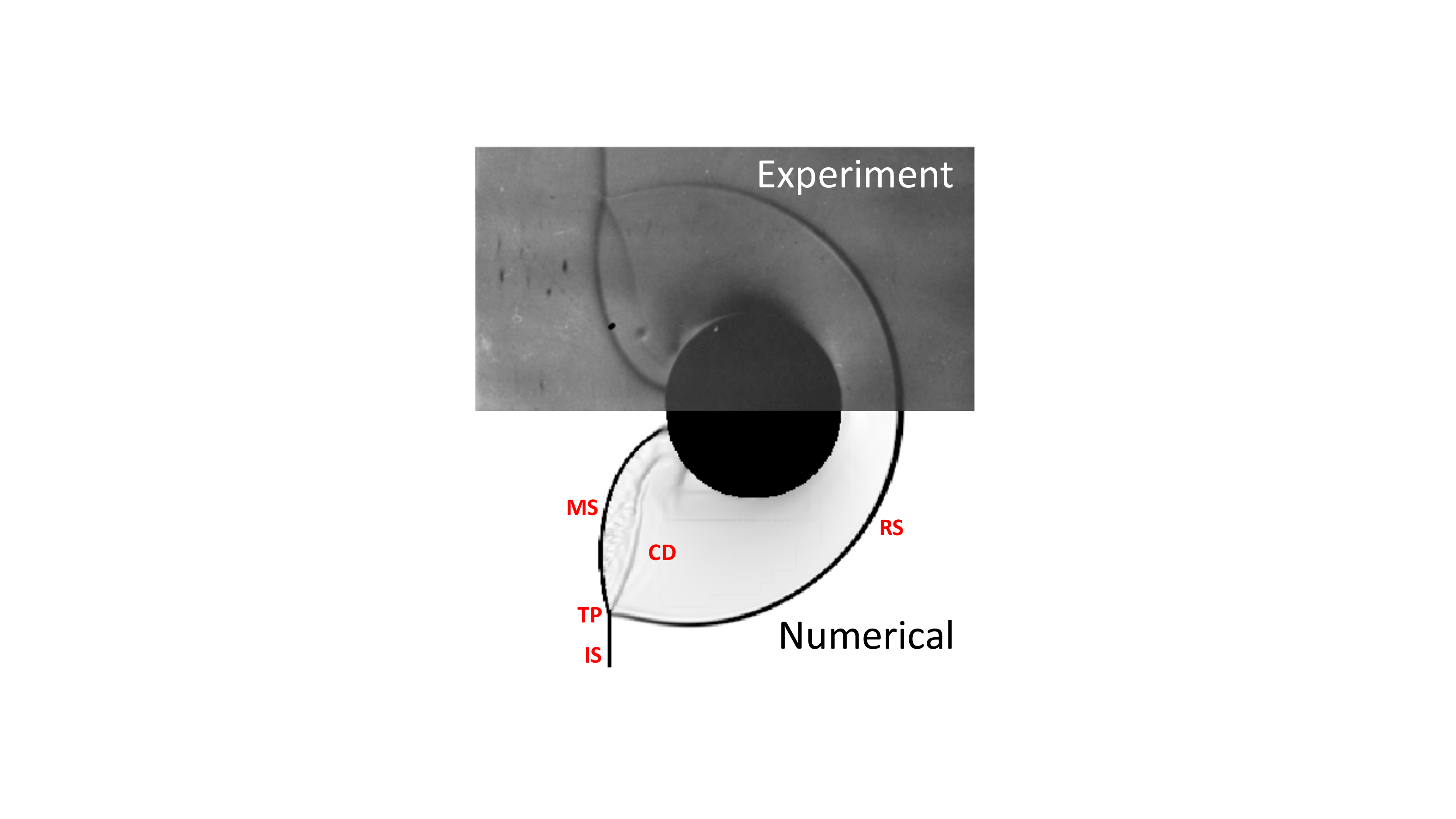}
\label{fig:ShockCylinder_Comparison}
}
\end{center}
\caption{The numerical Schlieren ($\vert\nabla\rho\vert$) at (a) $t=66$ $\mu$s, (b) $t=131$ $\mu$s, (c) $t=198$ $\mu$s and (d) the adaptive 5-level mesh (the coarsest level is completely refined and hence not seen at this instant) and (e) comparison of the numerical Schlieren with the experiment of \citet{bryson1961diffraction} showing the features. IS: Incident shock, RS: Reflected shock, MS: Mach stem, CD: Contact discontinuity, TP: mach triple point.}

\label{fig:ShockCylinder}
\end{figure}

To further demonstrate the capability of the moving boundary algorithm we consider the shock-cylinder interaction problem, which has been studied experimentally 
by \citet{bryson1961diffraction}, and computationally by \citet{bennett2018moving}. This test case consists of a rigid circular cylinder of diameter $D=0.04$ m interacting with a stationary Mach 1.34 shock wave. The initial condition corresponds to a stationary shock wave 
located at $x=0.05$ m, characterized by the left and right-hand states given by $\rho_L=0.595$ kg/m$^3$, $u_L=459.54$ m/s, $p_L=5e4$ Pa, 
and $\rho_R=0.944$ kg/m$^3$, $u_R=289.86$ m/s, $p_R=9.6e4$ Pa. The cylinder has a constant horizontal velocity of $u=u_L=459.54$ m/s. The center of 
the cylinder is initially located at $x=0.028$ m. The computational domain is 0.18 m$\times$ 0.18 m with a base mesh size of 32 $\times$ 32, 
with four levels of refinement, which gives a resolution of 114 points across the diameter of the cylinder. The refinement criterion tags all cut-cells and has an additional gradient based detector for resolving 
high-gradient regions. The detector at a point $(i,j)$ is given by \citet{wong2016multiresolution} as
\begin{eqnarray}\label{eqn:ShockRefine}
\widetilde{w}=\frac{\sqrt{w_x^2+w_y^2}}{w_\mathrm{mean}+\epsilon},
\end{eqnarray}
where
\begin{eqnarray*}
w_x&=&\vert\rho_{i+1,j}-2\rho_{i,j}+\rho_{i-1,j}\vert\\\nonumber
w_y&=&\vert\rho_{i,j+1}-2\rho_{i,j}+\rho_{i,j-1}\vert\\\nonumber
w_\mathrm{mean}&=&\sqrt{(\rho_{i+1,j}+2\rho_{i,j}+\rho_{i-1,j})^2+(\rho_{i,j+1}+2\rho_{i,j}+\rho_{i,j-1})^2}.\\\nonumber
\end{eqnarray*}
If $\widetilde{w}>0.005$, cells are tagged for refinement. Fig.~\ref{fig:ShockCylinder} (a)-(d) shows the evolution of the numerical Schlieren ($\vert\nabla\rho\vert$)
as the cylinder interacts with the shock wave. Since the cylinder moves at the same speed as the surrounding fluid before encountering the shock wave, 
the flow-field should not change with time during this period. The absence of any waves in the domain during this time shows that this consistency check is 
satisfied by our moving-boundary method similar to \citet{bennett2018moving}. Fig.~\ref{fig:ShockCylinder_Schlieren_4} shows the mesh at $t=197.86 \mu$s, and the efficiency of the refinement criterion to 
resolve the high gradient regions is evident.\\
\par Interaction of a cylinder with a stationary, Mach 2.82 shock wave is done for comparison with experiments of \citet{bryson1961diffraction}. The density and pressure ratios are 
$\rho_R/\rho_L=3.68$, and $p_R/p_L=9.11$. The initial condition is a shock wave located at $x=0.05$ m characterized by the left and right states given by
 $\rho_L=0.595$ kg/m$^3$, $u_L=967.25$ m/s, $p_L=5e4$ Pa, and $\rho_R=0.944$ kg/m$^3$, $u_R=262.59$ m/s, $p_R=4.56e5$ Pa. The cylinder moves with velocity 
$u_L$. Fig.~\ref{fig:ShockCylinder_Comparison} shows the comparison of the numerical Schlieren with the experiment of 
\citet{bryson1961diffraction}. When the incident shock (IS) first impinges on the cylinder, a regular reflected shock (RS) is formed, and later, as the cylinder moves past the 
incident shock wave, a Mach stem (MS) and a slip surface are formed, that leads to a triple Mach point (TP). The flow features are well resolved with adaptive refinement, and demonstrate 
qualitatively good comparison with experiments.

\subsection{Shock-wedge interaction}\label{sec:shock-wedge}
\begin{figure}[htpb!]
\centering
\subfigure[]
{
\includegraphics[trim=4.5cm 3.0cm 0.0cm 1.0cm, clip=true,scale=0.22]{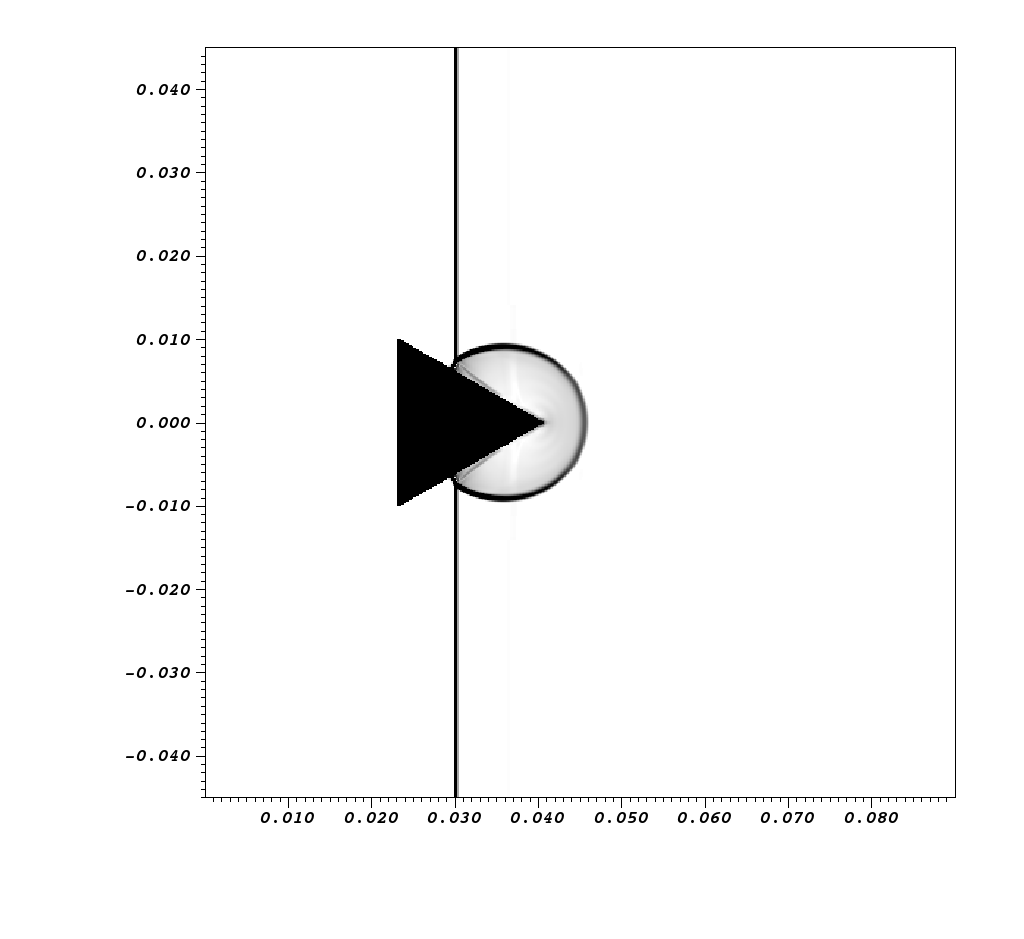}
\label{fig:ShockWedge_Schlieren_1}
}
\subfigure[]
{
\includegraphics[trim=4.5cm 3.0cm 0.0cm 1.0cm, clip=true,scale=0.22]{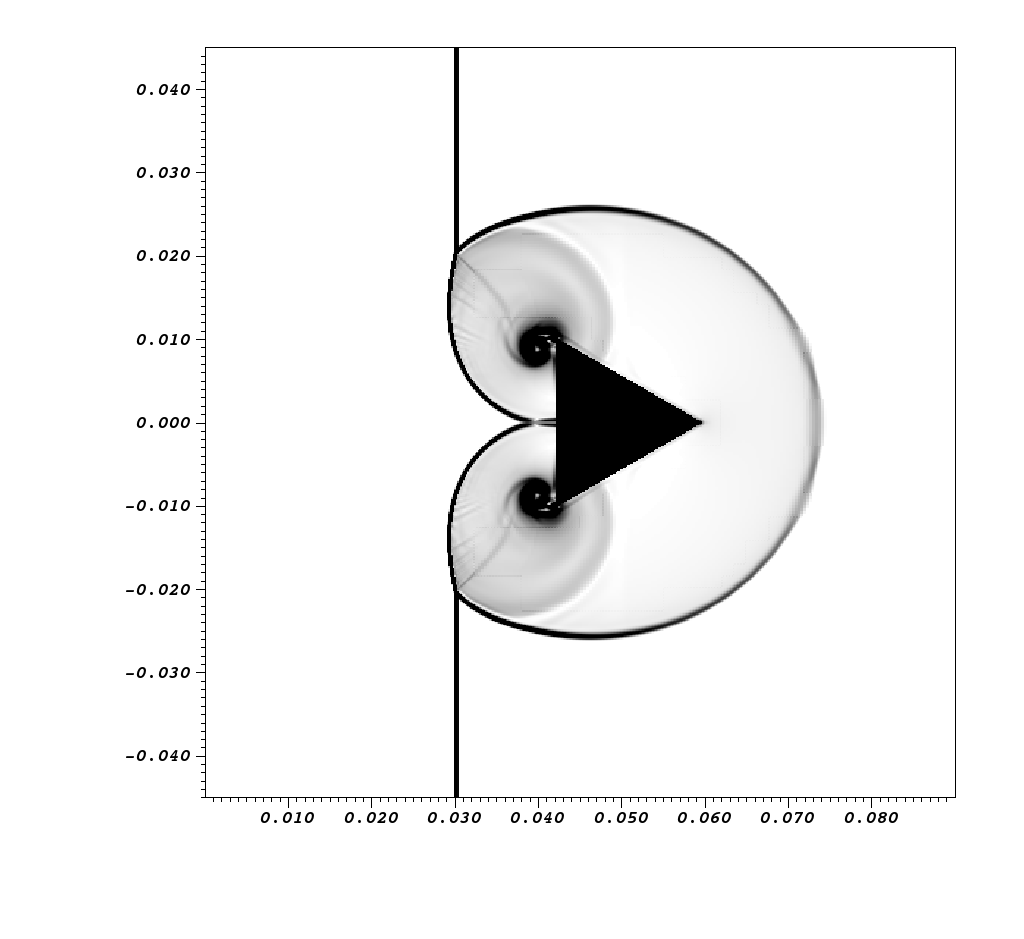}
\label{fig:ShockWedge_Schlieren_2}
}\\
\subfigure[]
{
\includegraphics[trim=4.5cm 3.0cm 0.0cm 1.0cm, clip=true,scale=0.22]{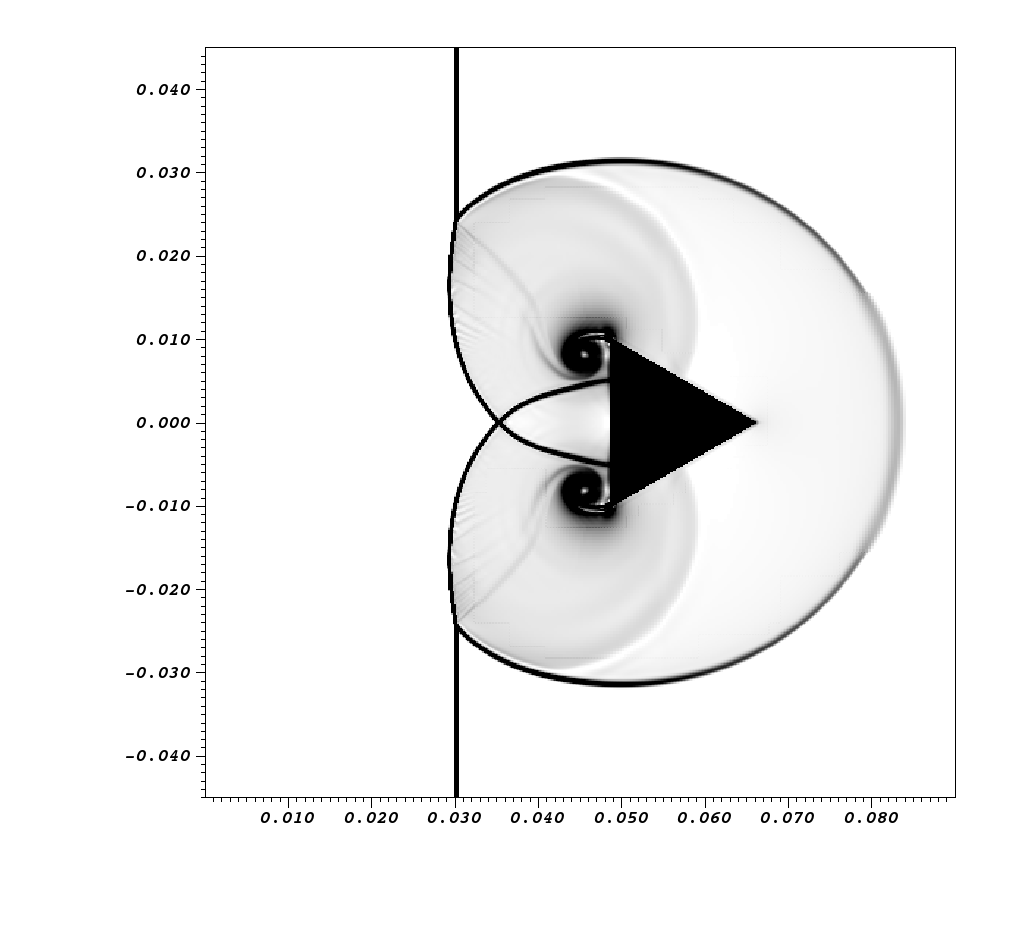}
\label{fig:ShockWedge_Schlieren_3}
}
\subfigure[]
{
\includegraphics[trim=4.5cm 3.0cm 0.0cm 1.0cm, clip=true,scale=0.22]{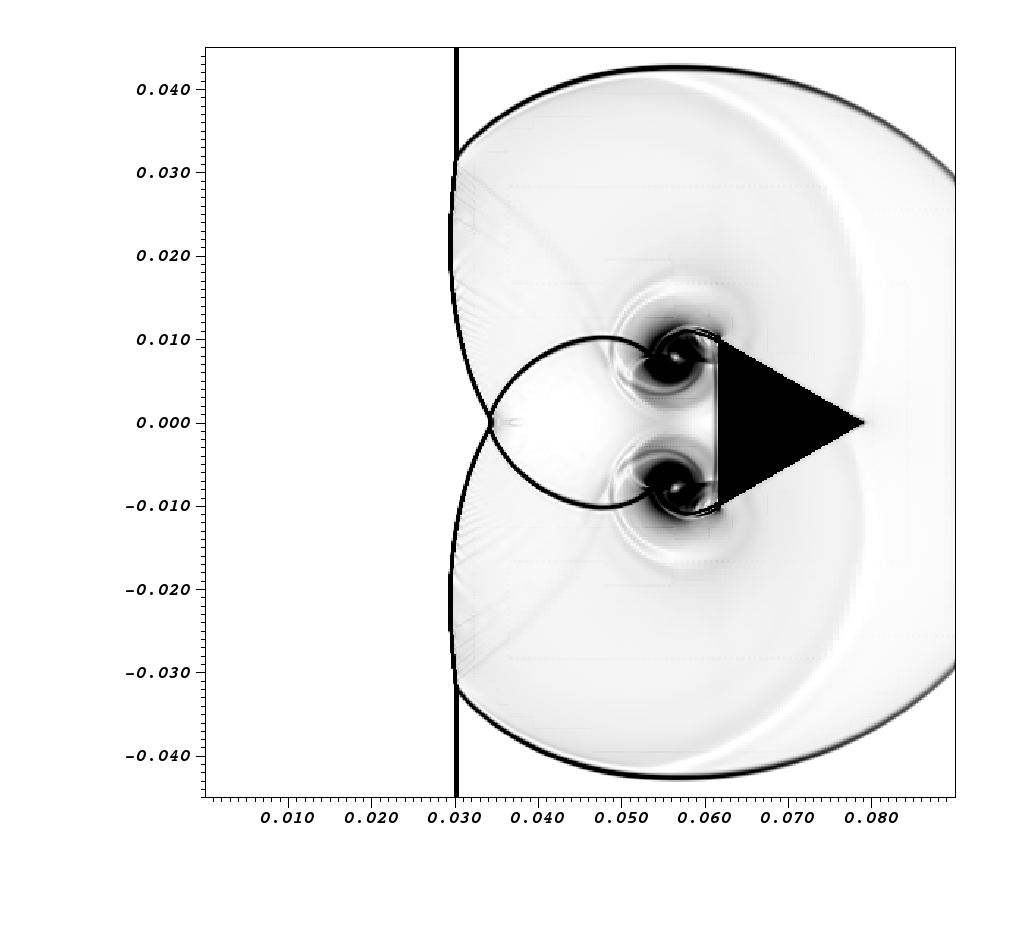}
\label{fig:ShockWedge_Schlieren_4}
}\\
\subfigure[]
{
\includegraphics[trim=4.5cm 3.0cm 0.0cm 1.0cm, clip=true,scale=0.22]{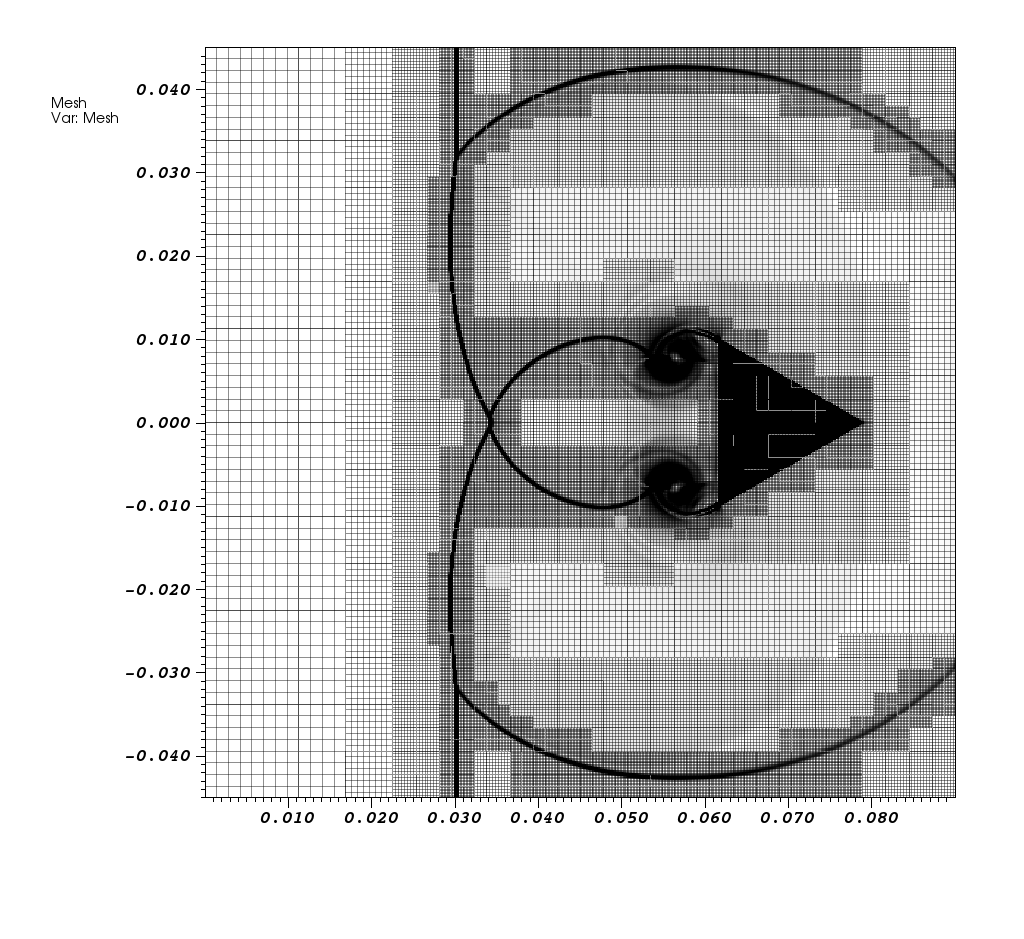}
\label{fig:ShockWedge_Schlieren_5}
}
\subfigure[]
{
\includegraphics[trim=10.5cm 3.5cm 10.5cm 3.5cm, clip=true,scale=0.52]{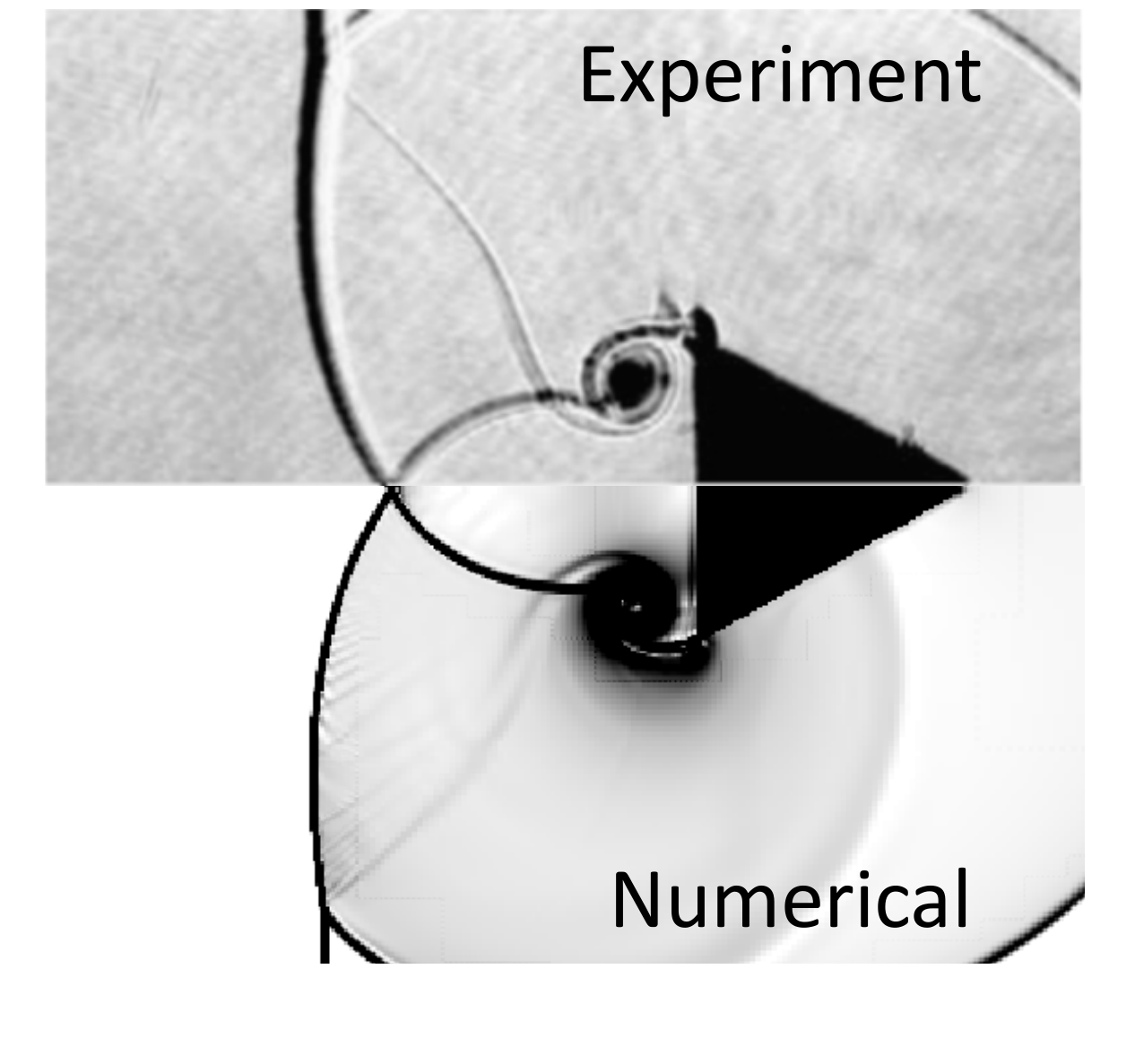}
\label{fig:ShockWedgeExpComparison}
}
\caption{The numerical Schlieren ($\vert\nabla\rho\vert$) at (a) $t=24$ $\mu$s, (b) $t=66$ $\mu$s, (c) $t=80$ $\mu$s, (d) $t=107$ $\mu$s, (e) the adaptive 5-level mesh (the coarsest level is completely refined and hence not seen at this instant) and (f) comparison of the numerical Schlieren with the experiments of \citet{chang2000shock}}
\end{figure}

In this section, we consider an experimental test case studied by \citet{chang2000shock}, known as Schardin's problem \citep{schardin1957high} -- a Mach 1.34 shock interaction with a triangular wedge. This test case demonstrates the capability of the algorithm to handle high-speed flows around sharp corners. This test case consists of a rigid, equilateral triangular wedge with side length
$L=0.02$ m interacting with a stationary Mach 1.34 shock wave. The initial condition corresponds to a stationary shock wave
located at $x=0.03$ m, characterized by the left and right-hand states given by $\rho_L=0.595$ kg/m$^3$, $u_L=459.54$ m/s, $p_L=5e4$ Pa,
and $\rho_R=0.944$ kg/m$^3$, $u_R=289.86$ m/s, $p_R=9.6e4$ Pa. The cylinder has a constant horizontal velocity of $u=u_L=459.54$ m/s. The center of
the vertical side of the wedge is initially located at $(x,y)=(0.012,0.0)$ m. The computational domain is 0.09 m$\times$ 0.09 m with a base mesh size of 32 $\times$ 32,
with four levels of refinement.  The refinement criterion tags all cut-cells and has an additional gradient based detector for resolving
high-gradient regions as described in Section~\ref{sec:shock-cylinder}. Fig.~\ref{fig:ShockWedge_Schlieren_1}-\ref{fig:ShockWedge_Schlieren_4} show the temporal evolution of the numerical Schlieren ($\vert\nabla\rho\vert$). Fig.~\ref{fig:ShockWedge_Schlieren_5} shows the 5-level mesh (coarser level is fully refined and hence not seen) showing the effectiveness 
of the refinement criterion in resolving the high-gradient regions. 
\par
\begin{figure}
\centering
\includegraphics[trim=6.0cm 4.0cm 5.0cm 3.0cm, clip=true,scale=0.55]{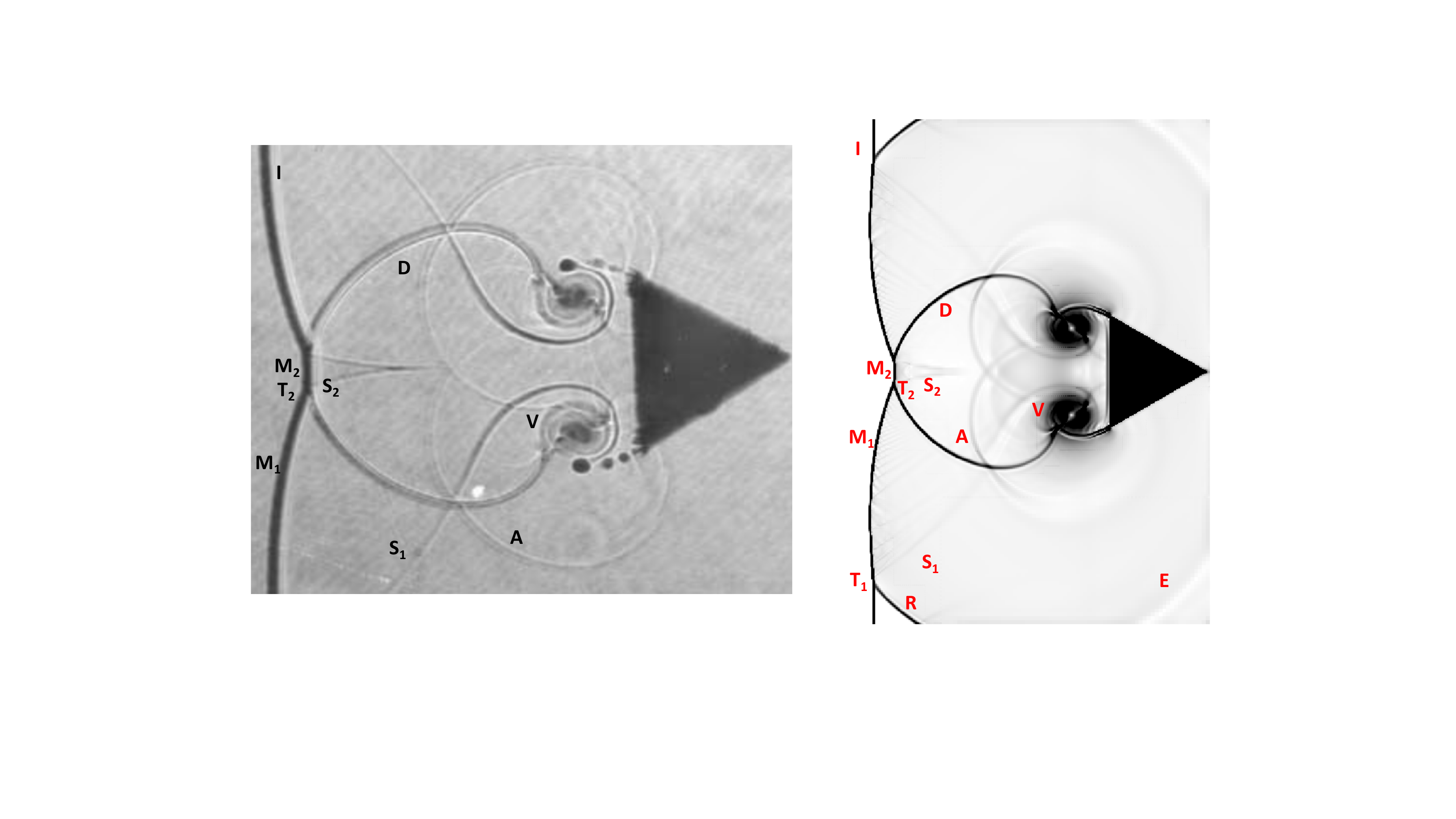}
\caption{Schlieren image from the experiment \citep{chang2000shock} (left), and simulation showing the comparison of the features. A: accelerated shock, D: decelerated shock, E: Expansion fan, R: Reflected shock, I: Incident shock T$_1$, T$_2$: Mach triple points, M$_1$, M$_2$: Mach stems , S$_1$, S$_2$: Slip lines (right).}
\label{fig:ShockWedgeExpComparison_Features}
\end{figure}

 Fig.~\ref{fig:ShockWedgeExpComparison} shows the comparison of the experimental \citep{chang2000shock} and numerical Schlieren  
images. It can be seen that the various features of the flow are well-resolved and qualitatively match the experimental results. A more detailed Schlieren image of the various features 
at a later time is shown in Fig.~\ref{fig:ShockWedgeExpComparison_Features}. As the wedge interacts with the incident shock wave, 
a regular reflected shock wave (R) and an expansion fan E are formed initially. As the wedge passes the incident shock the sharp corners lead  to the formation of 
strong vortices (V) and the symmetric decelerated shock wave pattern (D) . At a later time, Mach stems (M$_1$) form on the top and bottom, leading to a triple mach point T$_1$. As the 
wedge moves forward another Mach  stem (M$_2$) originates  leading to another triple point (T$_2$). 

\subsection{Pitching NACA 0012 airfoil}
The transonic buffet phenomenon over a NACA 0012 airfoil is widely studied both experimentally \citep{landon1982naca} and computationally \citep{venkatakrishnan1996implicit,mumtaz2017computational,kirshman2006flutter,schneiders2013accurate}. This test case consists of flow of air ($\gamma=1.4$, $R=287.0$ J/kgK) at Mach 0.755 over a pitching NACA 0012 airfoil with free-stream conditions of $\rho=1.226$ kg/m$^3$, $u=256.9$ m/s and $p=101325.0$ Pa. The pitching motion of the airfoil is about the quarter-chord point $x/c=0.25$, and is defined by the temporally varying angle of attack $\phi(t)=0.016+2.51\sin(\omega t)$ degrees, with $\omega=41.45$ rad/s. A polynomial representation is used to generate the NACA 0012 airfoil \citep{tools2015naca}. The domain size is 20 m $\times$ 10 m, and the base mesh size is 256 $\times$ 128 with three levels of refinement. The refinement criterion is defined to tag cut-cells and high gradient regions as defined by Eqn.\ref{eqn:ShockRefine}. \\ 
\par
\begin{figure}[htpb!]
\subfigure[]
{
\includegraphics[trim=0.0cm 3.0cm 3.0cm 3.0cm, clip=true,scale=0.22]{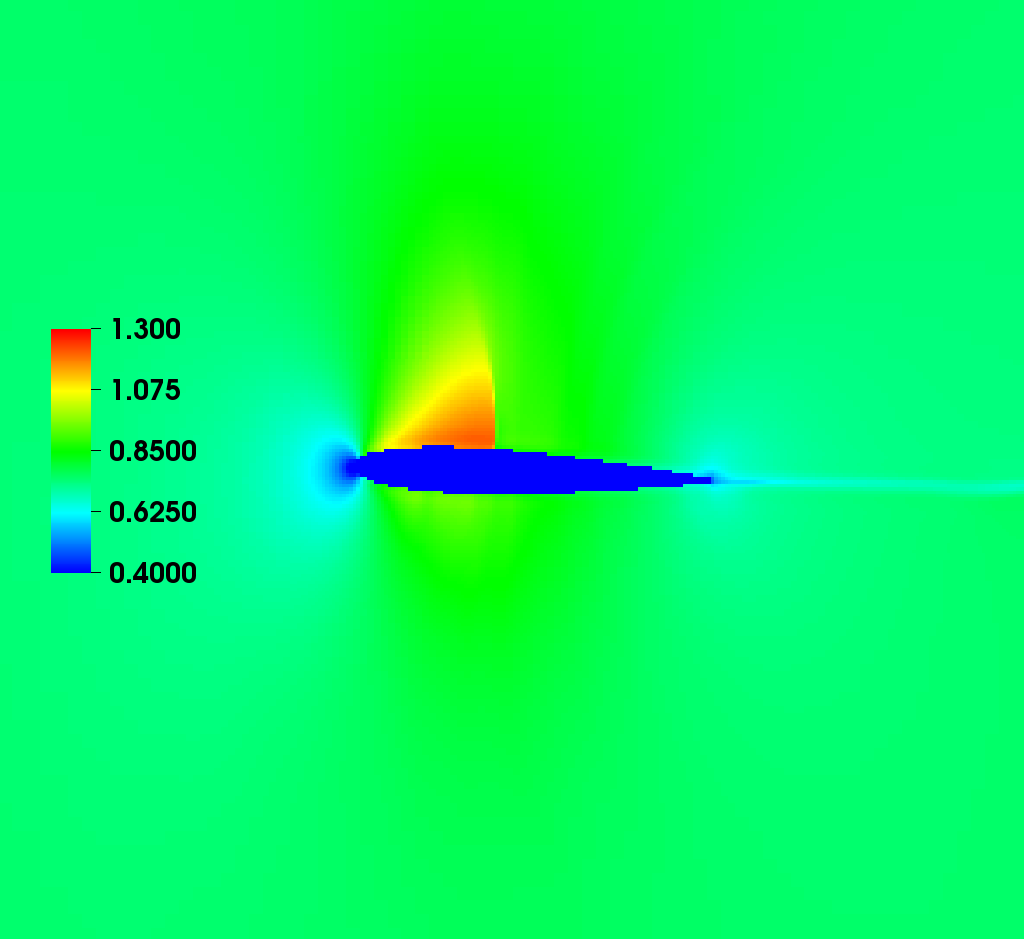}
\label{fig:PitchingAirfoil_Mach_1}
}
\subfigure[]
{
\includegraphics[trim=0.0cm 3.0cm 3.0cm 3.0cm, clip=true,scale=0.22]{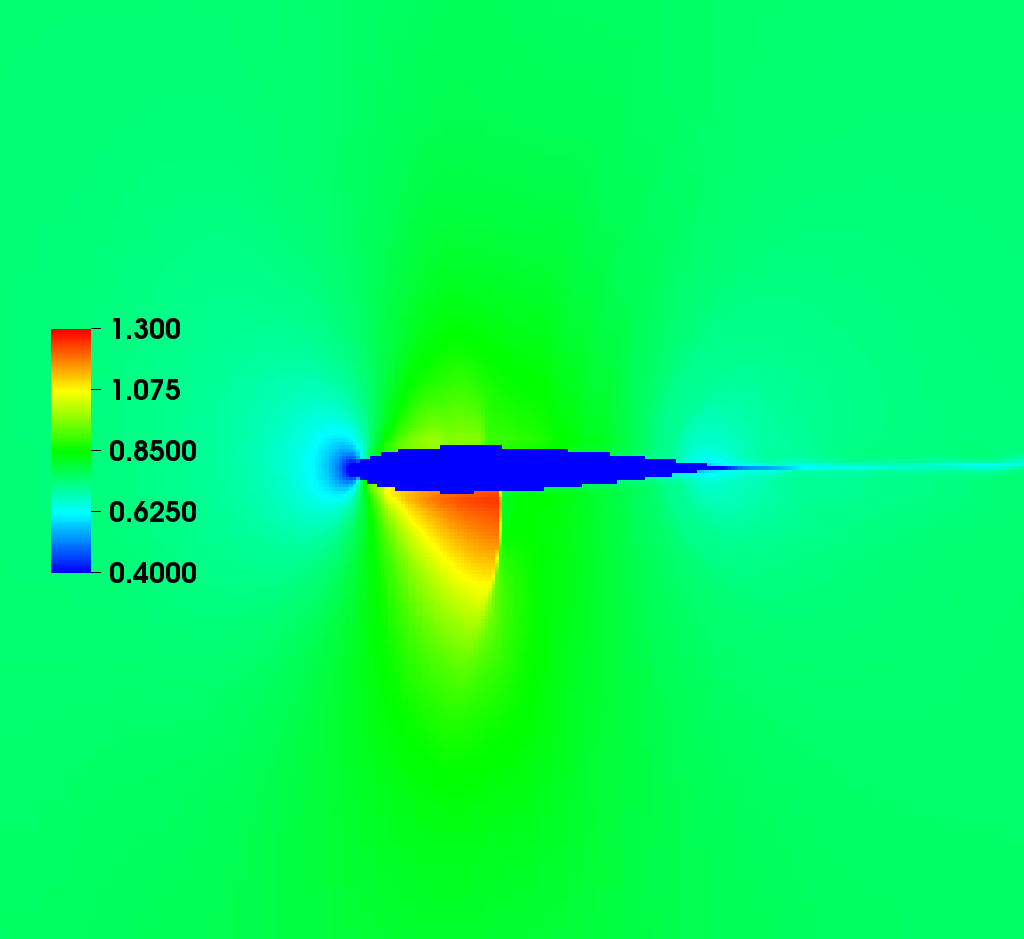}
\label{fig:PitchingAirfoil_Mach_2}
}\\
\subfigure[]
{
\includegraphics[trim=0.0cm 0.0cm 0.0cm 0.0cm, clip=true,scale=0.2]{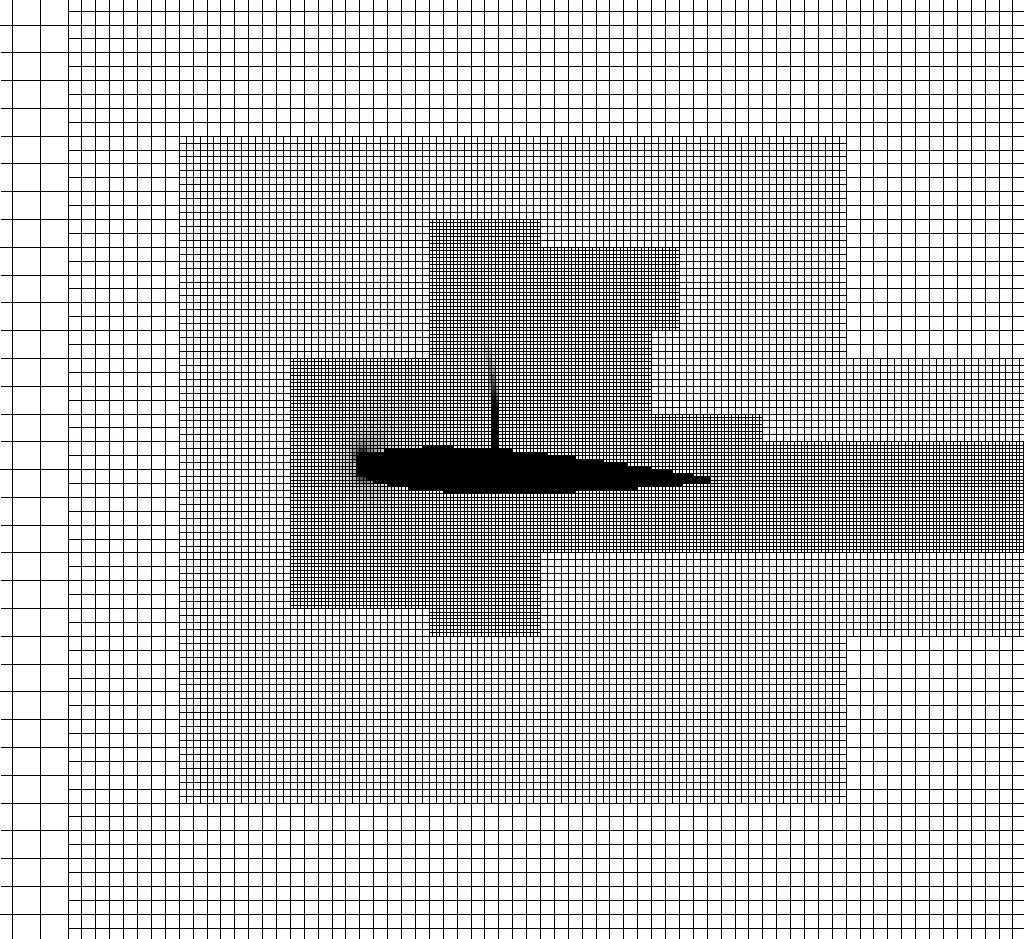}
\label{fig:PitchingAirfoil_Schlieren_1}
}
\subfigure[]
{
\includegraphics[trim=0.0cm 0.0cm 0.0cm 0.0cm, clip=true,scale=0.2]{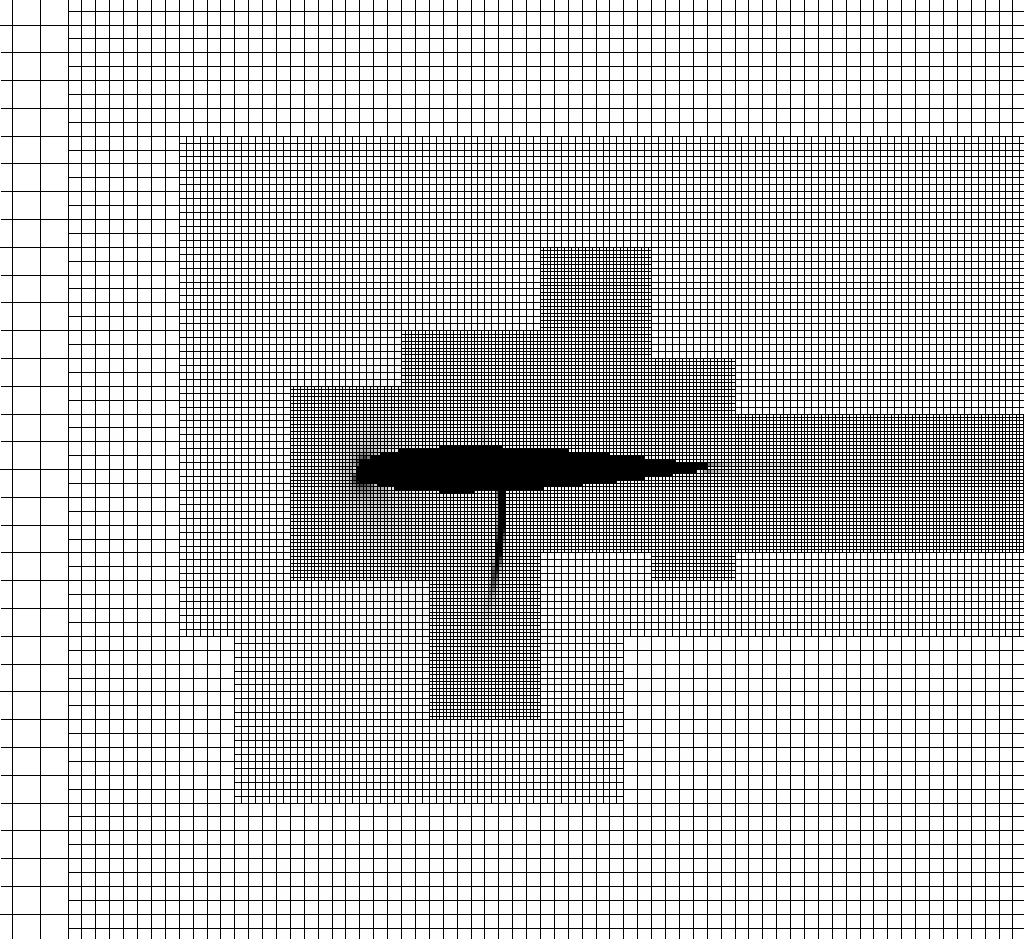}
\label{fig:PitchingAirfoil_Schlieren_2}
}\\
\subfigure[]
{
\includegraphics[trim=0.0cm 0.0cm 1.0cm 1.0cm, clip=true,scale=0.4]{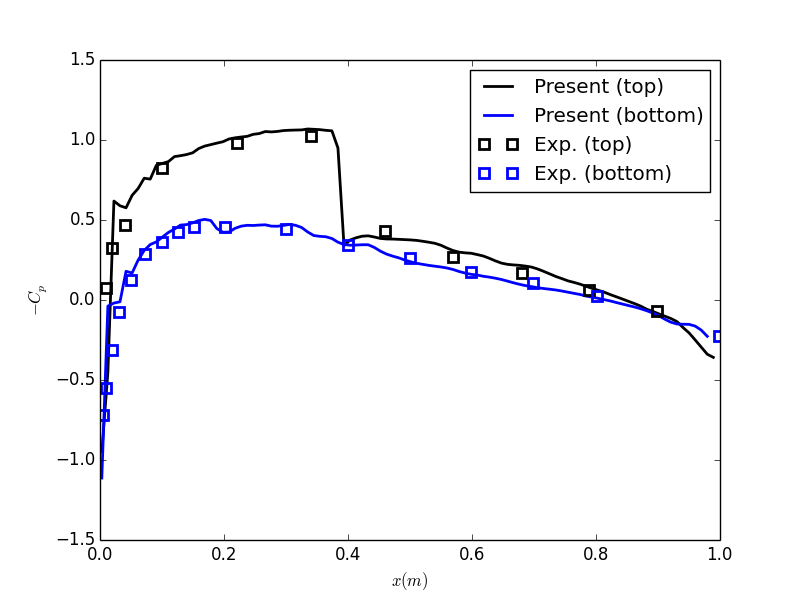}
\label{fig:CompareWithExp}
}
\subfigure[]
{
\includegraphics[trim=0.0cm 0.0cm 1.0cm 1.0cm, clip=true,scale=0.4]{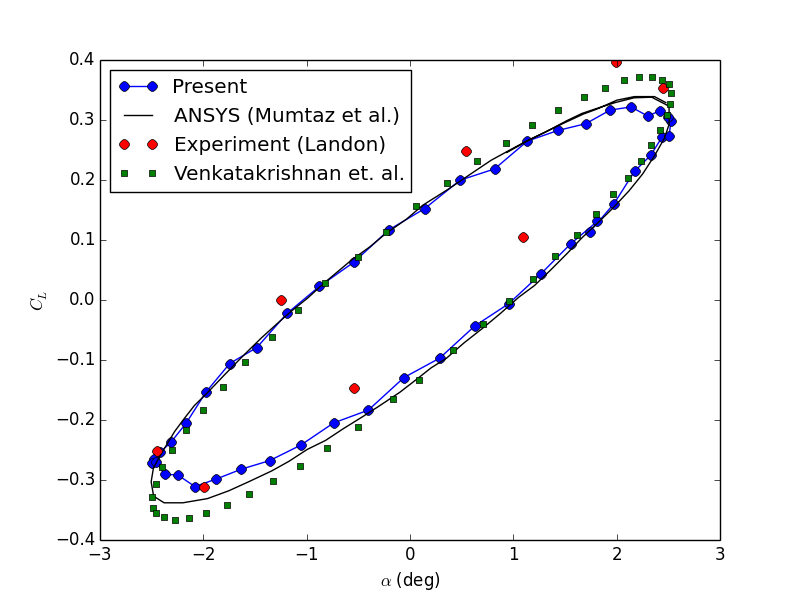}
\label{fig:cldeg}
}
\caption{The contours of Mach number at (a) $\phi=2.34^o$ and (b)  $\phi=-0.54^o$, the 4-level mesh and the numerical Schlieren at (c) $\phi=2.34^o$ and (d)  $\phi=-0.54^o$, (e) comparison of the numerical and experimental \citep{landon1982naca} pressure coefficient for the top and bottom surfaces of the airfoil at $\phi=2.34^o$, and (f) comparison of the numerical variation of the lift coefficient as a function of the angle of attack, with the experiment \citep{landon1982naca} and other results in literature \citep{venkatakrishnan1996implicit,mumtaz2017computational}.}
\label{fig:NACA 0012}
\end{figure}

 The flow over the airfoil is transonic as it accelerates over the surface and becomes supersonic, and forms a shock wave which makes it subsonic. Due to the pitching of the airfoil, the shock wave location is unsteady and an oscillating shock wave pattern known as the transonic buffet can be observed over the airfoil. This causes the pressure over the top and bottom surfaces of the airfoil to fluctuate with time, and leads to a cyclic variation of the lift coefficient, and as a result, the lift coefficient has different values as the airfoil encounters the same angle of attack during the upward and downward motion. The transonic buffet phenomenon can be seen in Fig.~\ref{fig:PitchingAirfoil_Mach_1} and \ref{fig:PitchingAirfoil_Mach_2}, which show the contours of Mach number at an angle of attack $\phi=2.34^o$ and $\phi=-0.54^o$ respectively. Fig.~\ref{fig:PitchingAirfoil_Schlieren_1} and \ref{fig:PitchingAirfoil_Schlieren_2} show the numerical Schlieren ($\vert\nabla\rho\vert$) and the 4-level mesh at $\phi=2.34^o$ and $\phi=-0.54^o$. Fig.~\ref{fig:CompareWithExp} shows the comparison of the numerical and experimental \citep{landon1982naca} pressure coefficient $C_p=\cfrac{p-p_\infty}{1/2\rho U_\infty^2}$ on the top and bottom surfaces of the airfoil at an angle of attack of $\phi=2.34^o$ . Fig.~\ref{fig:cldeg} shows the variation of the lift coefficient as a function of the angle of attack. Good quantitative agreement is observed with the experiments of \citet{landon1982naca}, the numerical simulation of \citet{venkatakrishnan1996implicit}, and the result from a commercial solver (ANSYS) \citep{mumtaz2017computational}.

\subsection{Reciprocating piston in an engine-like geometry}
\begin{figure}[htpb!]
\subfigure[]
{
\includegraphics[trim=8.0cm 6.0cm 9.0cm 5.0cm, clip=true,scale=0.35]{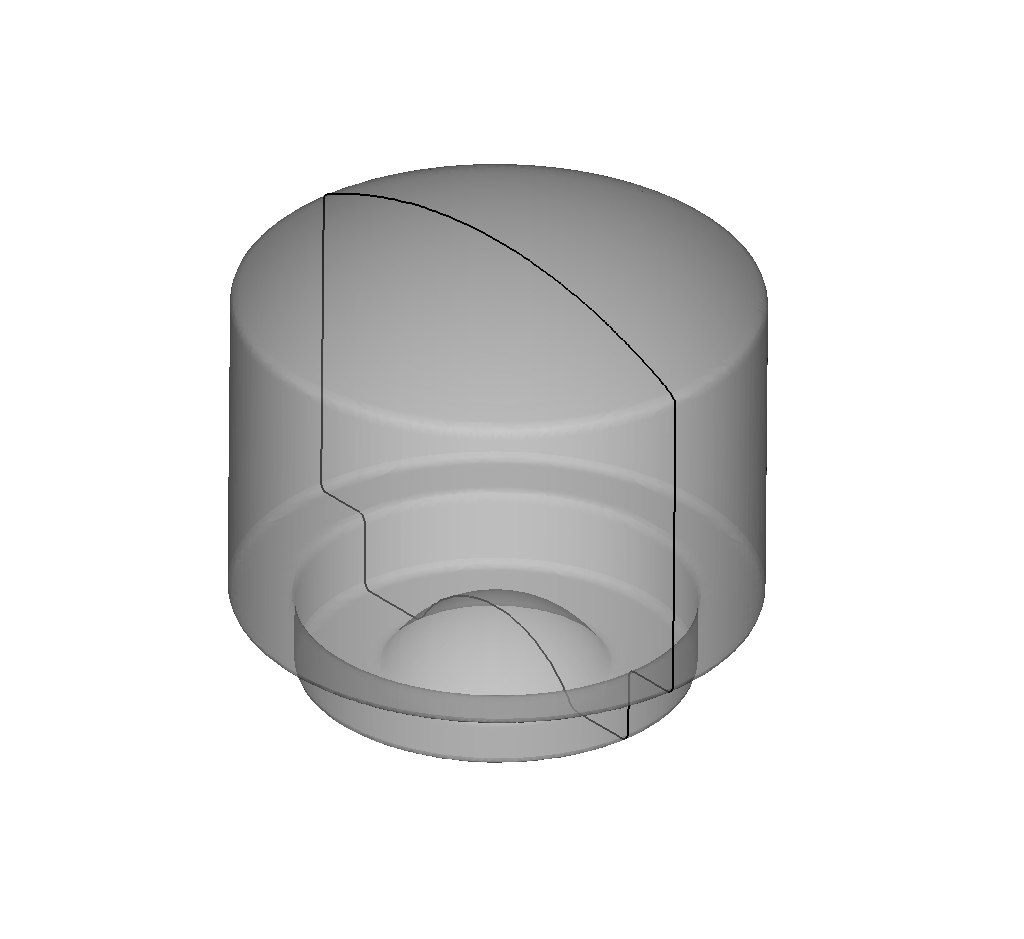}
\label{fig:ICEGeometry_1}
}
\hspace{3cm}
\subfigure[]
{
\includegraphics[trim=10.0cm 1.0cm 8.5cm 0.0cm, clip=true,scale=0.40]{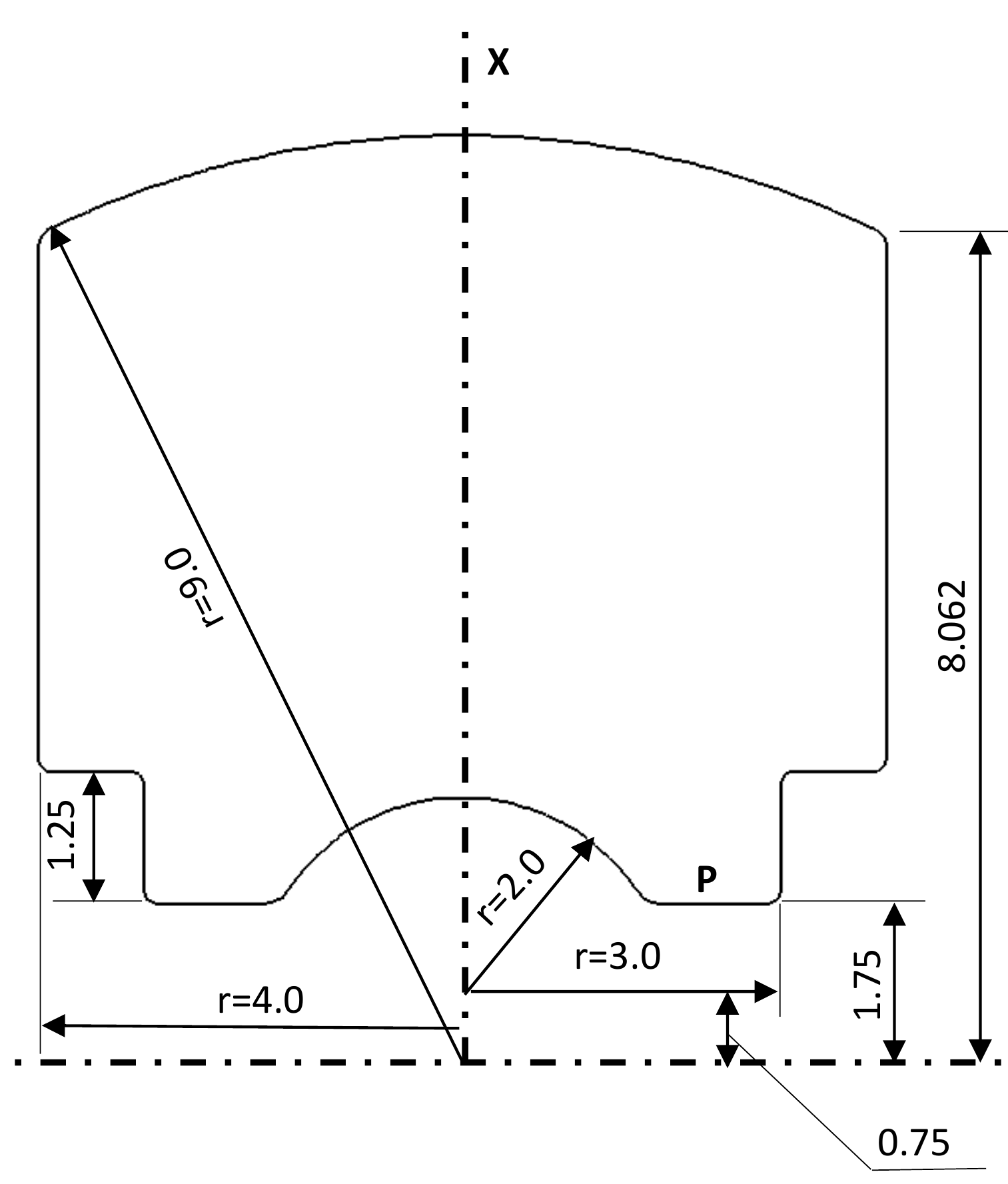}
\label{fig:ICEGeometry_2}
}
\caption{(a) The 0.5 isocontour of the volume fraction and the cross section of the geometry at the bottom dead center and (b) the cross-section of the geometry at the bottom dead center showing the dimensions in centimeters (not to scale).}
\label{fig:ICEGeometry}
\end{figure}

To demonstrate the capability of the algorithm to simulate flows with three dimensional complex, moving geometries, we compute the flow 
inside the geometry shown in Fig.~\ref{fig:ICEGeometry_1}, which mimics an internal combustion engine (without valves). The cross section 
with the dimensions is shown in Fig.~\ref{fig:ICEGeometry_2}. This is a closed system, and hence the mass within the geometry should remain 
constant with time, which makes this a good case for testing the conservative nature of the scheme. The piston P is initially located at $x(0)=0.0575$ m, 
and has a prescribed oscillatory motion given by $x(t) = 0.0375+0.02\cos(375t)$ m, which creates a flow within the geometry. The computational domain has 
dimensions 0.1 m $\times$ 0.1 m $\times$ 0.1 m, with a base mesh size of 32 $\times$ 32 $\times$ 32, with two levels of refinement, and a constant time step 
$\Delta t=1.5\times10^{-6}$ s is used. The refinement criterion tags all cut-cells for refinement.
\par
\begin{figure}[htpb!]
\subfigure[]
{
\includegraphics[trim=2.0cm 1.0cm 8.0cm 1.0cm, clip=true,scale=0.22]{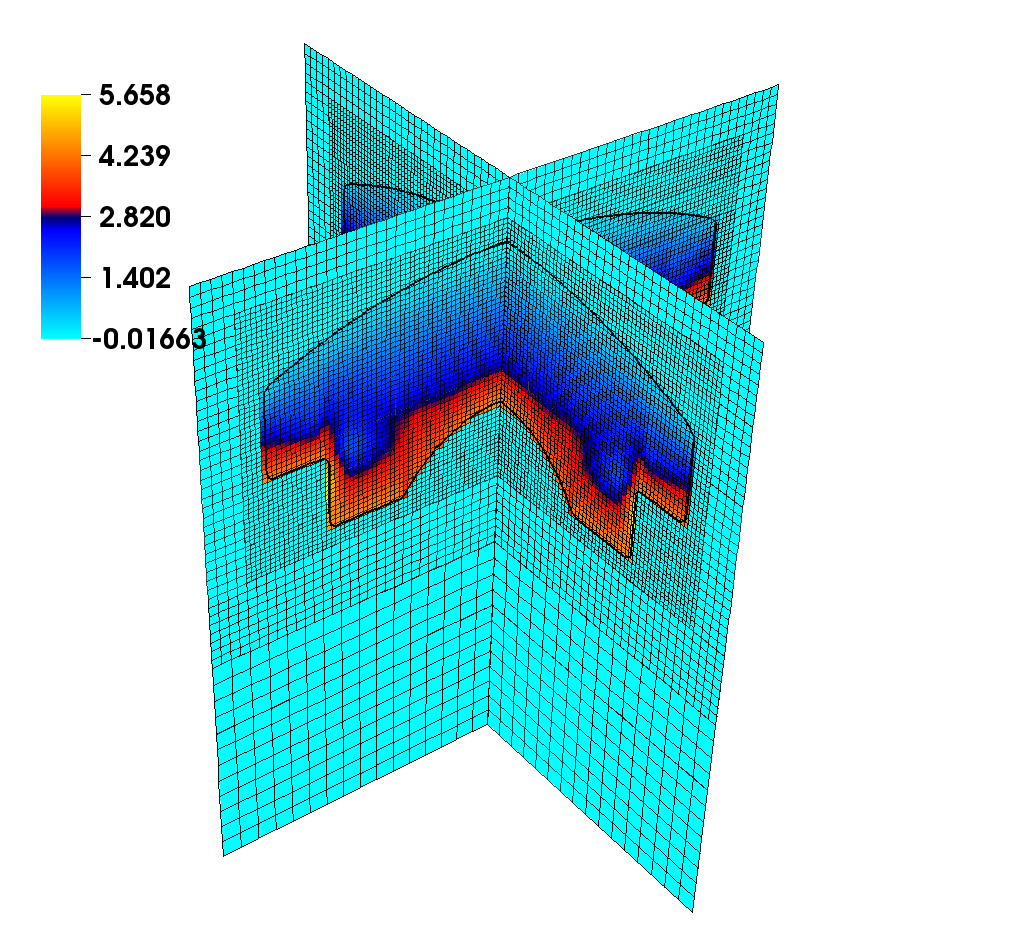}
\label{fig:ICEVelocityMesh_1}
}
\hspace{3cm}
\subfigure[]
{
\includegraphics[trim=2.0cm 1.0cm 8.0cm 1.0cm, clip=true,scale=0.22]{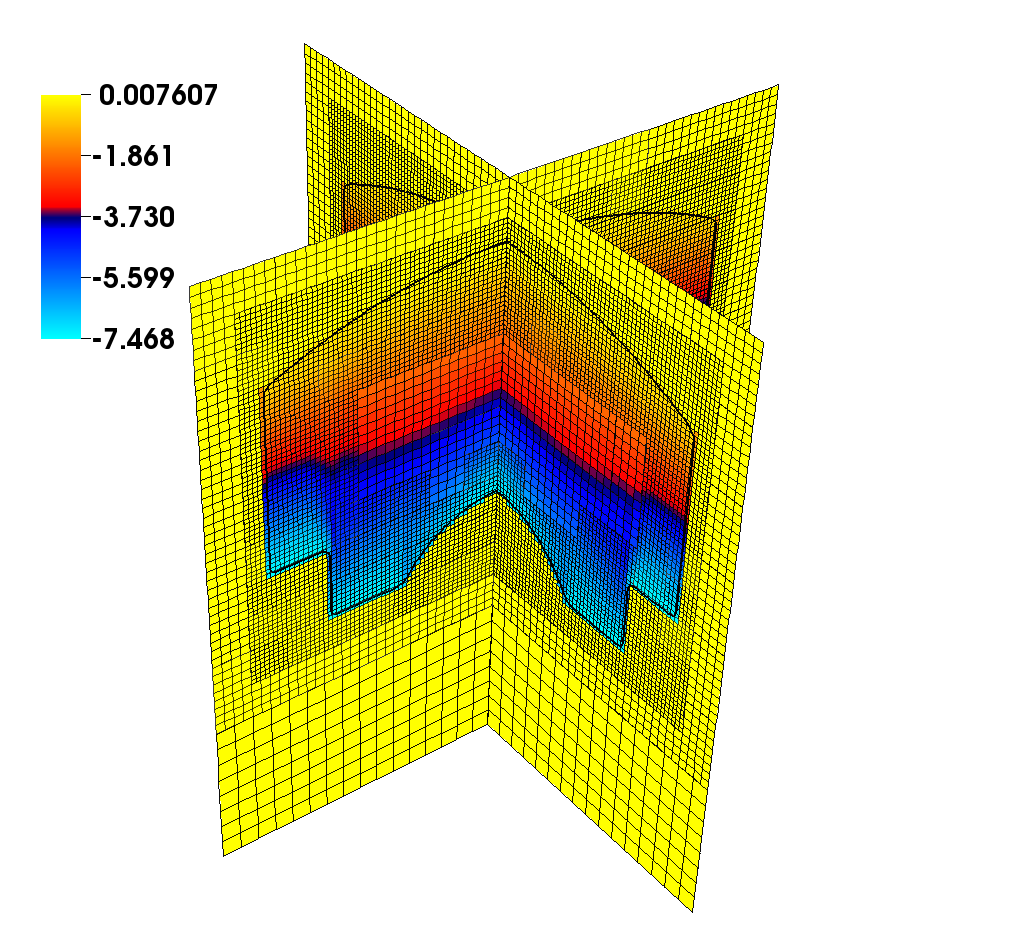}
\label{fig:ICEVelocityMesh_2}
}\\
\subfigure[]
{
\includegraphics[trim=2.0cm 1.0cm 8.0cm 1.0cm, clip=true,scale=0.22]{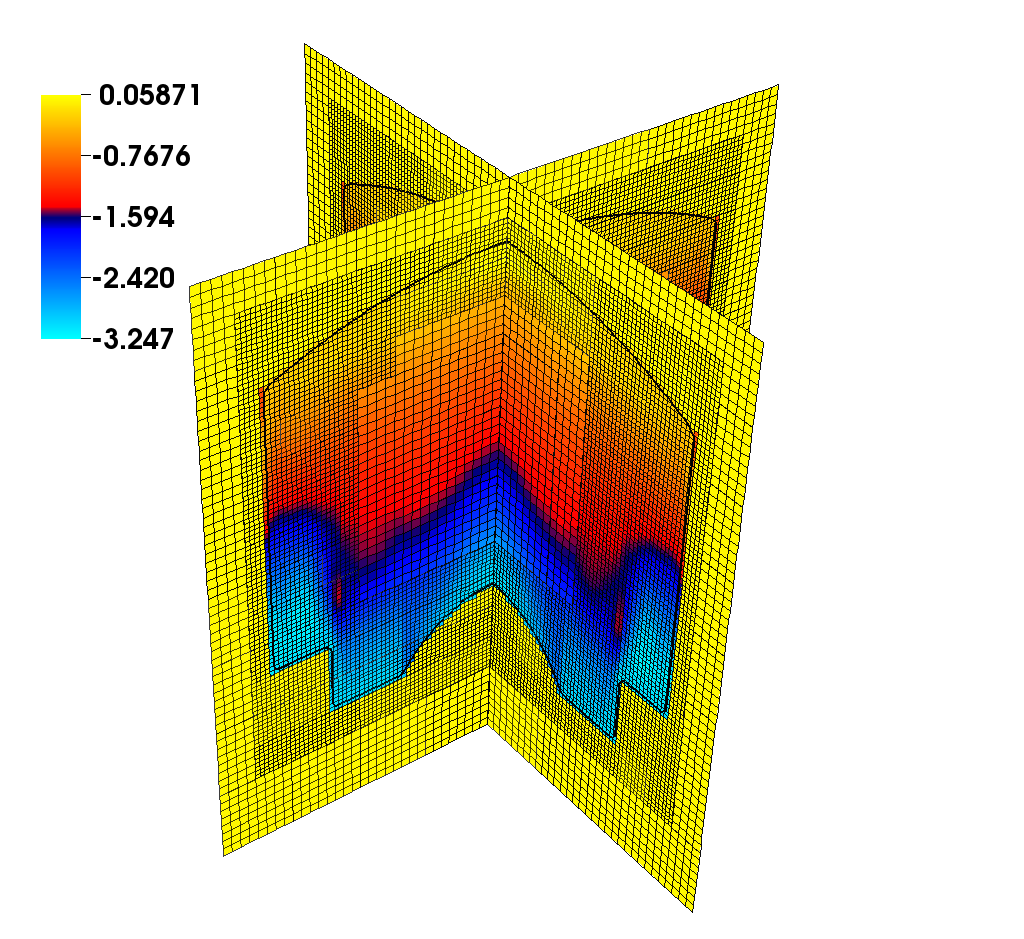}
\label{fig:ICEVelocityMesh_3}
}
\hspace{3cm}
\subfigure[]
{
\includegraphics[trim=2.0cm 1.0cm 8.0cm 1.0cm, clip=true,scale=0.22]{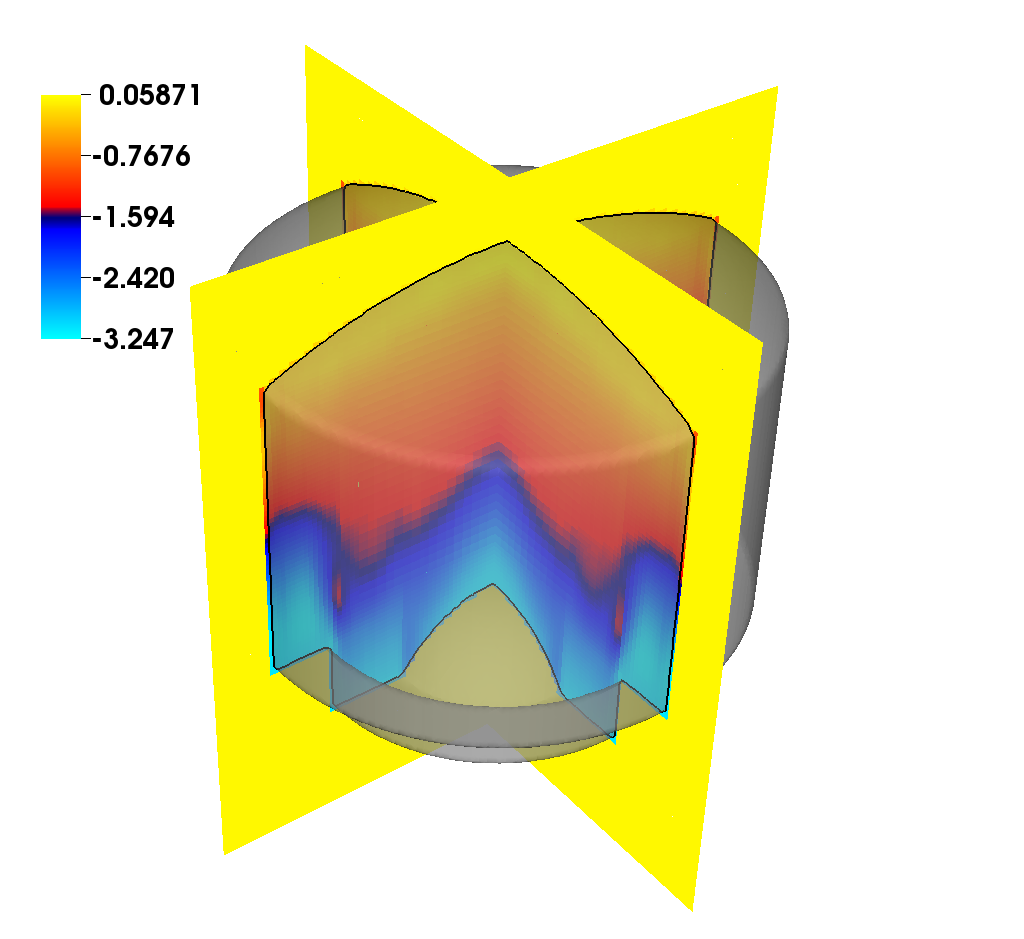}
\label{fig:ICEVelocityMesh_4}
}\\
\subfigure[]
{
\includegraphics[trim=0.0cm 0.0cm 1.0cm 1.0cm, clip=true,scale=0.4]{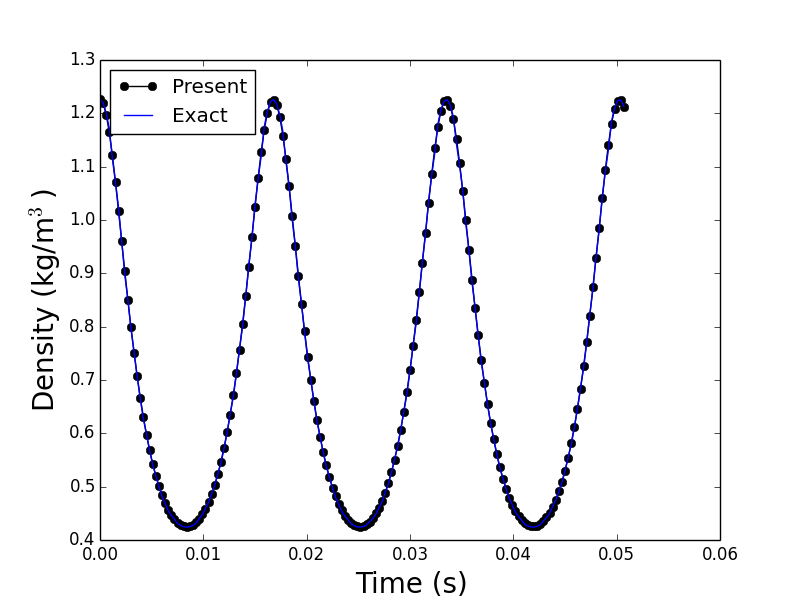}
\label{fig:DensityVsTime}
}
\subfigure[]
{
\includegraphics[trim=0.0cm 0.0cm 1.0cm 1.0cm, clip=true,scale=0.4]{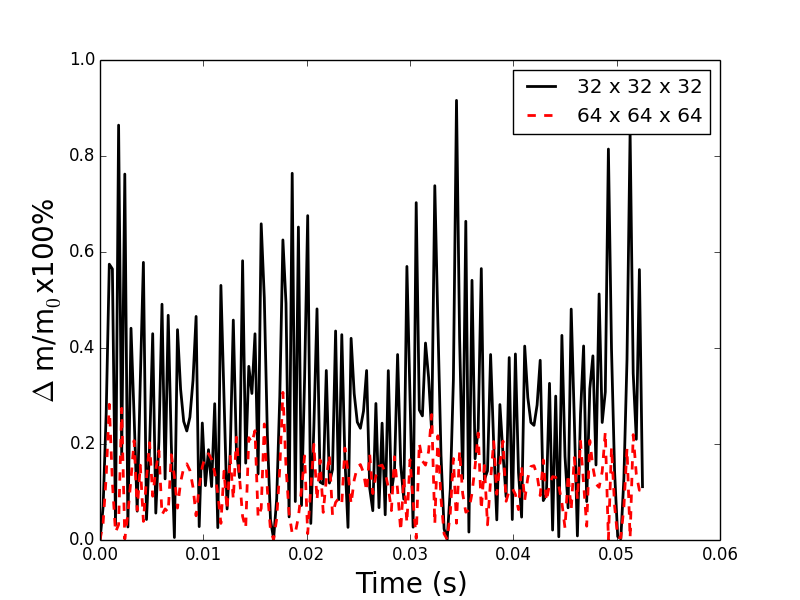}
\label{fig:DeltaMVsTime}
}
\caption{Contours of axial velocity (m/s) and the 3-level mesh at (a) $t=0.015$ s, (b) $t=0.021$ s and (c) $t=0.024$ s, (d) contours of axial velocity and 0.5 isocontour of volume fraction showing the surface of the geometry, (e) average density as a function of time and (f) the percentage error in mass as a function of time for mesh sizes $32^3$ and $64^3$.}
\label{fig:ICE}
\end{figure}

 Fig.~\ref{fig:ICEVelocityMesh_1}-\ref{fig:ICEVelocityMesh_3} show two perpendicular slices with the contours of axial 
velocity, and the 3-level mesh for different time instants. It can be seen that the geometry is always enclosed within the finest level of refinement. 
 Fig.~\ref{fig:ICEVelocityMesh_4} shows the slices of axial velocity and the 0.5 isocontour of the volume fraction at $t=0.024$ s.
 As the piston oscillates, the fluid in the geometry is compressed and expanded, and hence the density changes continuously. Since mass 
remains constant, the average exact density in the geometry at any time can be computed as $\rho_\mathrm{exact}(t)=m(0)/V(t)$, where $m(0)$ is the initial mass in 
the geometry, and $V(t)$ is the volume enclosed by a geometry at time $t$. Fig.~\ref{fig:DensityVsTime} shows a comparison of the computed and exact average density within the geometry 
as a function of time over a time period of three cycles of oscillation, which shows good quantitative comparison. To test the conservative nature of 
the scheme, the percentage error in mass defined as $\Delta m (\%) = \frac{m(t)-m(0)}{m(0)}\times100$, where $m(t) = \int\limits_{V(t)}\rho(t)\,dV$, is 
computed as a function of time for two mesh sizes -- $32^3$ and $64^3$, and shown in Fig.~\ref{fig:DeltaMVsTime}. The maximum percentage error is for 
the $32^3$ mesh is $0.92\%$, and reduces to $0.31\%$ for the $64^3$ mesh. Since the error in mass reduces with refinement, 
it shows that the algorithm ensures conservation.

\subsection{Shock-cone interaction}
To demonstrate the capability of the numerical algorithm to simulate high-speed flows with complex geometries and sharp corners in three dimensions, 
we extend Schardin's problem described in Section~\ref{sec:shock-wedge} to three dimensions, similar to \citet{bennett2018moving}. This test case consists 
of a rigid right circular cone of radius $R=0.02$ m and height $h=0.02$ m, interacting with a stationary Mach 1.34 shock wave. The domain size is 0.1 m $\times$ 
0.1 m $\times$ 0.05 m with a base mesh size 32 $\times$ 64 $\times$ 64, with three levels of refinement. The nose of the cone located at $(x,y,z)=(0.0,0.0,-0.02)$ m at $t=0$. 
The initial condition corresponds to a stationary shock wave located at $z=-0.02$ m, characterized by the left and right-hand states given by $\rho_L=0.595$ kg/m$^3$, $u_L=459.54$ m/s, $p_L=5e4$ Pa,
and $\rho_R=0.944$ kg/m$^3$, $u_R=289.86$ m/s, $p_R=9.6e4$ Pa. The cone has a constant horizontal velocity of $u=u_L=459.54$ m/s.
. The refinement criterion tags all cut-cells, and has an additional gradient based detector for resolving high-gradient regions.  
\par
\begin{figure}[htpb!]
\subfigure[]
{
\includegraphics[trim=3.2cm 5.0cm 2.0cm 7.0cm, clip=true,scale=0.28]{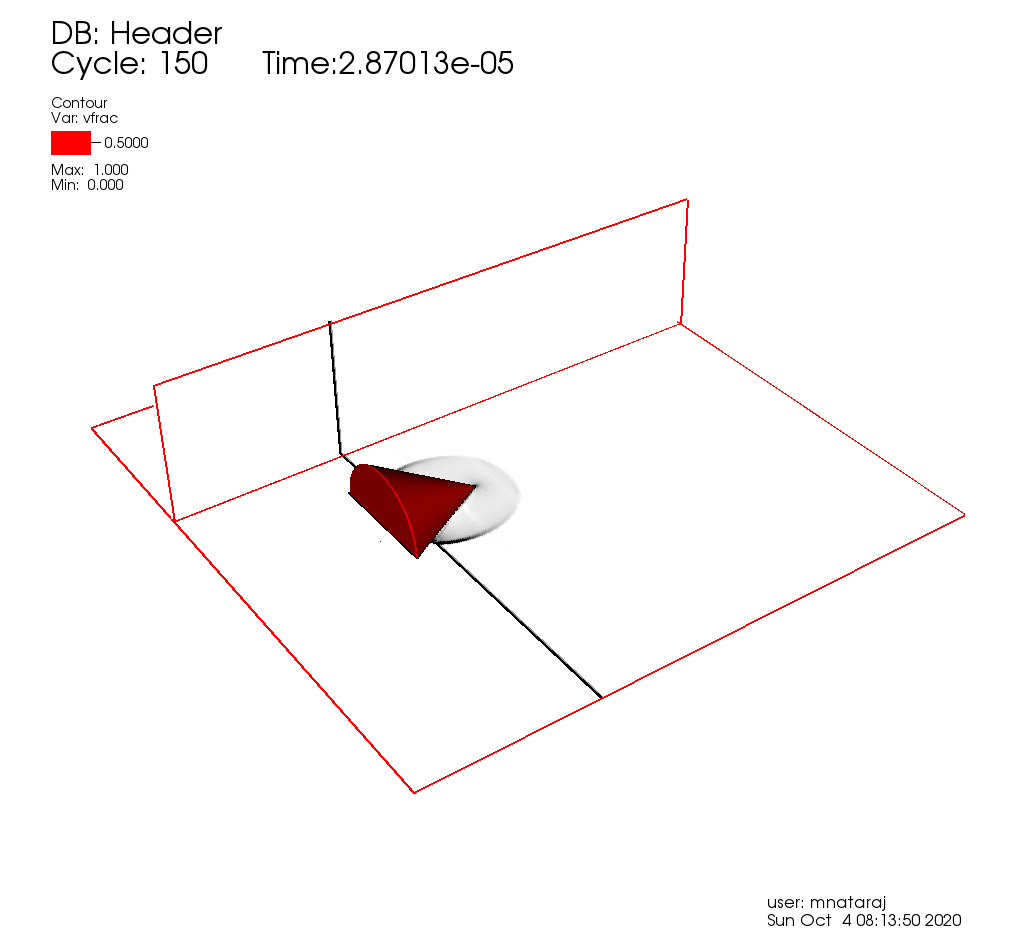}
\label{fig:MovingCone_Schlieren_1}
}
\subfigure[]
{
\includegraphics[trim=3.2cm 5.0cm 2.0cm 7.0cm, clip=true,scale=0.28]{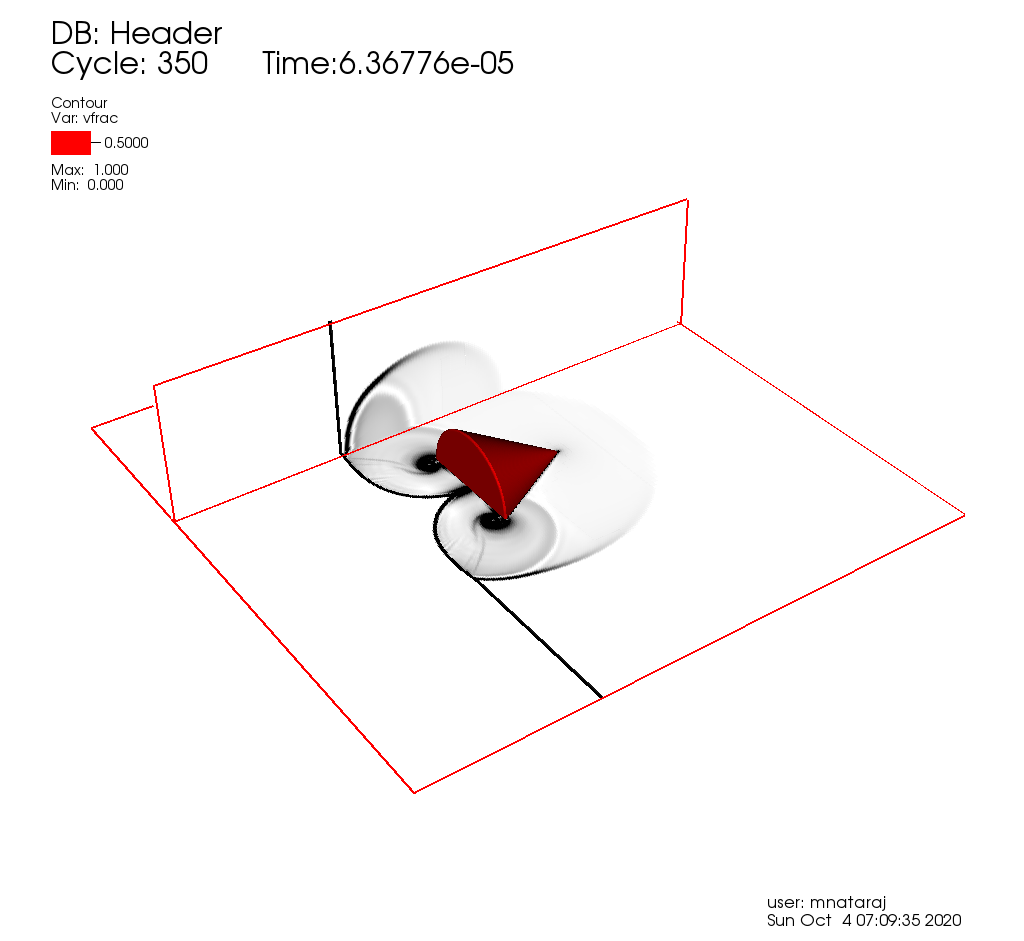}
\label{fig:MovingCone_Schlieren_3}
}
\subfigure[]
{
\includegraphics[trim=3.2cm 5.0cm 2.0cm 7.0cm, clip=true,scale=0.28]{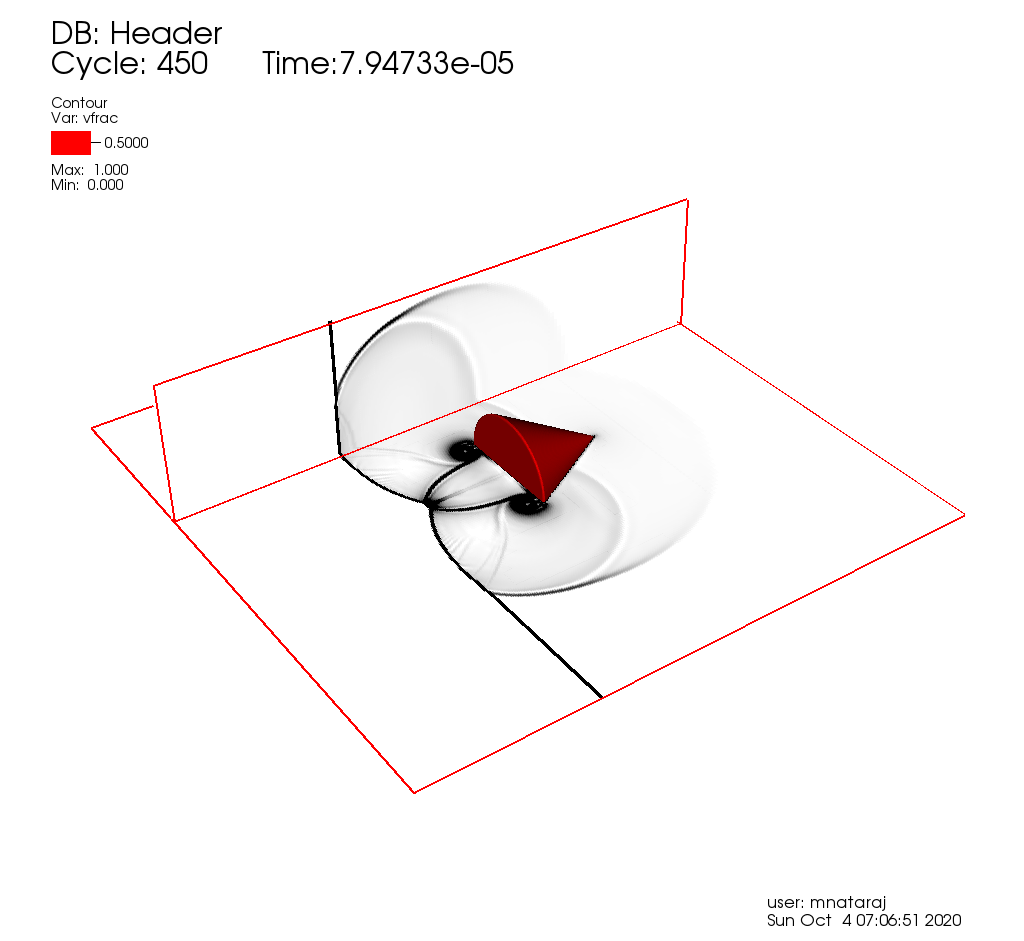}
\label{fig:MovingCone_Schlieren_4}
}
\subfigure[]
{
\includegraphics[trim=3.2cm 5.0cm 2.0cm 7.0cm, clip=true,scale=0.28]{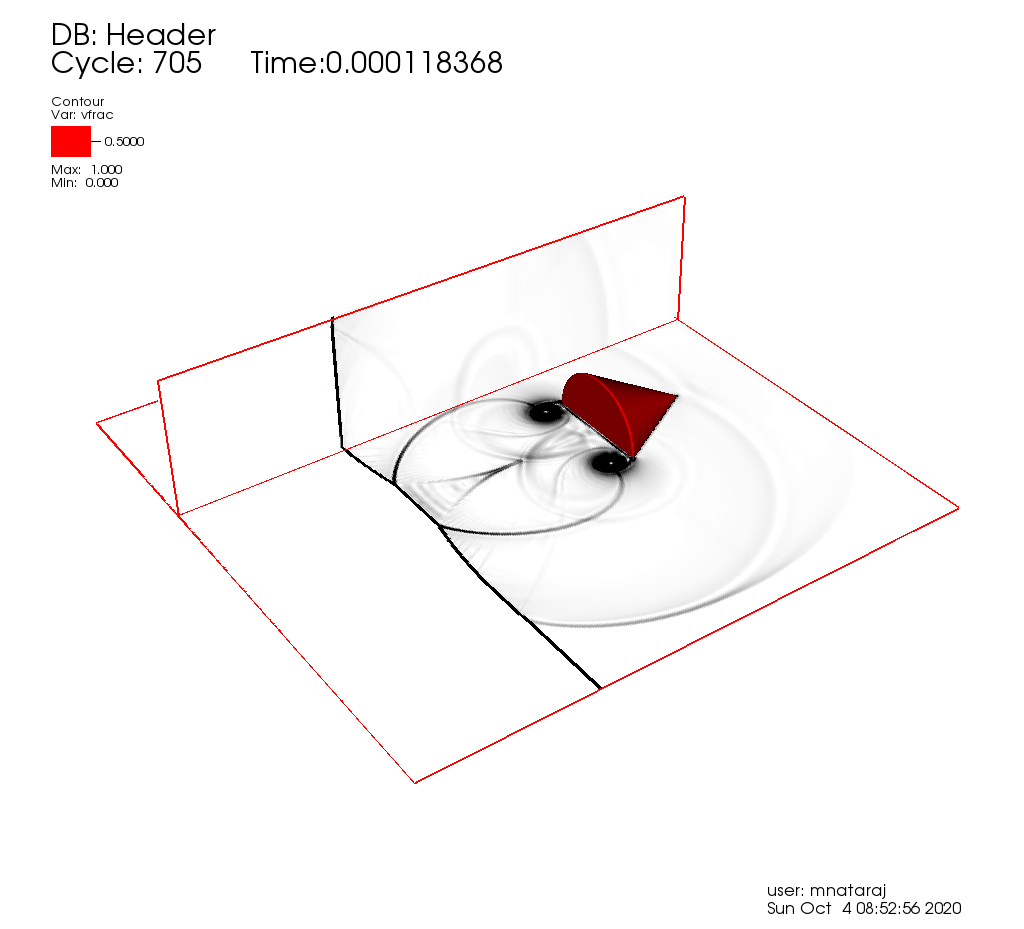}
\label{fig:MovingCone_Schlieren_6}
}
\subfigure[]
{
\includegraphics[trim=3.2cm 5.0cm 2.0cm 7.0cm, clip=true,scale=0.28]{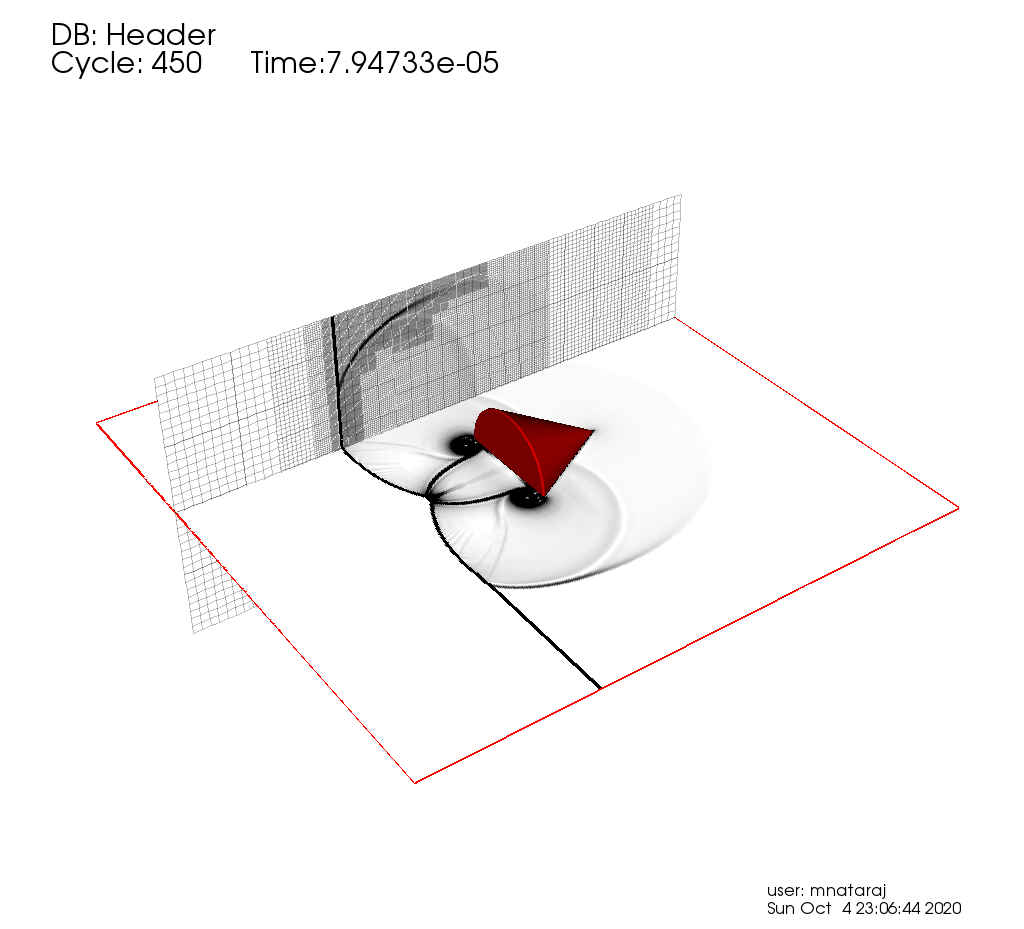}
\label{fig:MovingCone_Schlieren_Mesh}
}
\subfigure[]
{
\includegraphics[trim=3.5cm 5.0cm 2.0cm 7.0cm, clip=true,scale=0.28]{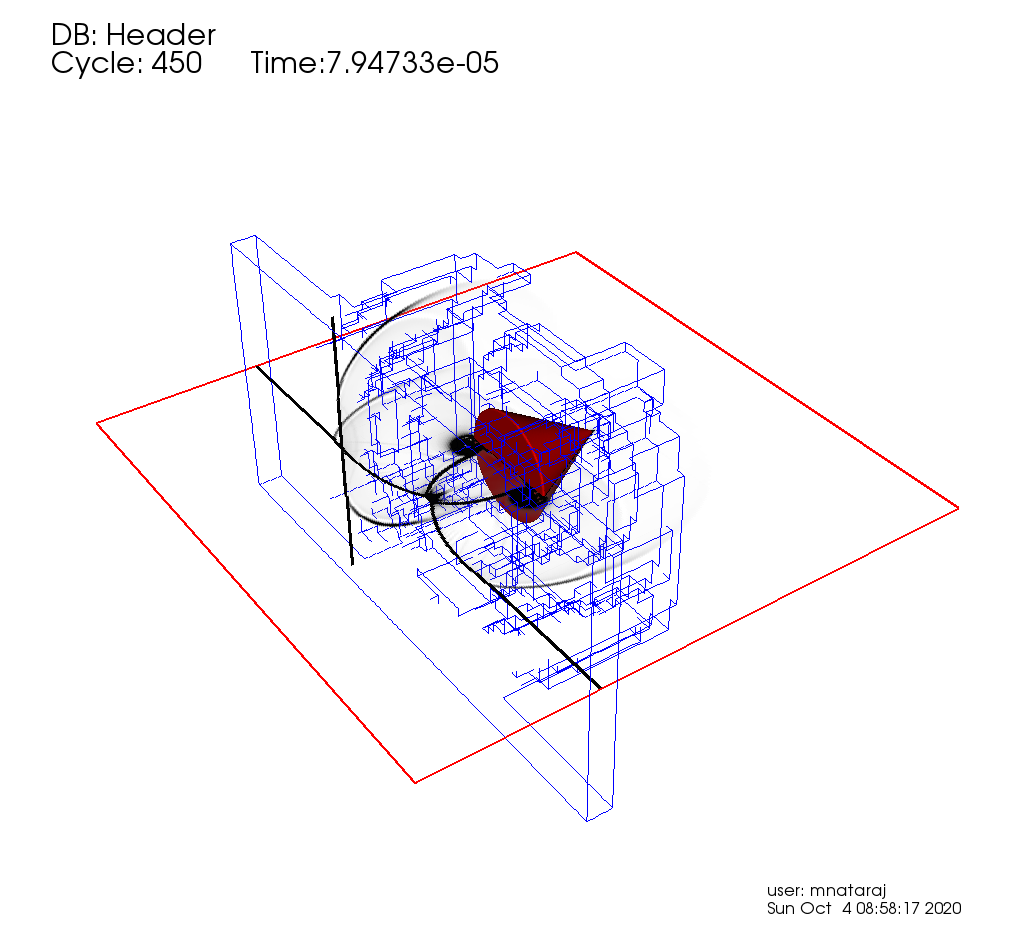}
\label{fig:MovingCone_Schlieren_FineLevel}
}
\caption{The 0.5 isocontour of volume fraction and the numerical Schlieren ($\vert\nabla\rho\vert$) images on two perpendicular planes at (a) $t=28.7$ $\mu$s, (b) $t=63.7$ $\mu$s, (c) $t=79.5$ $\mu$s, (d) $t=118$ $\mu$s, (e) the 4-level mesh shown on the vertical plane, and (f) the finest level of refinement showing the high-gradient regions of the flow.}
\label{fig:MovingCone}
\end{figure}

 Fig.~\ref{fig:MovingCone_Schlieren_1}-\ref{fig:MovingCone_Schlieren_6} show the temporal evolution of the numerical Schlieren ($\vert\nabla\rho\vert$) 
on two perpendicular slices. As the cone interacts with the shock, the flow features are noticeably different 
compared to the shock-wedge interaction case. The high-gradient features evolve spherically, and the interaction of these regions with the strong vortices at the rear of the cone lead to the creation of multiple weak shocks as shown in Fig.~\ref{fig:MovingCone_Schlieren_6}. Fig.~\ref{fig:MovingCone_Schlieren_Mesh} shows the 4-level mesh on the vertical slice showing the effectiveness
of the refinement criterion in resolving the high-gradient regions. Fig.~\ref{fig:MovingCone_Schlieren_FineLevel} shows an instantaneous image of the finest level of refinement, 
showing the three-dimensional spherical-like nature of the high-gradient regions. 

{\color{black}
\subsection{Horizontally moving cylinder in initially quiescent flow}
This test case consists of a horizontally moving cylinder in initially quiescent ambient fluid with $\rho_\infty=1.226$ kg/m$^3$, $\mu_\infty=0.613$ kg/ms, at a Reynolds number 
$Re=\rho_\infty U_c D/\mu_\infty=40$, based on the cylinder diameter $D=0.4$ m and cylinder velocity $U_c=50$ m/s. To evaluate the performance of the moving EB numerical scheme, the pressure 
and skin friction coefficients over the surface of the cylinder are computed and compared with the results in the 
literature. The pressure coefficient over the surface of the cylinder is given by $C_p = (p-p_\infty)/(1/2\rho_\infty U_c^2)$, and the skin friction coefficient is given by 
$C_f = \tau_f/(1/2\rho_\infty U_c^2)$, where $\tau_f$ is the shear stress tangential to the surface given by
\begin{eqnarray}\label{eqn:tauf}
\tau_f = (\tau_{yx}n_x+\tau_{yy}n_y)n_x-(\tau_{xx} n_x + \tau_{xy} n_y)n_y,
\end{eqnarray}
where $n_x$ and $n_y$ are the components of the surface normal on the body (pointing towards the wall). The domain size is 20 m $\times$ 10 m, with a base mesh size of 512 $\times$ 256, 
with 3 levels of refinement, which gives a resolution of $\sim$82 points in the cylinder diameter. At $t=0$, the center of the cylinder is located at $(x,y)=(2.0,0.0)$. 
A geometric refinement criterion is used to track the cylinder and its vicinity, which tags all cells in the domain which satisfies $x_c(t)-3.0 < x < x_c(t)+3.0$ and $y_c(t)-0.25<y<y_c(t)+0.25$, 
where $(x_c(t),y_c(t))$ are the coordinates of the center of the cylinder at any instant of time. The comparison is performed at a non-dimensional time of $t^*= tU_c/D = 31.25$. 
Fig.~\ref{fig:MovingCylinder_Re40}(a)-(c) show the instantaneous contours of velocity magnitude at $t=0.075$ s, 0.15 s, and 0.3 s. Fig.~\ref{fig:MovingCylinder_Pressure} and 
\subref{fig:MovingCylinder_SkinFriction} show the comparison of the pressure coefficient and skin friction coefficient over the surface of the cylinder respectively. The angle $\theta$ 
is measured from the stagnation point of the cylinder. Good quantitative comparison is observed, although minor oscillations can be seen in the surface data as has been 
observed by \citet{al2017versatile}. 

\begin{figure}[htpb!]
\centering
\subfigure[]
{
\includegraphics[trim=1.0cm 15.0cm 1.0cm 11.0cm, clip=true,scale=0.4]{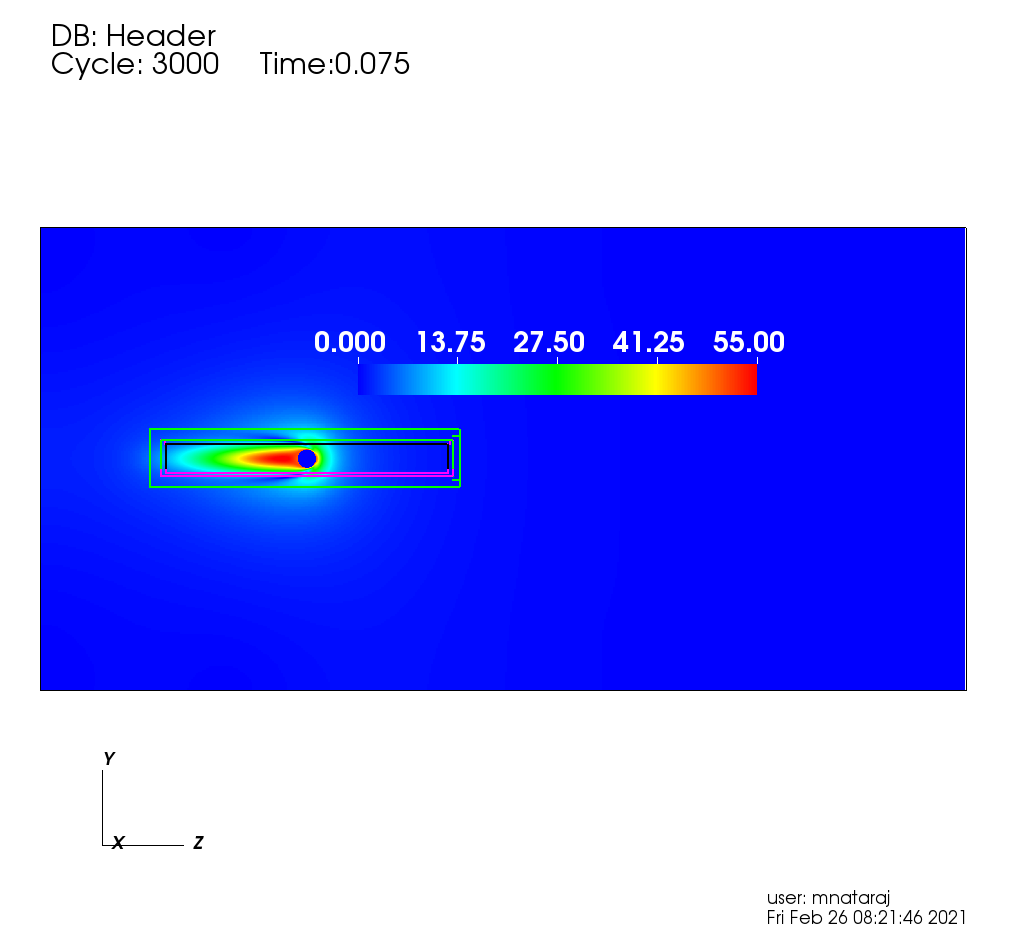}
\label{fig:MovingCylinder_p075}
}\\
\subfigure[]
{
\includegraphics[trim=1.0cm 15.0cm 1.0cm 11.0cm, clip=true,scale=0.4]{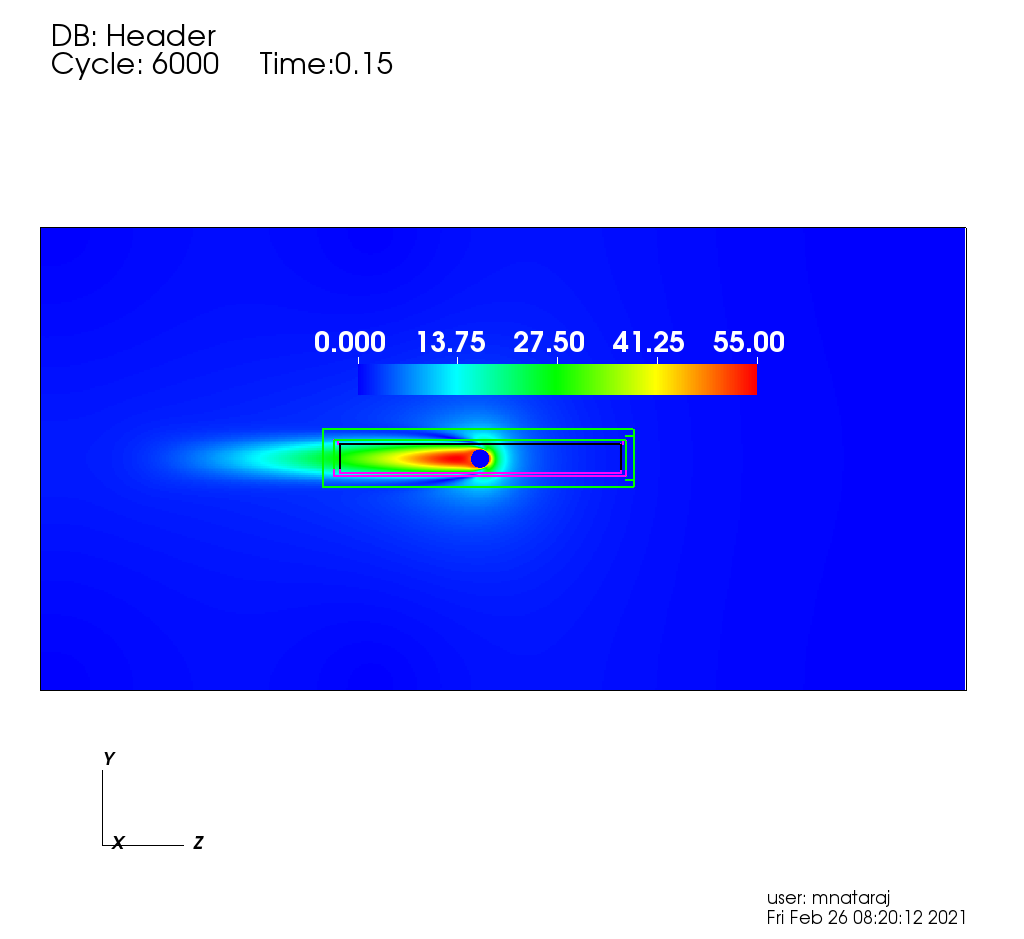}
\label{fig:OscillatingCylinder_SkinFriction_feq1p2f0_HOLS}
}\\
\subfigure[]
{
\includegraphics[trim=1.0cm 15.0cm 1.0cm 11.0cm, clip=true,scale=0.4]{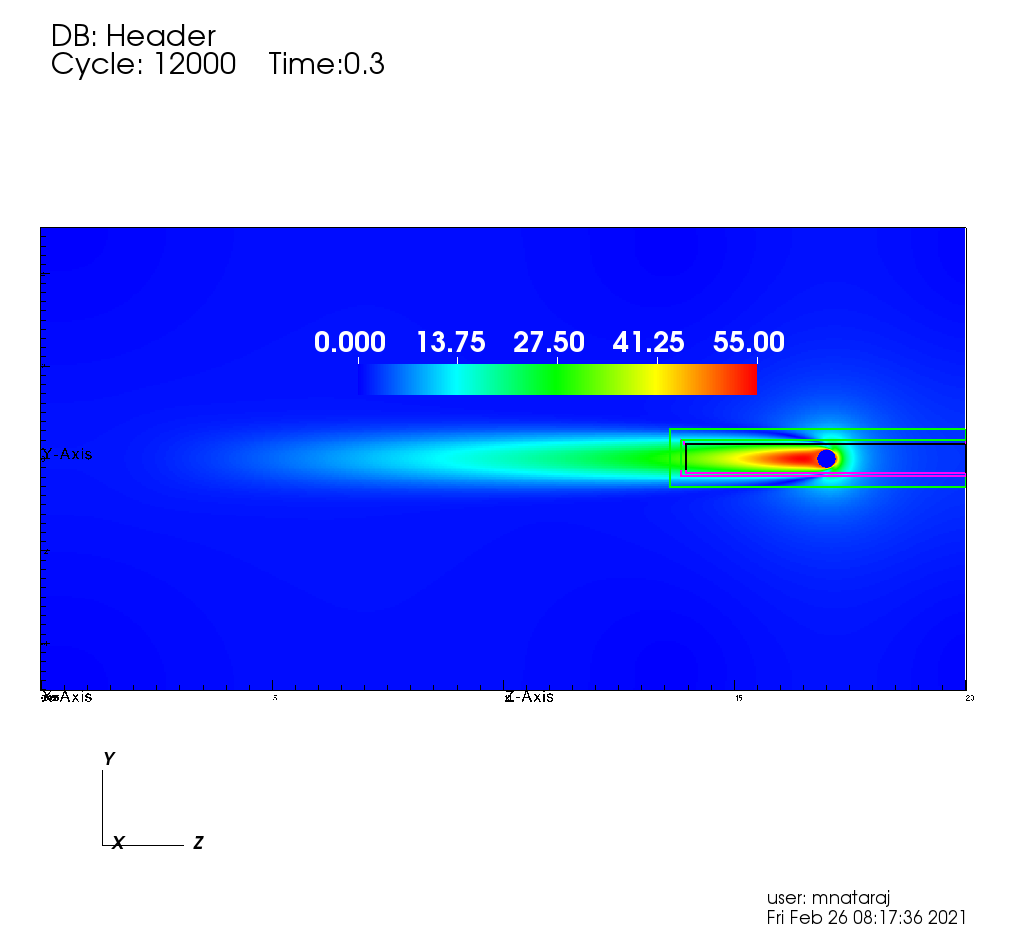}
\label{fig:OscillatingCylinder_SkinFriction_feqp8f0_HOLS}
}
\caption{Instantaneous contours of velocity magnitude (m/s) for the horizontally moving cylinder at $Re=40$. The boxes show the three levels of mesh refinement above the base level.}
\label{fig:MovingCylinder_Re40}
\end{figure}

\begin{figure}[htpb!]
\subfigure[]
{
\includegraphics[trim=0.5cm 4.5cm 14.0cm 7.0cm, clip=true,scale=0.6]{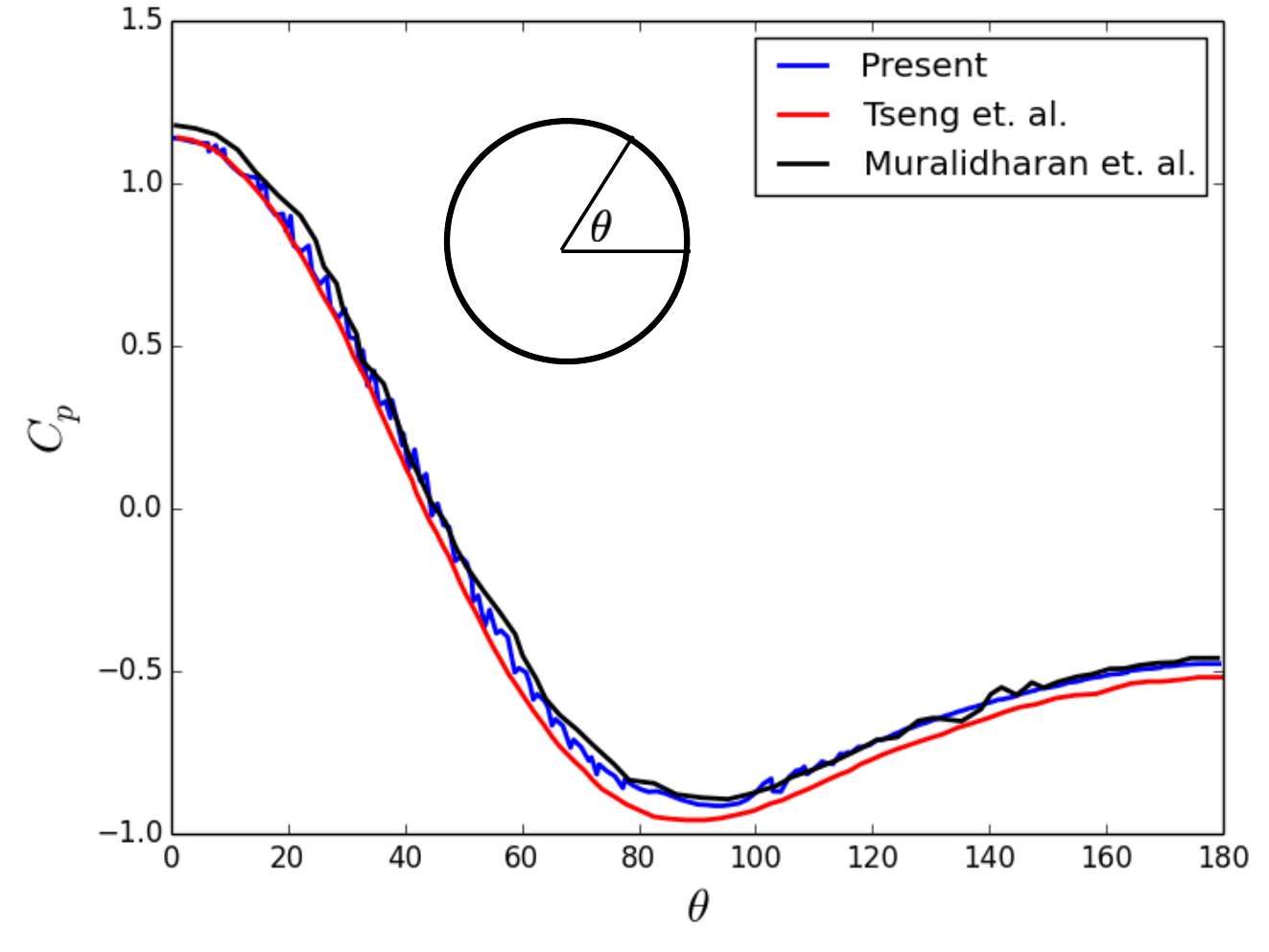}
\label{fig:MovingCylinder_Pressure}
}
\subfigure[]
{
\includegraphics[trim=14.0cm 4.5cm 0.5cm 7.0cm, clip=true,scale=0.6]{MovingCylinder_SurfaceData}
\label{fig:MovingCylinder_SkinFriction}
}
\caption{Comparison of (a) pressure coefficient and (b) skin friction coefficient on the surface of the cylinder with the results of \citet{tseng2003ghost} and \citet{muralidharan2018simulation}.}
\end{figure}

\subsection{Inline oscillating cylinder in initially quiescent flow}
A horizontally oscillating cylinder in initially quiescent flow has been widely studied in the literature both experimentally and numerically \citep{dutsch1998low,iliadis1998viscous,al2017versatile}. The test case consists of a cylinder of diameter D in a quiescent fluid with an imposed oscillatory motion given by
\begin{eqnarray*}
x(t) = -A_e\sin(2\pi f_e t),
\end{eqnarray*}
where $A_e$ is the amplitude of oscillation with frequency $f_e$. The relevant non-dimensional parameters are the Reynolds number $Re = \rho_\infty U_\mathrm{max}D/\mu_\infty$, and the 
Keulegan--Carpenter number $KC = U_\mathrm{max}/f_eD$, where $U_\mathrm{max}=2\pi f_e A_e$, is the velocity amplitude attained by the cylinder during the oscillatory motion. Consistent with the test case 
of \citet{dutsch1998low}, we use $KC=5$, which gives $A_e = 5D/2\pi$ m, and $Re=100$, which gives $f_e = 100\mu_\infty/(5D^2\rho_\infty)$ cycles per second. In the current case, $D=0.4$ m, $\rho_\infty = 1.226$ kg/m$^3$, and $\mu_\infty = 0.3$ kg/ms, which give $A_e = 0.3183$ m, and $f_e = 30.58727$ cycles per second. The domain size is 20 m $\times$ 10 m, with a base mesh size of 512 $\times$ 256,
with 2 levels of refinement, which gives a resolution of $\sim$41 points in the cylinder diameter. Since the amplitude of oscillation is small compared to the size of the domain, a static refinement criterion is used,  which tags all cells in the domain which satisfies $9.0 < x < 11.0$ and $-0.5<y<0.5$. Fig.~\ref{fig:InlineOscillatingCylinder_VortAndPres} shows 
the instantaneous contours of vorticity and pressure at various phase positions $\theta=2\pi f_e t = 0^0, 96^0, 192^0, 288^0$. As the cylinder oscillates, symmetric vortices develop, and when the 
direction of oscillation is reversed, the vortex pair gets separated, and a new pair of vortices are formed resulting in a wake reversal as has been observed by \citet{dutsch1998low}. 
Fig.~\ref{fig:InlineOscillatingCylinder_Velocities} shows the comparison of the normalized velocities in the horizontal and vertical directions at four different streamwise locations given by 
$x/D = -0.6, 0.0, 0.6, 1.2$ for different phase positions $\theta=2\pi f_e t = 180^0, 210^0, 330^0$. The total drag force on the cylinder in the streamwise direction is given by the streamwise component of the force vector 
\begin{eqnarray*}
F_D = x~\mathrm{component~of}\int\limits_S (p\bm{n}- \bm{\tau}\cdot\bm{n})\,dA = \int\limits_S (pn_x - (\tau_{xx}n_x+\tau_{xy}n_y))\,dA,
\end{eqnarray*}
where $S$ denotes the surface of the cylinder. Fig.~\ref{fig:DragHistory_InlineOscillatingCylinder} shows the comparison of the drag force over the cylinder as a function of time. The 
drag force has been normalized to match the results of \citet{dutsch1998low}. Good quantitative comparison is observed for all quantities.

\begin{figure}[htpb!]
\centering
\subfigure[]{
\includegraphics[trim=4.0cm 9.0cm 2.0cm 5.0cm, clip=true,scale=0.25]{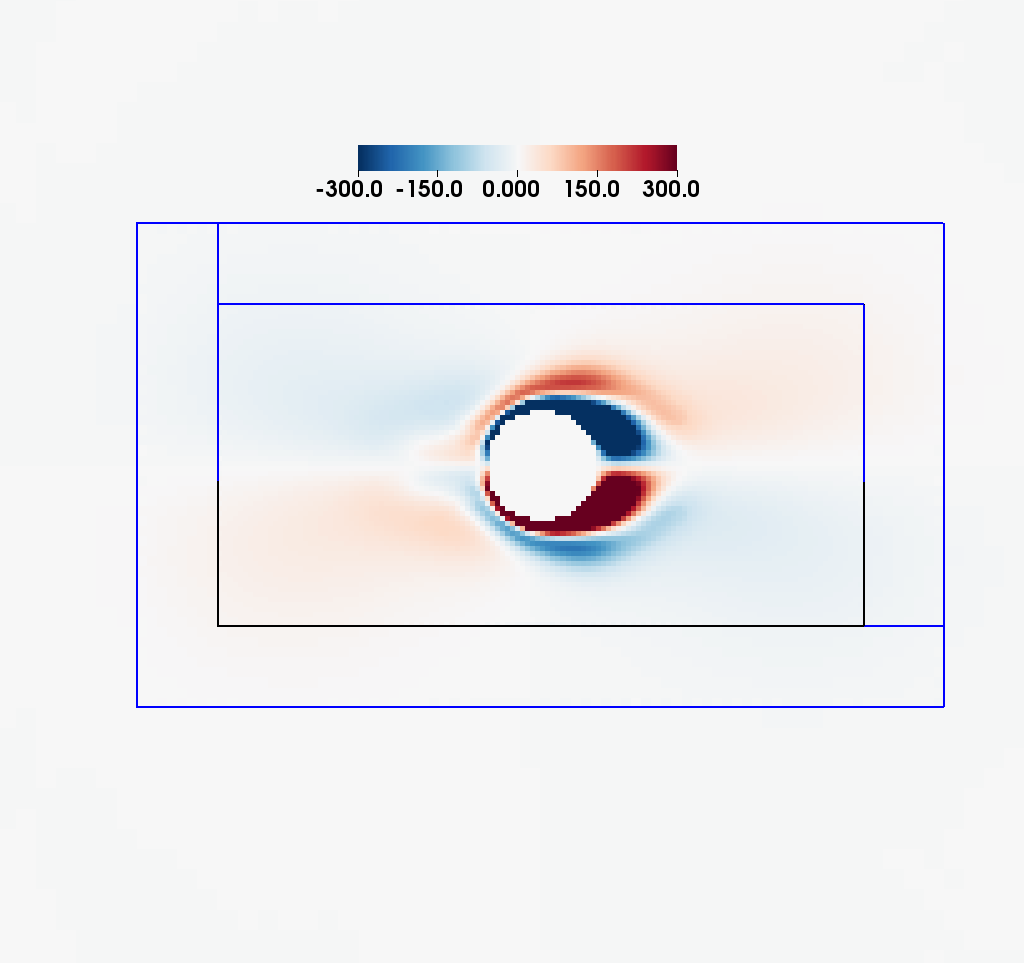}
}
\subfigure[]{
\includegraphics[trim=4.0cm 9.0cm 2.0cm 5.0cm, clip=true,scale=0.25]{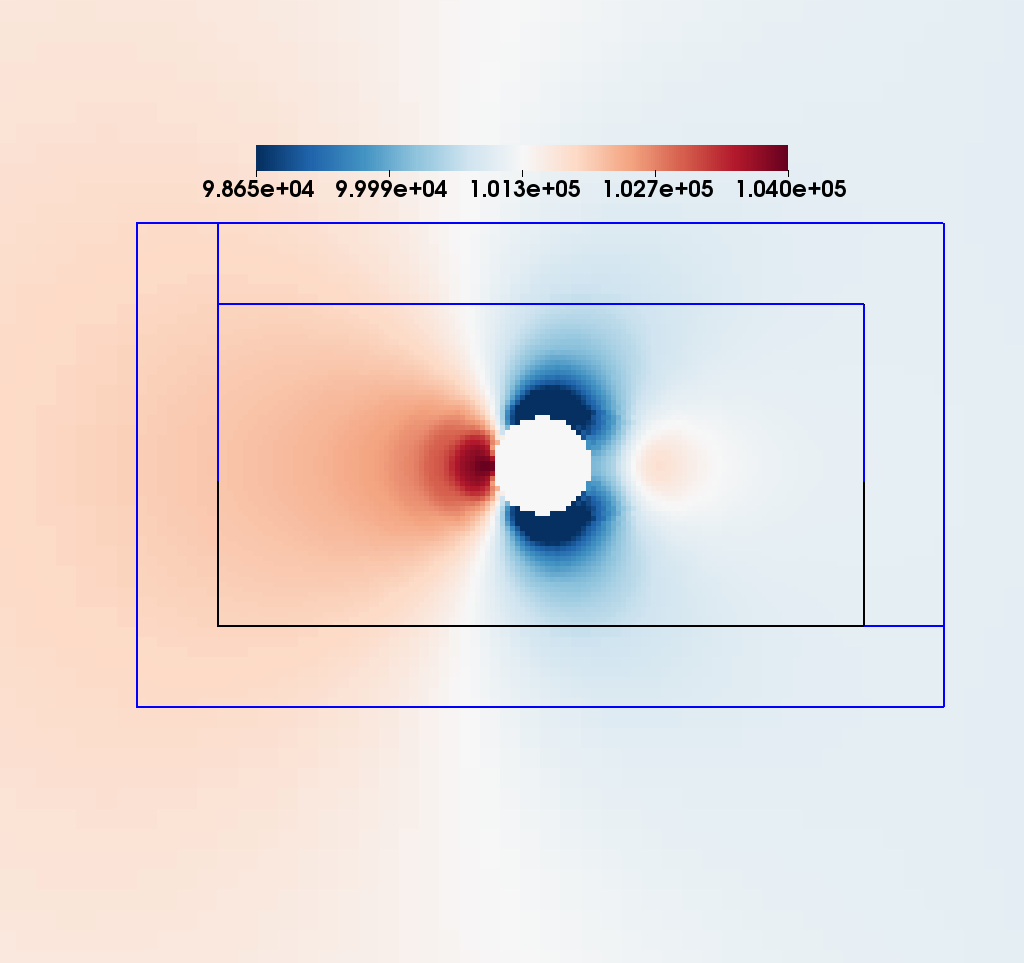}
}\\
\subfigure[]{
\includegraphics[trim=4.0cm 9.0cm 2.0cm 7.0cm, clip=true,scale=0.25]{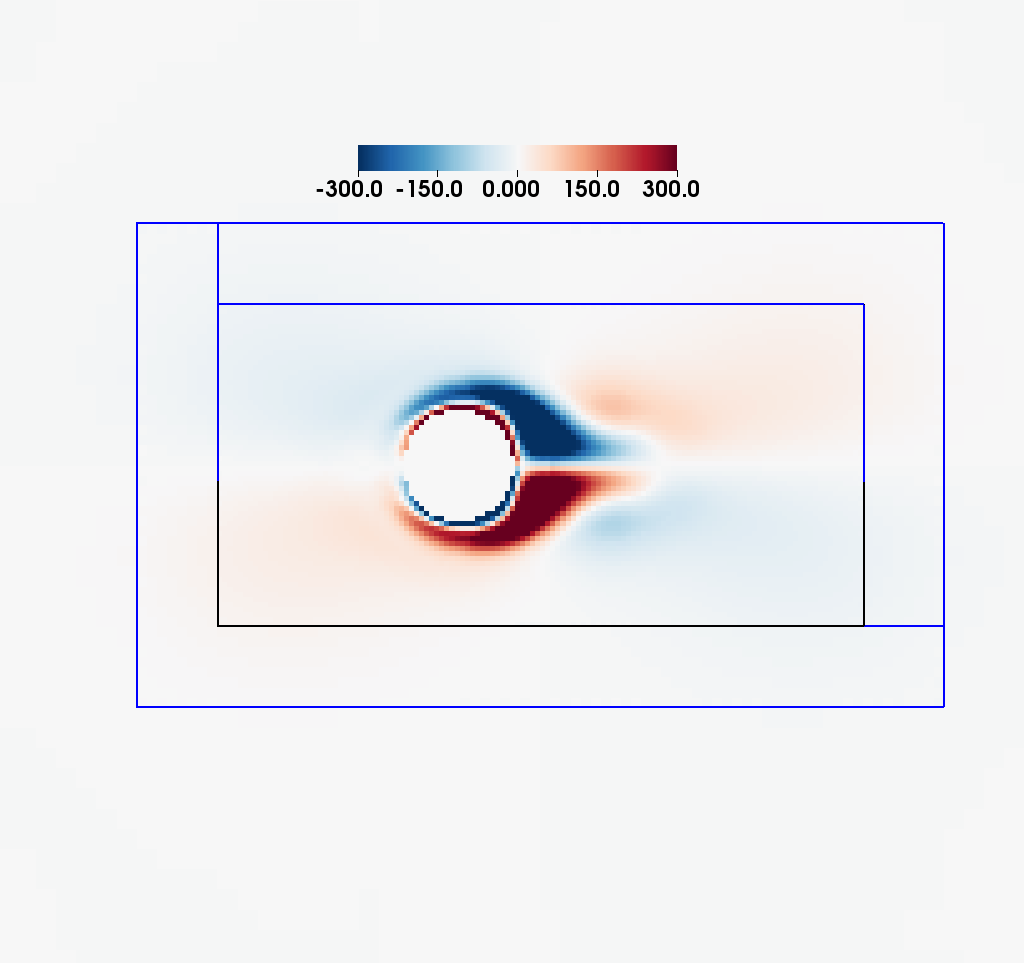}
}
\subfigure[]{
\includegraphics[trim=4.0cm 9.0cm 2.0cm 7.0cm, clip=true,scale=0.25]{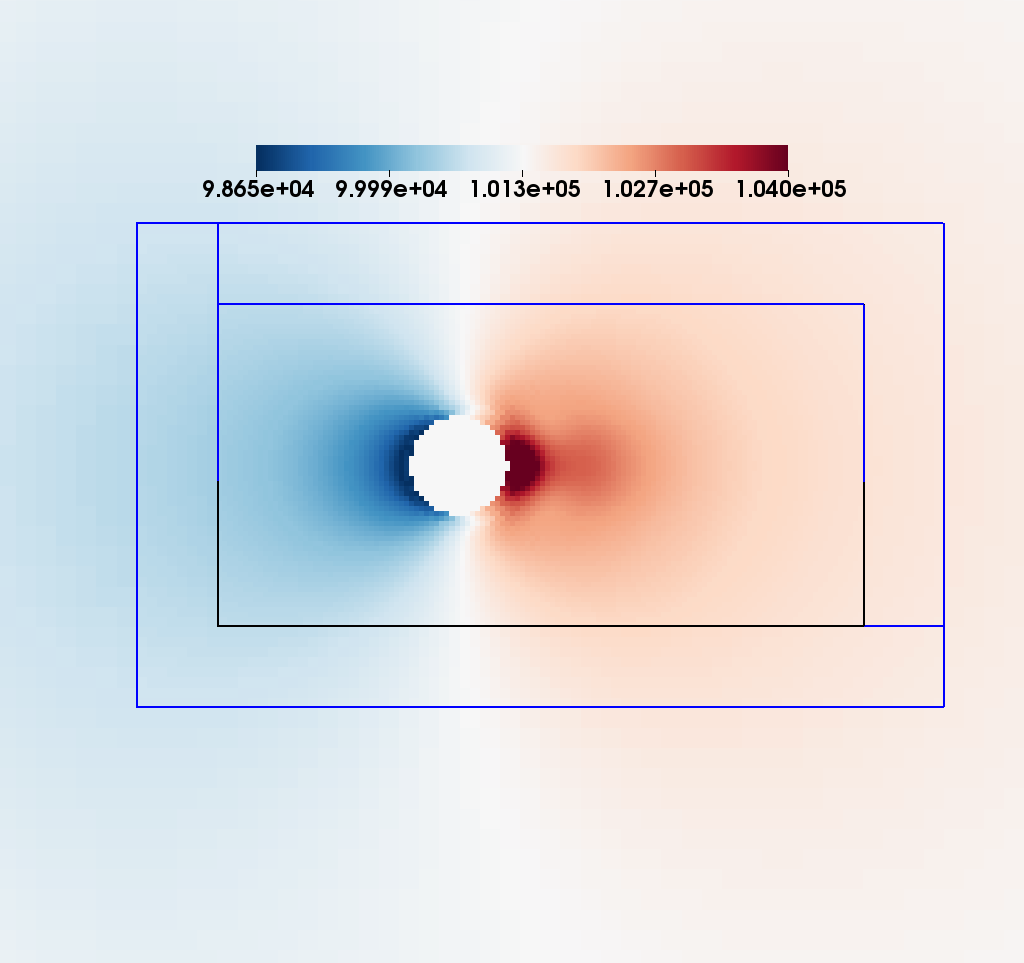}
}\\
\subfigure[]{
\includegraphics[trim=4.0cm 9.0cm 2.0cm 7.0cm, clip=true,scale=0.25]{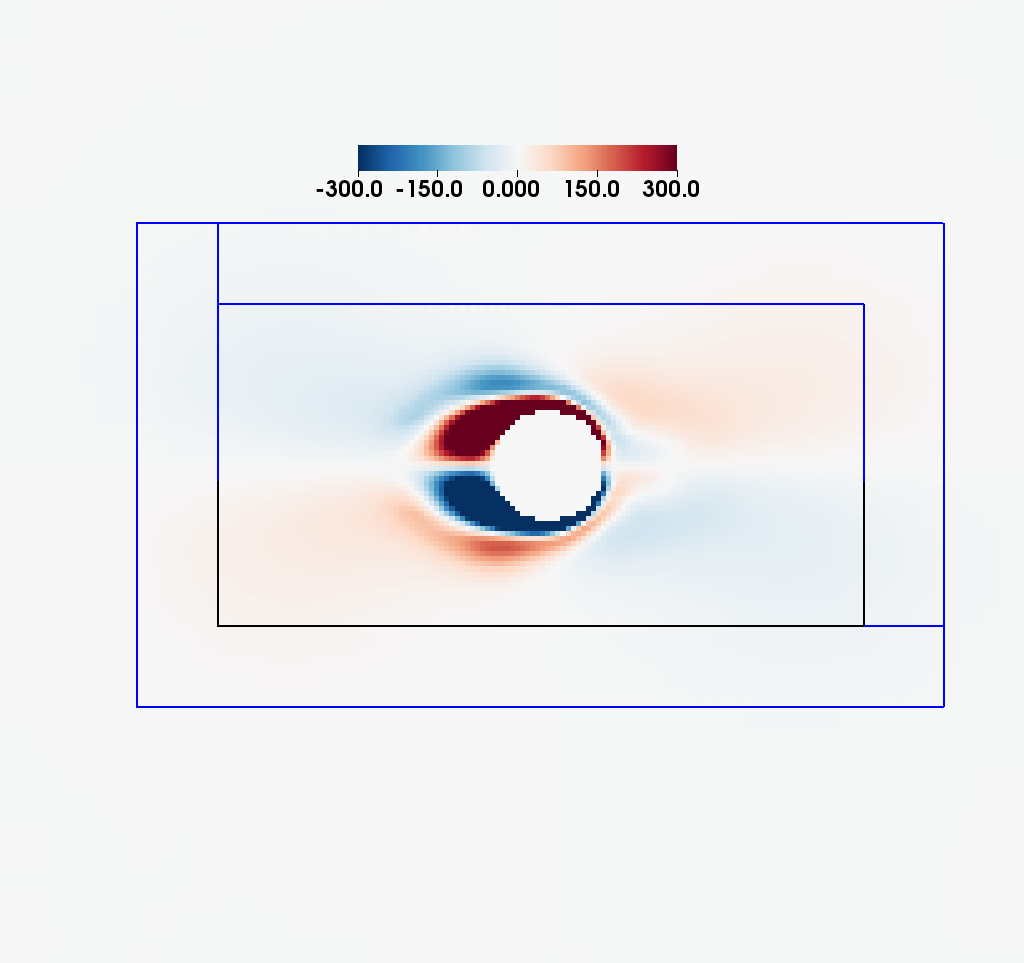}
}
\subfigure[]{
\includegraphics[trim=4.0cm 9.0cm 2.0cm 7.0cm, clip=true,scale=0.25]{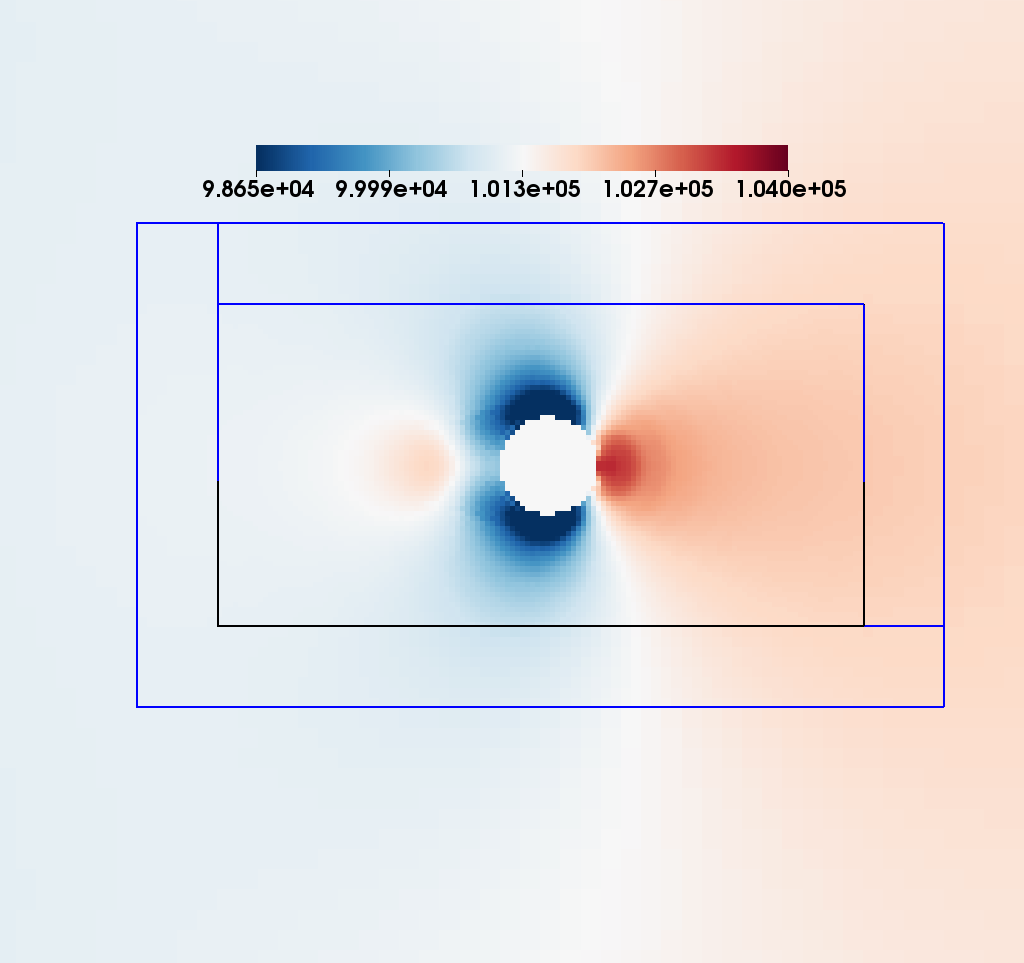}
}\\
\subfigure[]{
\includegraphics[trim=4.0cm 9.0cm 2.0cm 7.0cm, clip=true,scale=0.25]{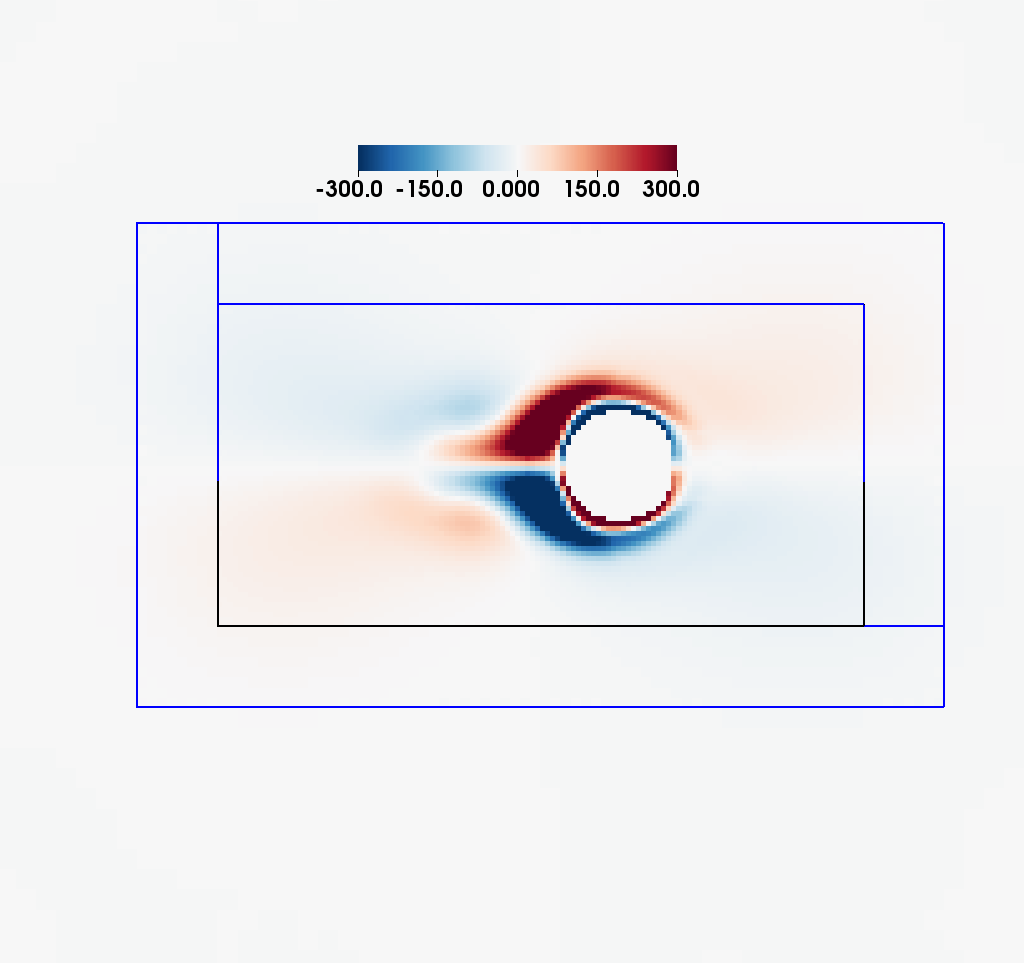}
}
\subfigure[]{
\includegraphics[trim=4.0cm 9.0cm 2.0cm 7.0cm, clip=true,scale=0.25]{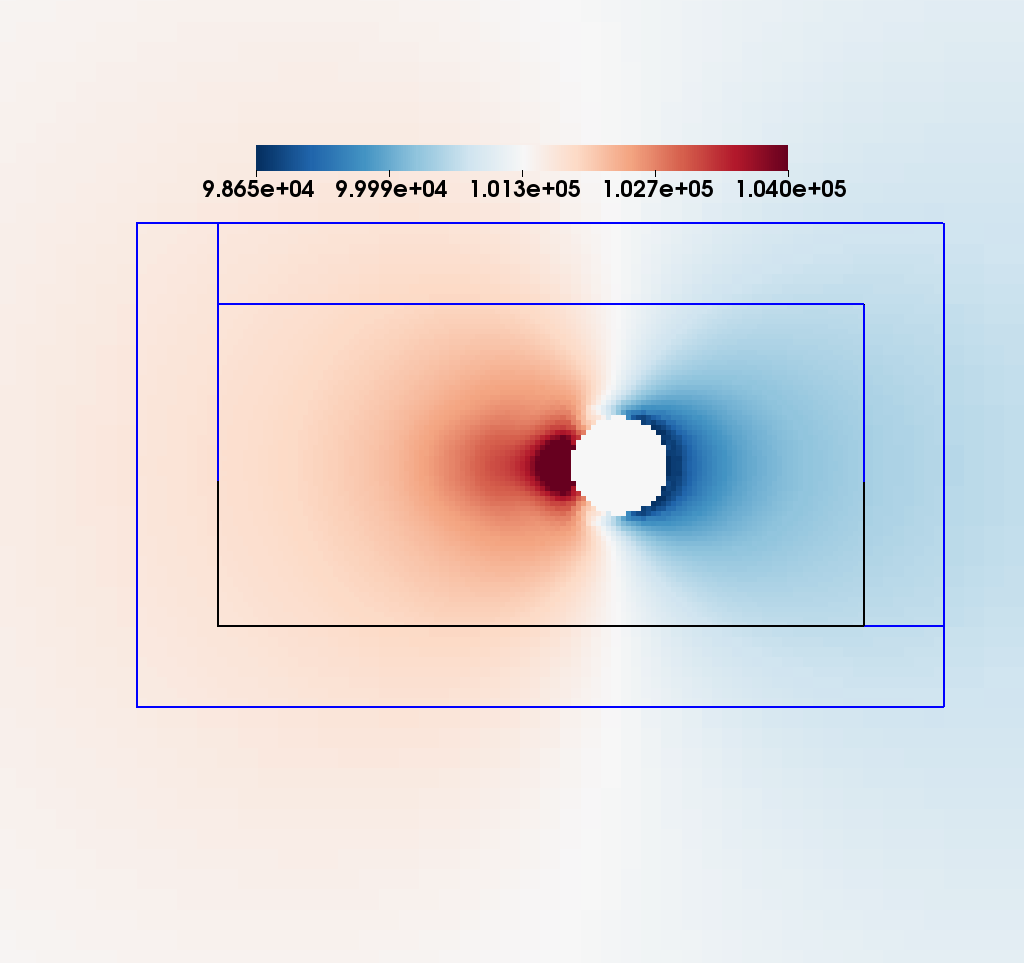}
}
\caption{Instantaneous contours of (a,c,e,g) vorticity ($s^{-1}$) and (b,d,f,h) pressure (N/m$^2$) for the inline oscillating cylinder at $Re=100$ and $KC=5$ for different phase positions 
 $\theta = 2\pi f t = 0^0, 96^0, 192^0, 288^0$ (top to bottom). The boxes show the two levels of mesh refinement above the base level. } 
\label{fig:InlineOscillatingCylinder_VortAndPres}
\end{figure}

\begin{figure}[htpb!]
\centering
\subfigure[]{
\includegraphics[trim=0.0cm 0.0cm 0.5cm 0.5cm, clip=true,scale=0.35]{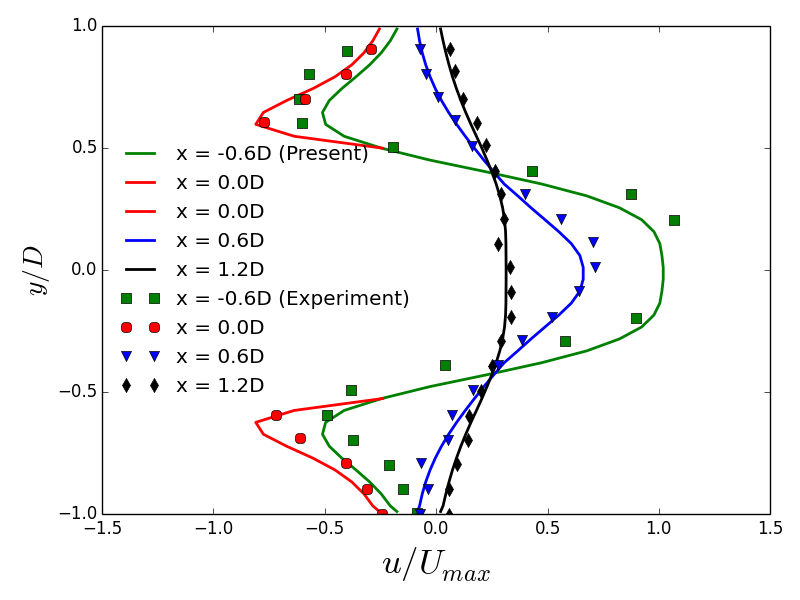}
}
\subfigure[]{
\includegraphics[trim=0.0cm 0.0cm 0.5cm 0.5cm, clip=true,scale=0.35]{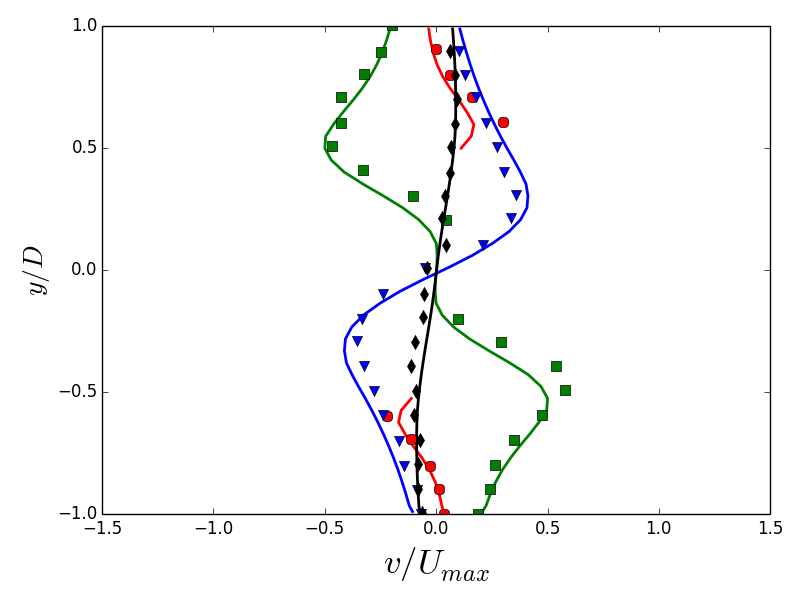}
}\\
\subfigure[]{
\includegraphics[trim=0.0cm 0.0cm 0.5cm 0.5cm, clip=true,scale=0.35]{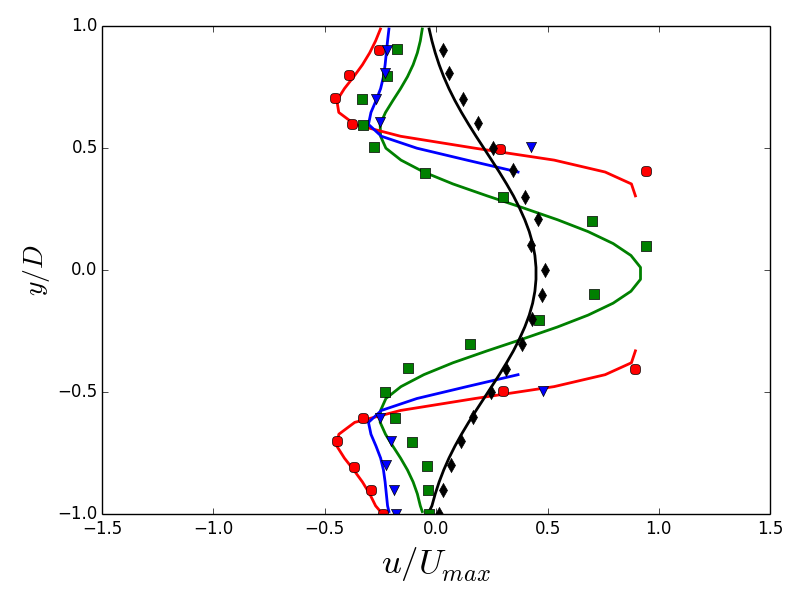}
}
\subfigure[]{
\includegraphics[trim=0.0cm 0.0cm 0.5cm 0.5cm, clip=true,scale=0.35]{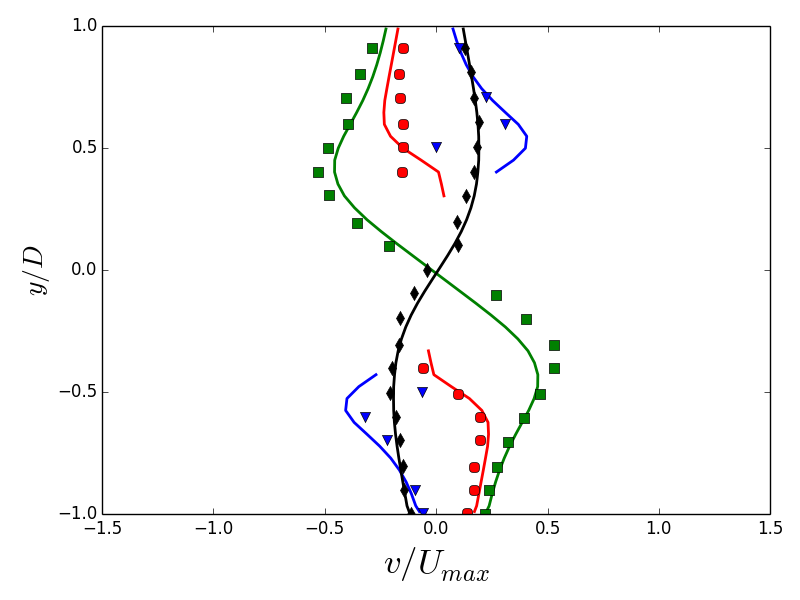}
}\\
\subfigure[]{
\includegraphics[trim=0.0cm 0.0cm 0.5cm 0.5cm, clip=true,scale=0.35]{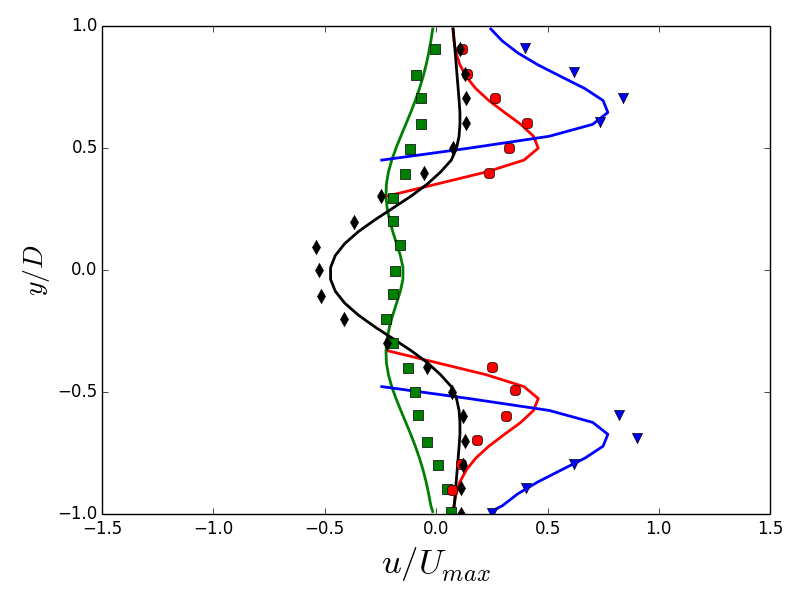}
}
\subfigure[]{
\includegraphics[trim=0.0cm 0.0cm 0.5cm 0.5cm, clip=true,scale=0.35]{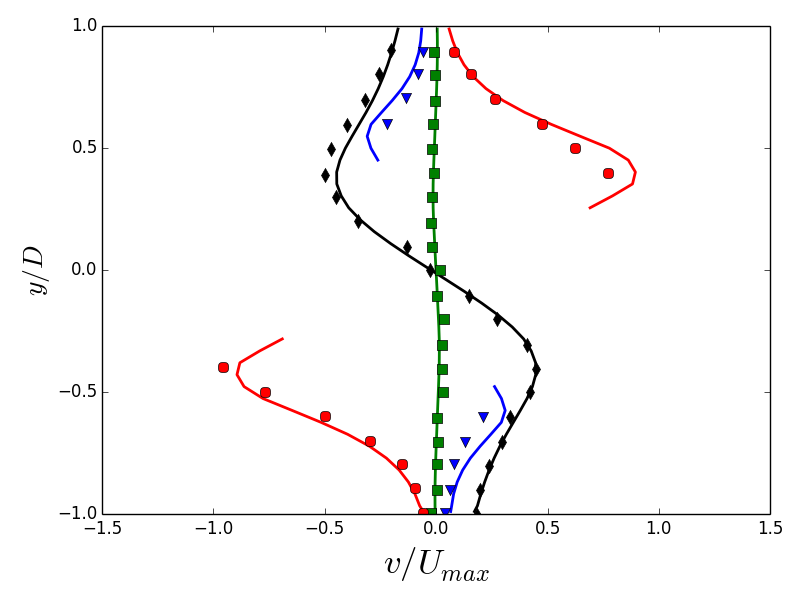}
}
\caption{Comparison of the horizontal (left) and vertical (right) components of normalized velocity at different streamwise locations as a function of the normalized vertical coordinate 
at different phase positions $\theta =  2\pi f_e t = 180^0, 210^0, 330^0$ (top to bottom), with the symbols showing the experimental results of \citet{dutsch1998low}.}
\label{fig:InlineOscillatingCylinder_Velocities}
\end{figure}

\begin{figure}[htpb!]
\centering
\includegraphics[trim=0.0cm 0.0cm 1.0cm 1.0cm, clip=true,scale=0.45]{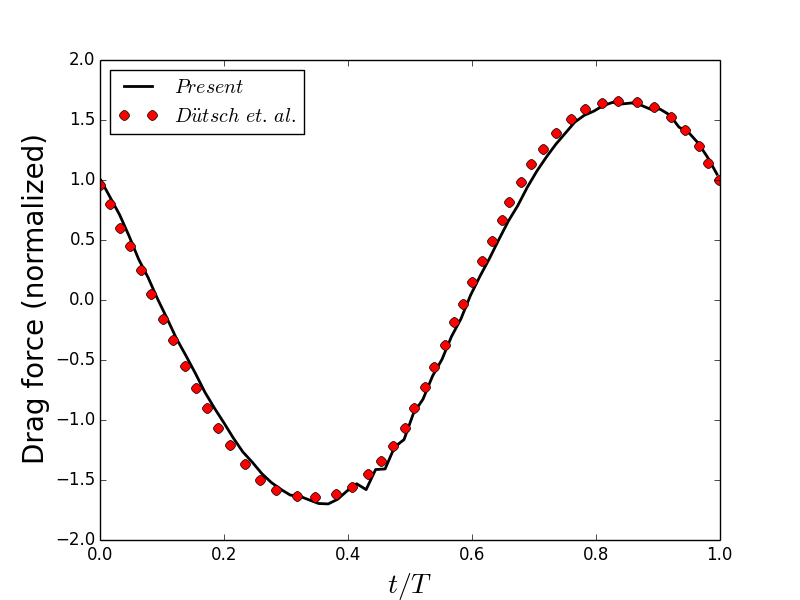}
\caption{The normalized drag force on the inline oscillating cylinder for one full cycle of oscillation.}
\label{fig:DragHistory_InlineOscillatingCylinder}
\end{figure}

\subsection{Transversely oscillating cylinder in quiescent flow at \emph{\textbf{Re}} = 185}
A transversely oscillating cylinder in a free stream of initially uniform velocity at various frequencies is a test case that has been widely studied \citep{guilmineau2002numerical}. 
The vertical position of the cylinder as a function of time is given by 
\begin{eqnarray*}
y(t) = A_e\cos(2\pi f_e t),
\end{eqnarray*}
where $A_e=0.2D$ is the amplitude of oscillation with frequency  $f_e$. The cylinder diameter is $D=0.4$ m, and the free-stream has
pressure $p=101325.0$ N/m$^2$, density $\rho_\infty=1.226$ kg/m$^3$, flow velocity $U_\infty=50$ m/s and viscosity $\mu_\infty=0.13254$ kg/ms, which gives the Reynolds number 
$Re=\rho_\infty U_\infty D/\mu_\infty=185$. The test case is performed for two different oscillation frequencies given by $f_e/f_0=0.8$ and 1.2, where $f_0$ is the natural frequency of vortex 
shedding from the cylinder. For $Re=185$, the natural frequency of oscillation, $f_0$, corresponds to a  Strouhal number $St=f_0D/U_\infty=0.195$ \citep{williamson1998series},
which gives $f_0=24.375$
cycles per second. The domain size is 20 m $\times$ 10 m, with a base mesh size of 512 $\times$ 256, with 3 levels of refinement, which gives a resolution of $\sim$82 points in the cylinder diameter. 
 A geometric refinement criterion is used to the capture the flow features in the vicinity of the cylinder, which tags all cells in the domain which satisfies $7.0 < x < 12.0$ and $-0.25<y<0.25$. 
Fig.~\ref{fig:Trans_Oscillating_Cylinder_Vort}(a)-(d) show the contours of instantaneous vorticity 
as the cylinder oscillates. The oscillation leads to a cyclic variation of the surface quantities over the cylinder.  Fig.~\ref{fig:OscillatingCylinder_Pressure_HOLS} (a) and (b) show 
the comparison of the pressure coefficient $C_p = (p-p_\infty)/(1/2\rho_\infty U_\infty^2)$ over the surface of the cylinder when the cylinder is at the extreme upper position 
for oscillation frequencies of $f_e/f_0 =$ 0.8 and 1.2 respectively, with 
the body fitted results of \citet{guilmineau2002numerical}. Fig.~\ref{fig:OscillatingCylinder_SkinFriction_HOLS} (a) and (b) show the comparison of the skin friction coefficient $C_f = \tau_f/(1/2\rho_\infty U_\infty^2)$, where $\tau_f$ is the shear stress tangential to the surface given by Eqn.~\ref{eqn:tauf}. The angle $\theta$ is measured from the stagnation 
point of the cylinder. Good quantitative comparison is observed, although minor oscillations can be seen. 

\begin{figure}[htpb!]
\subfigure[]
{
\includegraphics[trim=1.0cm 12.0cm 0.0cm 11.0cm, clip=true,scale=0.2]{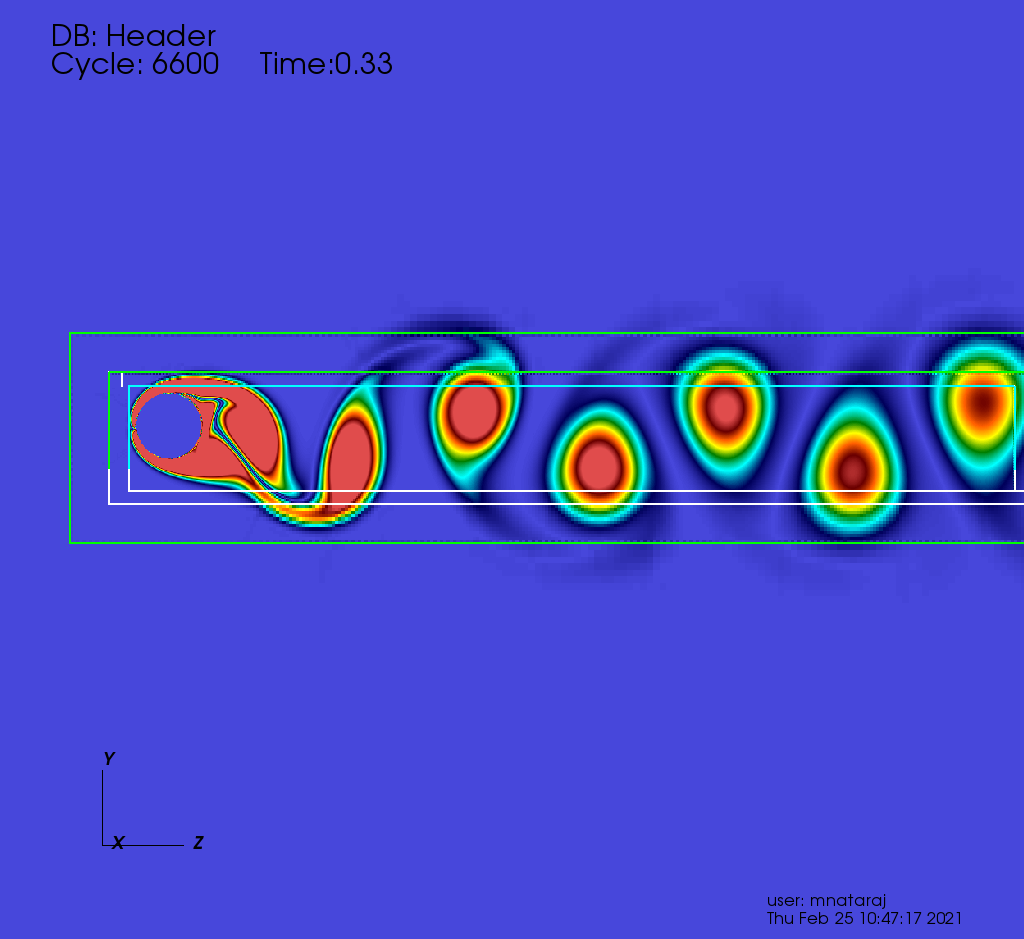}
}\quad\quad\quad\quad
\subfigure[]
{
\includegraphics[trim=1.0cm 12.0cm 0.0cm 11.0cm, clip=true,scale=0.2]{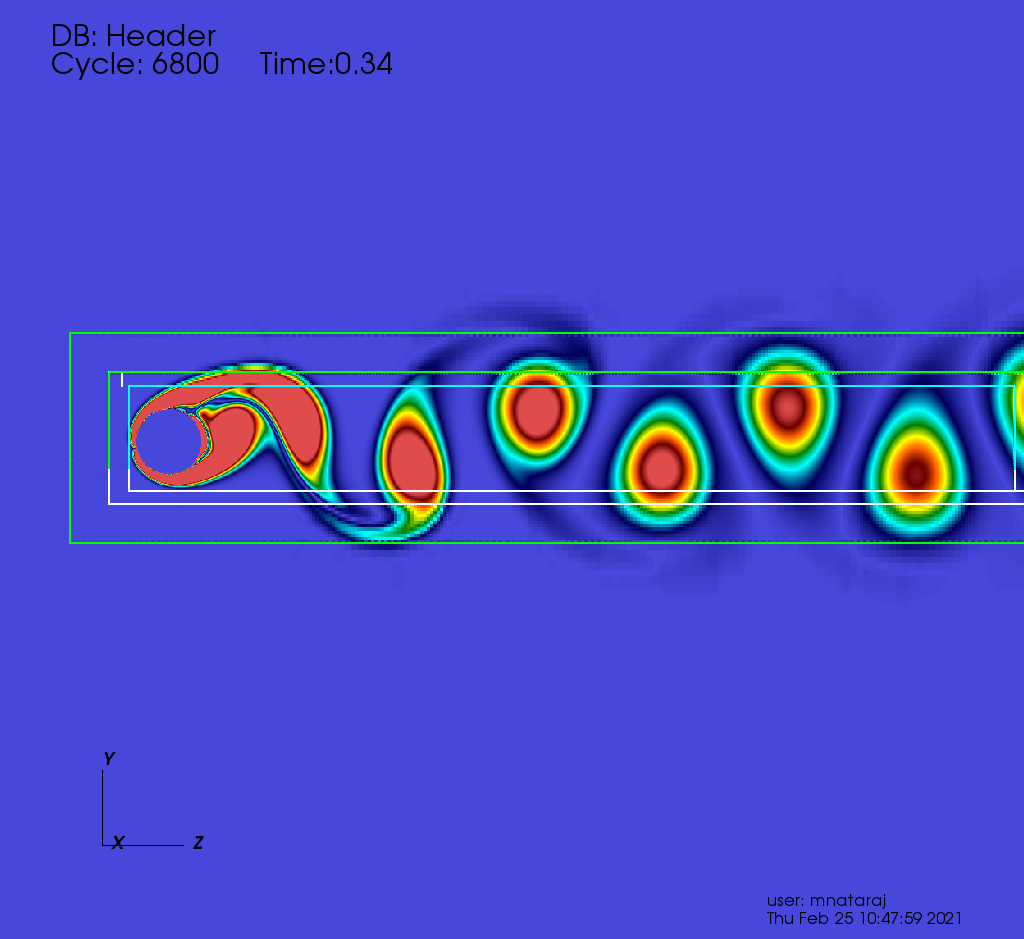}
}\\
\subfigure[]
{
\includegraphics[trim=1.0cm 12.0cm 0.0cm 11.0cm, clip=true,scale=0.2]{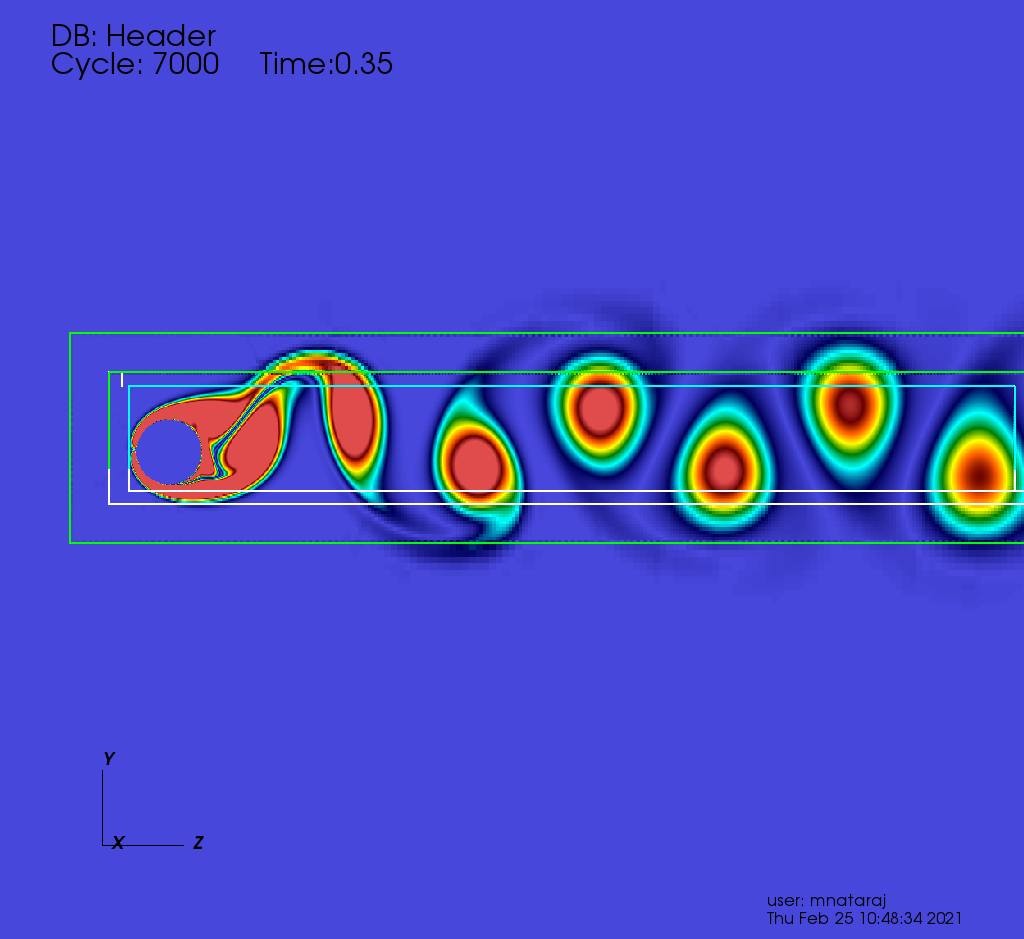}
}\quad\quad\quad\quad
\subfigure[]
{
\includegraphics[trim=1.0cm 12.0cm 0.0cm 11.0cm, clip=true,scale=0.2]{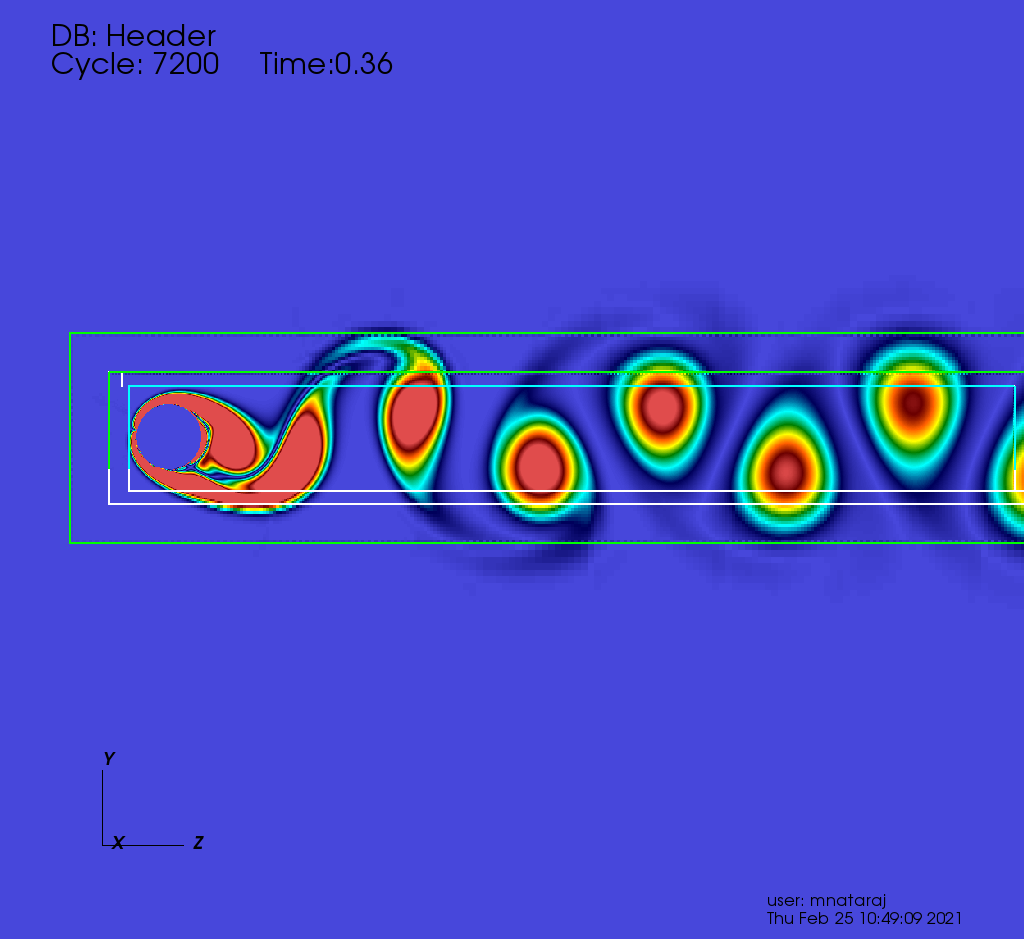}
}
\caption{Instantaneous vorticity contours for the transversely oscillating cylinder at $Re=185$. The boxes show the three levels of mesh refinement above the base level.}
\label{fig:Trans_Oscillating_Cylinder_Vort}
\end{figure}

\begin{figure}[htpb!]
\subfigure[]
{
\includegraphics[trim=5.5cm 5.8cm 6.5cm 5.5cm, clip=true,scale=0.54]{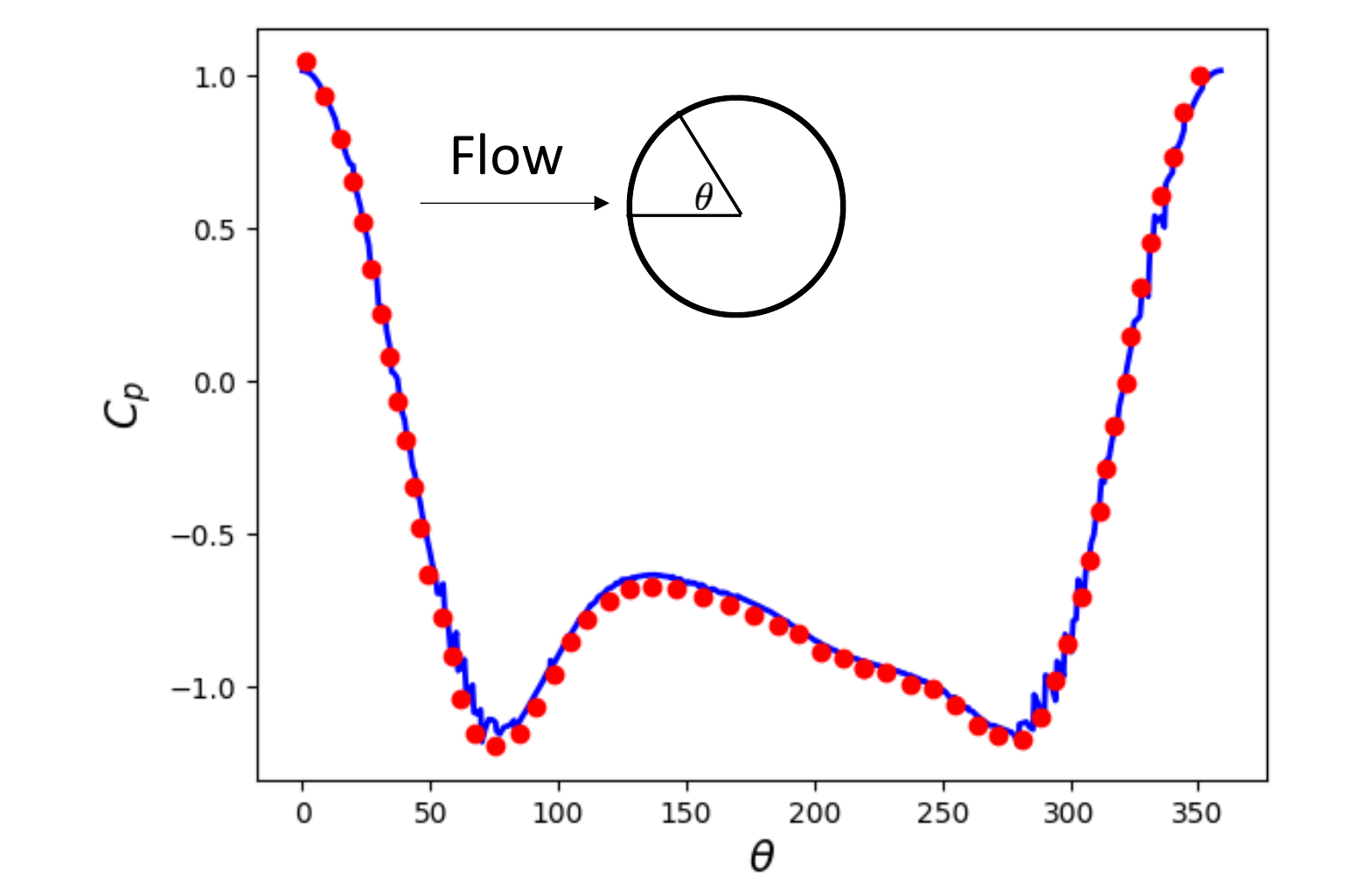}
\label{fig:OscillatingCylinder_Pressure_feqp8f0_HOLS}
}
\subfigure[]
{
\includegraphics[trim=0.1cm 0.1cm 1.0cm 1.41cm, clip=true,scale=0.5]{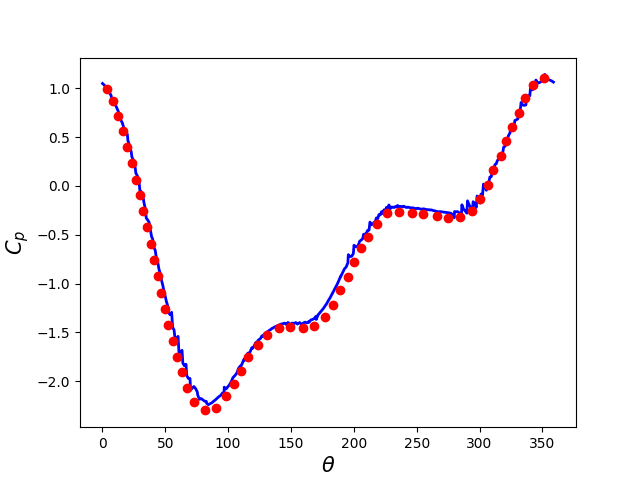}
\label{fig:OscillatingCylinder_Pressure_feq1p2f0_HOLS}
}
\caption{Comparison of the pressure coefficient on the surface of the cylinder when the cylinder is at the extreme upper position for (a) $f_e/f_0=0.8$ and (b) $f_e/f_0=1.2$. Comparison is made
with the results of \citet{guilmineau2002numerical} shown in symbols.}
\label{fig:OscillatingCylinder_Pressure_HOLS}
\end{figure}

\begin{figure}[htpb!]
\subfigure[]
{
\includegraphics[trim=6.5cm 3.5cm 8.0cm 3.5cm, clip=true,scale=0.45]{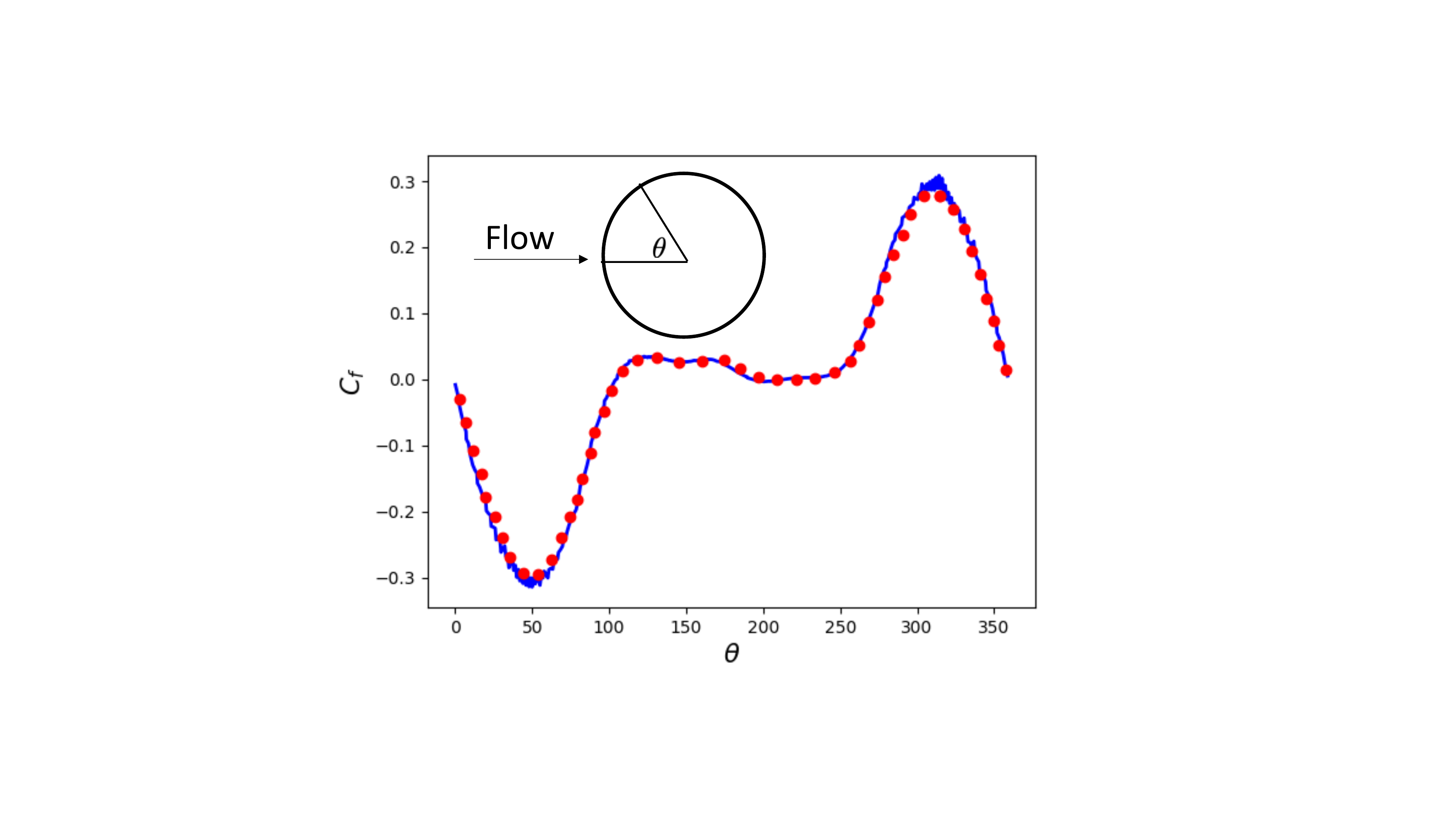}
\label{fig:OscillatingCylinder_SkinFriction_feqp8f0_HOLS}
}
\subfigure[]
{
\includegraphics[trim=0.1cm 0.1cm 1.0cm 1.41cm, clip=true,scale=0.5]{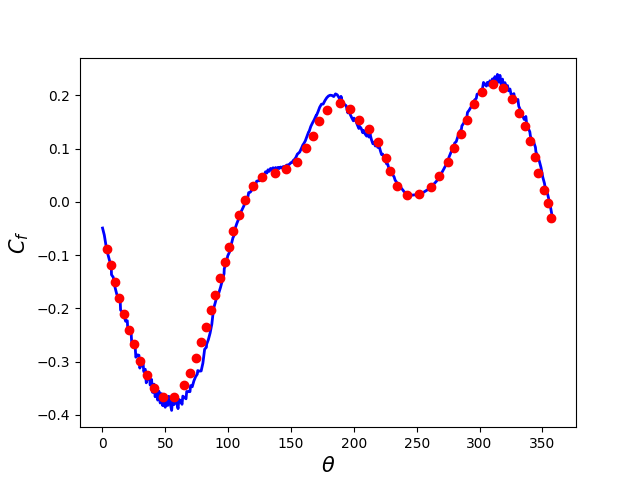}
\label{fig:OscillatingCylinder_SkinFriction_feq1p2f0_HOLS}
}
\caption{Comparison of the skin friction coefficient on the surface of the cylinder when the cylinder is at the extreme upper position for (a) $f_e/f_0=0.8$ and (b) $f_e/f_0=1.2$. Comparison is made
with the results of \citet{guilmineau2002numerical} shown in symbols.}
\label{fig:OscillatingCylinder_SkinFriction_HOLS}
\end{figure}

\section{Conclusions}\label{sec:conclusions}
A numerical framework has been developed and validated for the compressible, Navier-Stokes equations involving moving boundaries with an embedded boundary approach within the block-structured adaptive 
mesh refinement framework of AMReX. The flow solver is developed using a finite volume formulation with a conservative, unsplit, cut-cell approach, and a ghost-cell 
approach has been developed for computing the inviscid fluxes on the moving, embedded boundary faces. A 3$^\mathrm{rd}$ order least squares method was used to approximate 
the gradients of velocities at the EB faces in the computation of the viscous fluxes. The algorithm is validated against analytical and experimental results, and good quantitative 
comparison is observed. Simulations of shock-cylinder interaction and shock-wedge interaction with adaptive mesh refinement showed the capability of the algorithm to handle high-speed 
flows with high gradient regions such as shock waves, and flows with sharp corners. The transonic buffet phenomenon of an oscillating NACA 0012 airfoil was simulated, and the variation of the coefficient of lift with the angle of attack showed good quantitative comparison with previous results in literature. As a test of the conservative nature of the scheme, a closed system was simulated -- an  oscillating piston in a cylinder. The percentage error in mass inside the cylinder was found to decrease with refinement, demonstrating that the scheme is conservative. Viscous test cases of a horizontally moving cylinder, inline oscillating cylinder and a transversely oscillating cylinder were performed, and surface quantities -- pressure and skin friction coefficients, were computed and were observed to have good quantitative comparison with results in the literature. 

\section*{Acknowledgments}
This research was supported by the Exascale Computing Project (17-SC-20-SC), a collaborative effort of the U.S. Department of Energy Office of Science and the National Nuclear Security 
Administration. We also gratefully acknowledge the staff at NREL-HPC for the compute time on the Eagle supercomputer, and their continued support.

\clearpage

\section*{Appendix}
The idea behind computing gradients at a point (green square in Fig.~\ref{fig:EBGradient_LS}) using the least squares technique is to minimize the cumulative error in the fit to a function $\phi(x,y,z)$ over a chosen neighborhood region, with 
the minimization being done with respect to the gradient quantities at the point in consideration. The value of the function at any point in the neighborhood can be written using a 
Taylor series expansion about the point at which the gradient needs to be computed. This estimated value will differ from the actual value at the point and the difference between them is the error 
estimate. The order of the terms retained in the expansion will determine the order of the least squares approximation. For 
a 3$^\mathrm{rd}$ order least squares method, the cumulative error that needs to be minimized is given by 
\begin{eqnarray*}
E = \sum\limits_{i=1}^N \Big(\phi_i - \Big(\phi_0 &+& \phi_x \Delta x_i + \phi_y\Delta y_i + \phi_z\Delta z_i \\
                                                     &+& \phi_{xx}\Delta x_i^2 + \phi_{yy}\Delta y_i^2 + \phi_{zz}\Delta z_i^2\\
                                                     &+& \phi_{xy}\Delta x_i\Delta y_i + \phi_{yz}\Delta y_i\Delta z_i + \phi_{zx}\Delta z_i\Delta x_i\Big)\Big)^2,
\end{eqnarray*}
where $i$ loops over the neighborhood region. In Fig.~\ref{fig:EBGradient_LS}, the neighborhood region consists of the blue circles, which are the centroid of the fluid volumes (the cell containing the 
face on which the gradient is computed is avoided in the neighborhood region \citep{schwartz2006cartesian}). This error is now minimized with respect to each of the gradient quantities. For eg. 
$\cfrac{\partial E}{\partial \phi_x} = 0$ gives
\begin{eqnarray*}
\sum\limits_{i=1}^N (\phi_i -\phi_0)\Delta x_i - \Delta x_i&\times&\Big(\phi_x \Delta x_i + \phi_y\Delta y_i + \phi_z\Delta z_i \\
                                                     &+& \phi_{xx}\Delta x_i^2 + \phi_{yy}\Delta y_i^2 + \phi_{zz}\Delta z_i^2\\
                                                     &+& \phi_{xy}\Delta x_i\Delta y_i + \phi_{yz}\Delta y_i\Delta z_i + \phi_{zx}\Delta z_i\Delta x_i\Big) = 0,
\end{eqnarray*}
which can be rearranged to give
\begin{eqnarray*}
\sum\limits_{i=1}^N \Delta x_i&\times&\Big(\phi_x \Delta x_i + \phi_y\Delta y_i + \phi_z\Delta z_i \\
&+& \phi_{xx}\Delta x_i^2 + \phi_{yy}\Delta y_i^2 + \phi_{zz}\Delta z_i^2\\
&+& \phi_{xy}\Delta x_i\Delta y_i + \phi_{yz}\Delta y_i\Delta z_i + \phi_{zx}\Delta z_i\Delta x_i\Big) = \sum\limits_{i=1}^N (\phi_i -\phi_0)\Delta x_i.
\end{eqnarray*}
Repeating the above procedure for each of the 9 gradient quantities -- $\phi_x$, $\phi_y$, $\phi_z$, $\phi_{xx}$, $\phi_{yy}$, $\phi_{zz}$, $\phi_{xy}$, $\phi_{yz}$ and $\phi_{zx}$, gives a 
9$\times$ 9 system of equations as 
\begin{equation*}
\begin{pmatrix}
\sum\Delta x_i\times\Big[\Delta x_i\quad\Delta y_i\quad\Delta z_i\quad\Delta x_i^2\quad\Delta y_i^2\quad\Delta z_i^2\quad\Delta x_i\Delta y_i\quad\Delta y_i\Delta z_i\quad\Delta z_i\Delta x_i \Big]\\
\\
\sum\Delta y_i\times\Big[\Delta x_i\quad\Delta y_i\quad\Delta z_i\quad\Delta x_i^2\quad\Delta y_i^2\quad\Delta z_i^2\quad\Delta x_i\Delta y_i\quad\Delta y_i\Delta z_i\quad\Delta z_i\Delta x_i \Big]\\
\\
\sum\Delta z_i\times\Big[\Delta x_i\quad\Delta y_i\quad\Delta z_i\quad\Delta x_i^2\quad\Delta y_i^2\quad\Delta z_i^2\quad\Delta x_i\Delta y_i\quad\Delta y_i\Delta z_i\quad\Delta z_i\Delta x_i \Big]\\
\\
\sum\Delta x_i^2\times\Big[\Delta x_i\quad\Delta y_i\quad\Delta z_i\quad\Delta x_i^2\quad\Delta y_i^2\quad\Delta z_i^2\quad\Delta x_i\Delta y_i\quad\Delta y_i\Delta z_i\quad\Delta z_i\Delta x_i \Big]\\
\\
\sum\Delta y_i^2\times\Big[\Delta x_i\quad\Delta y_i\quad\Delta z_i\quad\Delta x_i^2\quad\Delta y_i^2\quad\Delta z_i^2\quad\Delta x_i\Delta y_i\quad\Delta y_i\Delta z_i\quad\Delta z_i\Delta x_i \Big]\\
\\
\sum\Delta z_i^2\times\Big[\Delta x_i\quad\Delta y_i\quad\Delta z_i\quad\Delta x_i^2\quad\Delta y_i^2\quad\Delta z_i^2\quad\Delta x_i\Delta y_i\quad\Delta y_i\Delta z_i\quad\Delta z_i\Delta x_i \Big]\\
\\
\sum\Delta x_i\Delta y_i\times\Big[\Delta x_i\quad\Delta y_i\quad\Delta z_i\quad\Delta x_i^2\quad\Delta y_i^2\quad\Delta z_i^2\quad\Delta x_i\Delta y_i\quad\Delta y_i\Delta z_i\quad\Delta z_i\Delta x_i \Big]\\
\\
\sum\Delta y_i\Delta z_i\times\Big[\Delta x_i\quad\Delta y_i\quad\Delta z_i\quad\Delta x_i^2\quad\Delta y_i^2\quad\Delta z_i^2\quad\Delta x_i\Delta y_i\quad\Delta y_i\Delta z_i\quad\Delta z_i\Delta x_i \Big]\\
\\
\sum\Delta z_i\Delta x_i\times\Big[\Delta x_i\quad\Delta y_i\quad\Delta z_i\quad\Delta x_i^2\quad\Delta y_i^2\quad\Delta z_i^2\quad\Delta x_i\Delta y_i\quad\Delta y_i\Delta z_i\quad\Delta z_i\Delta x_i \Big]\\
\end{pmatrix}
\begin{pmatrix}
\cfrac{\partial \phi}{\partial x}\\
\\
\cfrac{\partial \phi}{\partial y}\\
\\
\cfrac{\partial \phi}{\partial z}\\
\\
\cfrac{\partial^2\phi}{\partial x^2}\\
\\
\cfrac{\partial^2\phi}{\partial y^2}\\
\\
\cfrac{\partial^2\phi}{\partial z^2}\\
\\
\cfrac{\partial^2\phi}{\partial x\partial y}\\
\\
\cfrac{\partial^2\phi}{\partial y\partial z}\\
\\
\cfrac{\partial^2\phi}{\partial z\partial x}\\
\end{pmatrix}
=
\begin{pmatrix}
\sum\limits_{N} (\phi_i - \phi_0)\Delta x_i\\
\\
\sum\limits_{N} (\phi_i - \phi_0)\Delta y_i\\
\\
\sum\limits_{N} (\phi_i - \phi_0)\Delta z_i\\
\\
\sum\limits_{N} (\phi_i - \phi_0)\Delta x_i^2\\
\\
\sum\limits_{N} (\phi_i - \phi_0)\Delta y_i^2\\
\\
\sum\limits_{N} (\phi_i - \phi_0)\Delta z_i^2\\
\\
\sum\limits_{N} (\phi_i - \phi_0)\Delta x_i\Delta y_i\\
\\
\sum\limits_{N} (\phi_i - \phi_0)\Delta y_i\Delta z_i\\
\\
\sum\limits_{N} (\phi_i - \phi_0)\Delta z_i\Delta x_i\\
\end{pmatrix},
\end{equation*}
which is solved using LAPACK \citep{anderson1990lapack} to obtain the gradients at the point.

For a 2$^\mathrm{nd}$ order least squares method, the system of equations is given by
\begin{equation*}
\begin{pmatrix}
\sum\limits_{N} \Delta x_i^2 & \sum\limits_{N} \Delta x_i\Delta y_i & \sum\limits_{N} \Delta x_i\Delta z_i \\
& & \\
\sum\limits_{N} \Delta x_i\Delta y_i & \sum\limits_{N} \Delta y_i^2 & \sum\limits_{N} \Delta y_i\Delta z_i \\
& & \\
\sum\limits_{N} \Delta x_i\Delta z_i & \sum\limits_{N} \Delta y_i\Delta z_i & \sum\limits_{N} \Delta z_i^2
\end{pmatrix}
\begin{pmatrix}
\cfrac{\partial \phi}{\partial x}\\
\\
\cfrac{\partial \phi}{\partial y}\\
\\
\cfrac{\partial \phi}{\partial z}
\end{pmatrix}
=
\begin{pmatrix}
\sum\limits_{N} (\phi_i - \phi_0)\Delta x_i\\
\\
\sum\limits_{N} (\phi_i - \phi_0)\Delta y_i\\
\\
\sum\limits_{N} (\phi_i - \phi_0)\Delta z_i
\end{pmatrix}.
\end{equation*}
}

\clearpage
\bibliographystyle{plainnat}
\bibliography{Master}
\bibliographystyle{unsrt}






\end{document}